\documentclass[12pt]{iopart}
\pdfoutput=1

\usepackage[pdftex]{graphicx}
\usepackage{amsfonts}
\usepackage{color}
\usepackage{framed}
\usepackage{colortbl}

\newcommand{\addspan}[1]{#1}
\newcommand{\delspan}[1]{}

\usepackage{lscape}

\def\Izag{{\em Izanagi}}
\def\Izam{{\em Izanami}}
\def\Iwat{{\em Iwatsuchibiko}}
\def\Shin{{\em Shinatsuhiko}}
\def\Haya{{\em Hayaakitsuhime}}
\def\Parent{{\mathbb P}}
\def\parent{{\mathrm P}}

\def\RankD{{$[\ll]$     }}
\def\RankC{{$[\leq]$    }}
\def\RankB{{$[\simeq]$  }}
\def\RankA{{$[\gg]$     }}
\def\mathRankD{{[\ll]     }}
\def\mathRankC{{[\leq]    }}
\def\mathRankB{{[\simeq]  }}
\def\mathRankA{{[\gg]     }}

\definecolor{grey}{rgb}{0.8, 0.8, 0.8}

\begin{document}

\title[Paraiso]{Paraiso : An Automated Tuning Framework for Explicit Solvers of Partial Differential Equations}

\author{Takayuki Muranushi}

\address{
Hakubi Center for Advanced Research / 
Yukawa Institute for Theoretical Physics, Kyoto University
Kitashirakawa Oiwakecho, Sakyo-ku, Kyoto 606-8502 Japan}
\ead{muranushi.takayuki.3r@kyoto-u.ac.jp}
\begin{abstract}

We propose Paraiso, a domain specific language embedded in functional
programming language Haskell, for automated tuning of explicit solvers
of partial differential equations (PDEs) on Graphic Processing Units
(GPUs) as well as multicore CPUs. In Paraiso, one can describe PDE
solving algorithms succinctly using tensor equations notation.
Hydrodynamic properties, interpolation methods and other building
blocks are described in abstract, modular, re-usable and combinable
forms, which lets us generate versatile solvers from little set of
Paraiso source codes.

We demonstrate Paraiso by implementing a compressive hydrodynamics
solver.  A single source code less than 500 lines can be used to
generate solvers of arbitrary dimensions, for both multicore CPUs and
GPUs.  We demonstrate both manual annotation based tuning and
\addspan{evolutionary computing based} automated tuning of the program.

\end{abstract}

\maketitle

\section{Introduction}
Today, computer architectures are becoming more and more complex,
making it more and more difficult to predict the performance of a
program before running it.  Parallel architectures, GPUs being one
example and distributed memory machines another, forces us to
program in different ways and our programs tend to become longer.
Moreover, to optimize them we are asked to write many different
implementations of such programs, and the coding task becomes
time-consuming.  However, if we describe the problem we want to solve
in a domain-specific language (DSL) from which a lot of possible
implementations are generated and benchmarked, we can automate the
tuning processes.

Examples of such automated-tuning DSL approaches are found in fast
Fourier transformation library FFTW \cite{frigo_design_2005}, linear
algebra library ATLAS \cite{clint_whaley_automated_2001}, digital
signal processing library SPIRAL \cite{puschel_spiral:_2005}.  In
these works the authors regarded automated tuning not just as a tool to avoid
manual tuning, but as a necessary tool to have ``portable
performance'' --- the practical way of optimizing domain-specific
codes for various complicated architecture we have today and in
the future.

Now, our domain is explicit solvers of partial differential equations
(PDEs).  The main portions of such solvers consist of stencil
computations, and a few global reduction operations needed to
calculate Courant-Friedrichs-Lewy (CFL) conditions.  Stencil
computations are computations that update mesh structures, and the
next state of a mesh depends only on the states of its neighbor
meshes.  Due to its inherent and coherent parallelism, stencil
computation DSLs have been actively studied.  For example,
\cite{datta_stencil_2008} have demonstrated $\times 1.5 - \times 5.6$
speedup on various hardwares via hardware and memory-hierarchy aware
automated tuning.  \cite{maruyama_physis:_2011} have demonstrated
generating multi-GPU codes that weakly scales up to 256 GPUs.

Still, reports are limited to simple equations such as diffusion
equations and Jacobi solvers of Poisson equations, and to implement
solvers of more complicated equations such as compressive
hydrodynamics, magneto-hydrodynamics or general relativity, and their
higher-order versions, we are forced to manually decompose the
solving algorithms to imperative instructions that are often tens and
hundred thousands of lines. 

It is a problem common to all the languages that is designed to be
compatible with C or Fortran, such as OpenMP, CUDA, PGI Fortran and
OpenACC --- that it is hard, if not impossible at all, to avoid
repeating yourself in these languages. Despite all the efforts made so
far to extend these languages, the programs in those languages
fundamentally lack the ability to manipulate programs themselves and
other abstract concepts. Every time a new generation of language
appear, we are forced to make painful choices of porting a huge
amounts of legacy codes to the new language.  It is also a pain that
various numerical techniques are mixed in one code, and can hardly be
reused. If we want to make computers automatically combine such
numerical techniques, compose a variety of implementations, and search
for the better ones, the abstraction power is necessary.  A
complementary approach is needed here, that works together with the
parallel languages.

Our contribution, Paraiso ({\bf PAR}allel {\bf A}utomated {\bf
  I}ntegration {\bf S}cheme {\bf O}rganizer ), enables us describe
PDE-solving algorithms succinctly using algebraic concepts such as
tensors.  Paraiso also enables us to describe various manipulations on
algorithms such as introducing higher-order interpolations, in
composable and reusable manner. That is, once we define a certain
interpolation method in our language as a transformation of basic
solver to a higher-order one, we can reuse the transformation for any
equations. In this way Paraiso reduces the cost of rewriting when we
search for better discretizations or interpolations; enables us to
port the collections of algorithms to new parallel languages, without
changing the Paraiso source codes but by updating the common code
generator; allows yet another layer of automated tuning at translating
Paraiso codes to other parallel languages.

In writing Paraiso, we wanted to define arithmetic operations between
tensors and code generator fragments. We want our tensors to be
polymorphic, but we don't want to allow addition between tensors of
different dimensions.  We wanted to make generalized functional
applications. We wanted to traverse over data structures.  We need to
manage various contexts, like context of code generation and the
context of serialization from a program to a genome.

Programming language Haskell supports all of these.  Moreover, they
are supported {\em not} as built-in features that users cannot change;
they are libraries that Haskell users can freely combine or create
their own.  This flexibility of Haskell is based on its strong,
static, higher-order type system with type classes and type inference.
Thus Haskell essentially allows us to develop our own type-systems
within it, which gives it a unique advantage as a platform for
developing embedded DSLs. Many parallel and distributed programming
languages has been implemented using Haskell
\cite{trinder_parallel_2002}.
Nepal\cite{chakravarty_nepalnested_2001} and Data Parallel Haskell
\cite{jones_harnessing_2008} are implementations of NESL, a language
for operating nested arrays in Haskell.  Accelerate
\cite{chakravarty_accelerating_2011} and Nikola
\cite{mainland_nikola:_2010} are languages to manipulate arrays on
GPUs written in Haskell. Finally, Liszt \cite{devito_liszt:_2011} is a
DSL for solving mesh-based PDEs based on functional
programming language Scala.

Our contributions are the following:
\begin{itemize}
  \item A domain-specific language (DSL) embedded in Haskell, with which one
    can describe explicit solvers of partial differential equations (PDEs) in
    a succinct and organized manner, using tensor notations.
  \item A code-generation mechanism that takes the DSL and generates
    OpenMP and CUDA programs.
  \item A compressible Euler equations solver implemented in the DSL, and
    tuning experiments using the solver. Written in tensor notations,
    the solver can be applied to problems of arbitrary dimension
    without changing the source code at all but a single type
    declaration that sets the dimension of the solver.
  \item An annotation mechanism with which one can give hints to code
    generators, which makes it drastically easy to search for better
    implementations manually. By adding just one line of hint to the
    solver causes overall refactoring, adding 4 subroutines to the
    generated code, making it consume 1.3 times more memory but $6.42$
    times faster.
  \item An automated benchmarking and tuning mechanism based on
    parallel simulated annealing and genetic algorithms, which
    generates a lot of different implementations of the PDE solvers
    and search for faster implementations. It speeds up the
    unannotated solver by factor of $5.26$, and the annotated solver
    further by factor of $1.78$.
\end{itemize}

Just for a comparison, the Paraiso framework is about 5'000 lines of
code in Haskell, and the compressible Euler equations solver
implemented in Paraiso is less than 500 lines. From that, we have
generated more than 500'000 instances of the solver, each being 3'000
- 10'000 lines of code in CUDA. The automatically tuned codes are
faster than the manually tuned codes reported by other groups. Paraiso
is ready to optimize solvers of other equations, and all the solvers
written in Paraiso can migrate to new parallel languages and new
hardwares once the Paraiso code generator supports them. 

\delspan{%
  With Paraiso, we demonstrate the utility of DSL and automated
  tuning techniques in the domain of explicit solvers of PDEs, too.
}
\addspan{ 
  Our work, Paraiso is the first  consistent system that combine
  those previously studied techniques of 
  symbolic computations, DSLs, GPU computations and automated tuning.
  We demonstrate the utility of such a system in the domain of 
  explicit solvers of PDEs.  
}

This paper is organized as follows.
In section
\ref{sectionCodeGeneration}, we describe the overall design of Paraiso as well as its components.
In section
\ref{sectionAutoTuning}, we describe our automated tuning mechanism, 
which is a combination of  genetic algorithm and simulated annealing that is designed
to utilize the varying number of available nodes in a shared computer system.
In section \ref{sectionExperiment}, we introduce the compressible hydrodynamics solver
we choose as the tuning target, and describe the manual and automated tuning experiments.
In section \ref{sectionAnalysis}, we analyze the automated tuning history.
In section \ref{sectionConclusion} are concluding remarks and discussions.

\section{The Design of Paraiso} \label{sectionCodeGeneration}

\begin{figure}
  \begin{center}
    \includegraphics[width=10cm,angle=270]{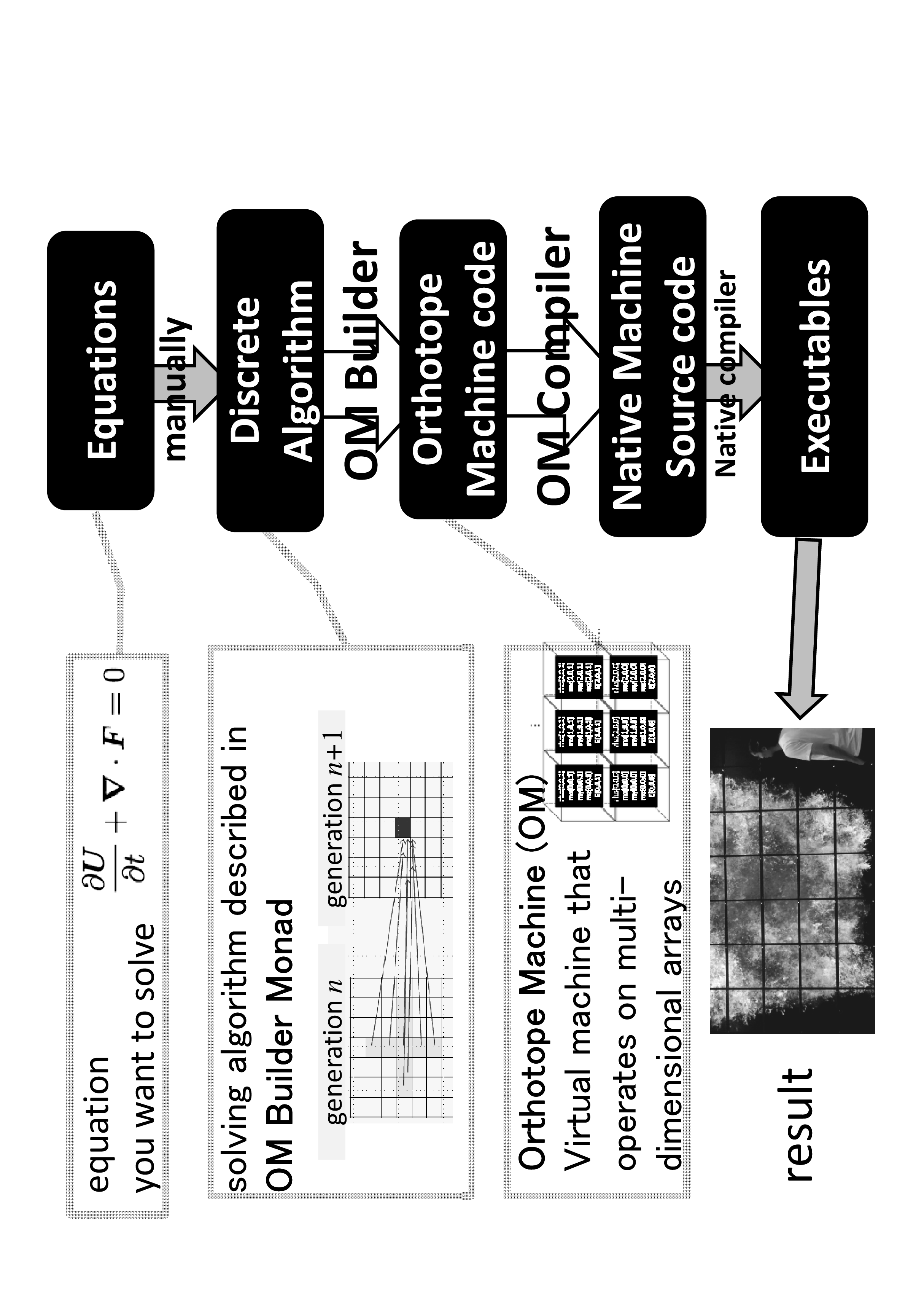}
    \caption {
      The Overall Picture.
    }
    \label{figureOverall}
  \end{center} 
\end{figure}

\subsection{Overall Design}

Paraiso is to tackle the ambitious problem of generating fast and
massively-parallel codes from human-friendly notations of algorithms.
To divide the problem into major components, which may be conquered by
different set of people, we set the overall design of Paraiso,
as illustrated in Fig. \ref{figureOverall}.

The users of Paraiso must manually invent a discretized algorithm for the partial
differential equations they want to solve.  The solving algorithms are
described in Paraiso using {\tt Builder Monad}s.  {\tt Builder Monad}s
generate programs for {\tt Orthotope Machine} (OM), a virtual parallel
machine designed to denote parallel computations on multi-dimensional
arrays. Then the back end OM compiler generates the native machine
source codes such as C++ and CUDA. Finally, native compilers translate
them to executable codes.

When \delspan{a physicist}someone translates a simulation algorithm in his mind, to
different languages such as Fortran, C and CUDA, he starts from
building mathematical notations in his mind, and then gradually
decomposes it to machine level. There should be the last common level
that is independent of the detail of the target languages or the
target hardwares. The level consist of very primitive operations on
arrays.  Orthotope Machine (OM) is designed to capture this level: The
instruction set of the OM is kept as compact as possible, while
maintaining all the parallelisms found in the solving algorithms.  In
forthcoming technical report we plan to provide the formal definition
of the Orthotope Machine. 
\delspan{%
as well as the translation of OM to lightweight markup
languages, so that OM will serve as the interface between numerical
simulation scientists and language/hardware scientists.
}

\begin{table}
  \begin{center}
    \small
    \begin{tabular}{|p{6.75cm}|p{8.25cm}|}
      \hline
      \multicolumn{1}{|c|}{Name and Module} &
      \multicolumn{1}{c|}{Description} \\
      \hline
      \begin{tabular}{l}
        {\tt \verb@data :~@}, {\tt data Vec}, \\
        {\tt class Vector}, {\tt class VectorRing}\\ 
        in {\tt Data.Tensor.TypeLevel}
      \end{tabular}
      &
      \begin{tabular}{l}
        Tensor algebra library that provides type level \\
        information for the tensor rank and dimension. \\
        c.f. \S \ref{sectionTLT}.
      \end{tabular}
      \\
      \hline
      \begin{tabular}{l}
        {\tt type Builder} in\\
         {\tt Language.Paraiso.OM.Builder}
      \end{tabular}
      &
      \begin{tabular}{l}
        The Builder Monad for constructing the data \\
        flow graphs for OMs. c.f. \S \ref{sectionBuilder}.
      \end{tabular}
      \\
      \hline
      \begin{tabular}{l}
        {\tt data OM} in\\
        {\tt Language.Paraiso.OM}
      \end{tabular}
      &
      \begin{tabular}{l}
        The Orthotope Machine(OM), a virtual machine \\
        with basic instructions for stencil computations\\
        and reductions. c.f. \S \ref{sectionOM}.
      \end{tabular}
      \\
      \hline
      \begin{tabular}{l}
        {\tt type Graph} in\\
        {\tt Language.Paraiso.OM.Graph}
      \end{tabular}
      &
      \begin{tabular}{l}
        The data-flow graph for the OM. c.f. \S \ref{sectionOM},\\
        Fig. \ref{figureDFG}, Fig. \ref{figLinearWaveOM}.
      \end{tabular}
      \\
      \hline
      \begin{tabular}{l}
        {\tt type Annotation} in\\
        {\tt Language.Paraiso.Annotation}
      \end{tabular}
      &
      \begin{tabular}{l}
        The collection of annotations that are added \\
        to each OM data-flow graph node. c.f. \S \ref{sectionBackend}, \\
        \S \ref{sectionTuningTargets}.
      \end{tabular}
      \\
      \hline
      \begin{tabular}{l}
        {\tt data Plan} in\\
        {\tt Language.Paraiso.Generator.Plan}
      \end{tabular}
      &
      \begin{tabular}{l}
        The fixed detail of the code to be generated\\
        such as amount of memories and what to do in \\
        each subroutine. c.f. Fig. \ref{figureBackend}. To see how a plan\\
        is fixed c.f. Fig. \ref{figureSubKernel}.
      \end{tabular}
      \\
      \hline
      \begin{tabular}{l}
        {\tt data Program} in\\
        {\tt Language.Paraiso.Generator.Claris}
      \end{tabular}
      &
      \begin{tabular}{l}
        The subset of C++ and CUDA syntax tree \\
        which is sufficient in generating codes in scope\\
        of Paraiso. c.f. Fig. \ref{figureBackend}. 
      \end{tabular}
      \\
      \hline
      \begin{tabular}{l}
        {\tt newtype Genome} in\\
        {\tt Language.Paraiso.Tuning.Genetic}
      \end{tabular}
      &
      \begin{tabular}{l}
        The set of annotations that belongs to an\\
         individual encoded as a string of letters, with \\
         which one can {\tt mutate}, {\tt cross} and  {\tt triangulate}.\\
         The evolution algorithm is in \S \ref{sectionPAGSA}.
      \end{tabular}
      \\
      \hline
    \end{tabular}
  \end{center}
  \addspan{
    \caption{List of concepts introduced by our work.} \label{tblListConcept}
  }
\end{table}

\addspan{
  In order to generate the OM data-flow graphs from tensor expressions, 
  and to translate them as native programs
  and further to apply manual and automated tuning over them,
  Paraiso introduces a number of abstract concepts centered around the OM. These components
  are quite orthogonal to each other,
  and some may even be useful outside the context of Paraiso.
  Table \ref{tblListConcept} summarizes those concepts, and provides pointers
  to the source code and the sections in this paper.
}

Orthotope Machine will endure the change in parallel languages and
hardwares, as long as there are needs for explicit solvers of PDEs.
Language/hardware designers can access to various applications for
test and practical purpose, once they support translation from
Orthotope Machine to their language.  With Orthotope Machine as an
interface, we can combine various stencil computation applications
with state-of-the-art techniques developed so far, such as
cache-friendly data structures \cite{datta_stencil_2008}, overlapping
communication with computation
\cite{shimokawabe_80-fold_2010,maruyama_physis:_2011} or heterogeneous
utilization of CPU/GPU \cite{shimokawabe_peta-scale_2011}. On the
other hand, various concepts of numerical simulations has been
decomposed to elemental calculations before the OM level, so the
problem of how to build a parallel computation from components such as
new spatial interpolations, time marching methods and approximate
Riemann solvers, can be addressed and developed separately from the
detail of the hardware.
\addspan{ To achieve such orthogonality is one of the aims of 
  the Paraiso project. }

\delspan{
Adding a physicist-friendly scripting language layer on top of
{\tt Builder Monad} is a future work, so that physicist not familiar with
Haskell can access Paraiso.
}

\subsection{Outline of The Orthotope Machine(OM)}  \label{sectionOM} 

The Orthotope Machine (OM) is a virtual machine much like vector
computers.  Each register of OM is multidimensional array of infinite
size. Arithmetic operations of OM work in parallel on each mesh, or
loads from neighbor cells. We have no intention of building a real
hardware: OM is a thought object to capture parallel algorithms to
data-flow graphs without losing parallelism.

The instruction set of OM resembles those of historical parallel
machines such as PAX computer \cite{hoshino_pax_1989}, and is subset
of partitioned global address space (PGAS) languages such as
XcalableMP \cite{lee_implementation_2010}.

Each instance of OM have a specific dimension (e.g. a
two-dimensional OM of size $N0 \times N1$).  The variables
of OM are either arrays of that dimension, or a scalar value.  All of
the arrays must have a common size. The actual numbers ($N0, N1$)
are fixed at native code generation phase. We say that arrays are variable
with {\tt Local Realm} , while the scalars are variable with {\tt Global Realm}.

OM has a set of {\tt Static} variables which denotes the current state
of the simulation.  Each {\tt Static} variable has a string name.  OM
has a set of {\tt Kernel}s --- they are subroutines for updating the
{\tt Static} variables.  Inside each {\tt Kernel}, you can generate
{\tt Temporal} values in static single assignment (SSA) manner.

To summarize, the lifetime of an OM variable is either {\tt Static} or
{\tt Temporal}, and the {\tt Realm} of an OM variable is either {\tt
  Local} or {\tt Global}.  {\tt Static} variables survive multiple
{\tt Kernel} calls, while {\tt Temporal} variables are limited to one
{\tt Kernel}.  {\tt Local} variables are arrays. {\tt Global}
variables are scalar values; in other words, they are arrays whose
elements are globally the same.

OM cannot handle array of structures; {\tt Local} variables may
contain only simple objects such as {\tt Bool} or {\tt Double}, but
not composite ones such as complex numbers or vectors.  Nevertheless
we can easily use such composite concepts in Paraiso programs at {\tt
  Builder Monad} level and translate them to OM instructions, for
example by using the applicative programming style
\cite{mcbride_functional_2008}.

\begin{figure}
  \begin{center}
    \includegraphics[width=10cm,angle=270]{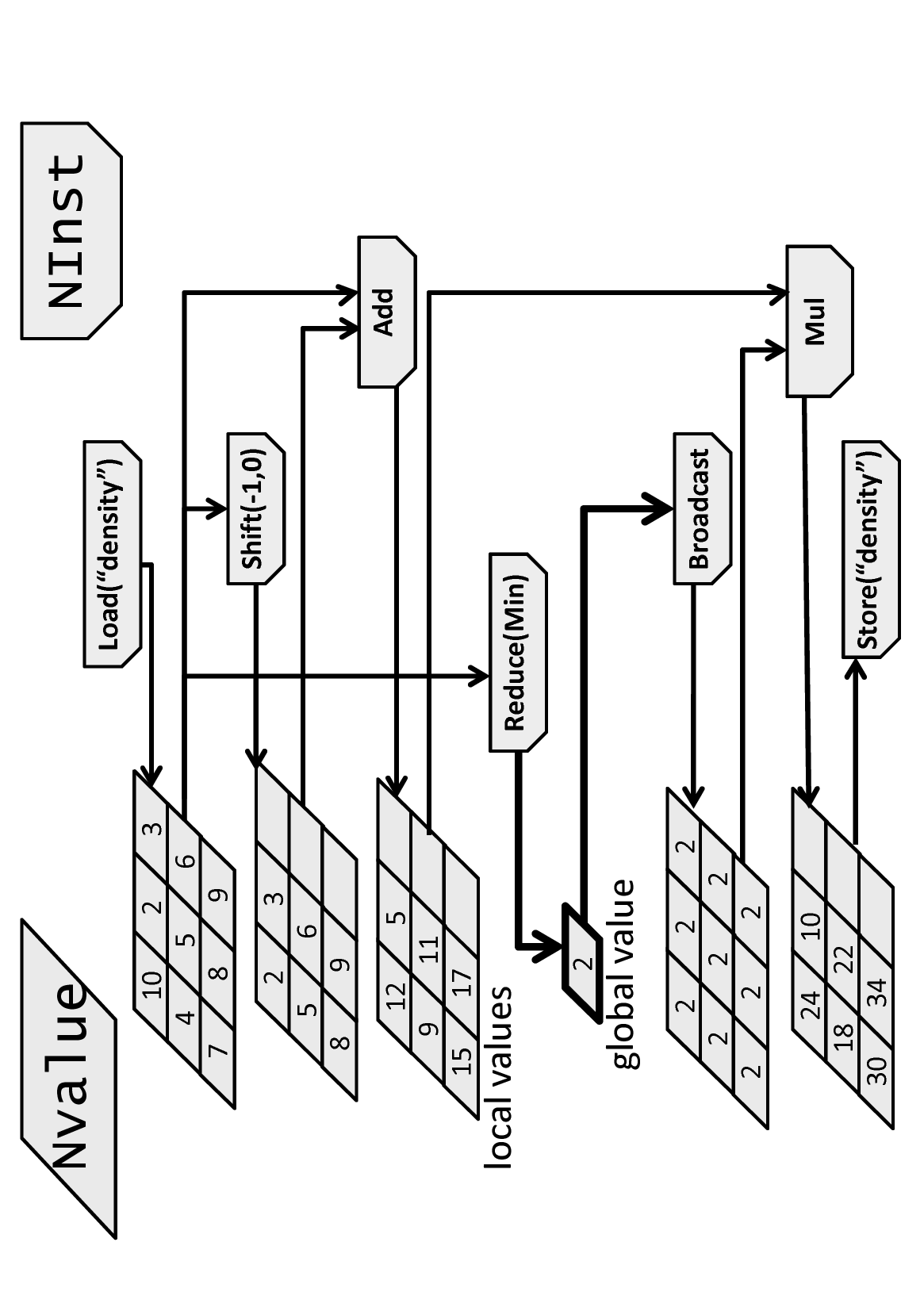}
    \caption {A data-flow graph for Orthotope Machine(OM).
       }
    \label{figureDFG}
  \end{center} 
\end{figure}

An OM Kernel is a directed bipartite graph consisting of {\tt NInst}
nodes and {\tt NValue} nodes, as illustrated in Fig. \ref{figureDFG}.
Each node has {\tt arity :: a -> (Int, Int)}, number of incoming and
outgoing edges. {\tt NValue} nodes have arity of (1,$n$), where $n$ is
the number of {\tt NInst} nodes that uses the value and can be an
arbitrary integer. Arities of {\tt NInst} nodes are inherited from the
instructions they carry.

OM has nine instructions:
\begin{verbatim}
data Inst vector gauge 
  = Imm Dynamic 
  | Load StaticIdx
  | Store StaticIdx
  | Reduce R.Operator 
  | Broadcast 
  | Shift (vector gauge) 
  | LoadIndex (Axis vector) 
  | LoadSize (Axis vector) 
  | Arith A.Operator 
\end{verbatim}

Here, we limit ourselves to C pseudo-code of OM instruction semantics.
The formal definition of the OM will be in forthcoming 
technical reports. Note that we do not
translate the OM instructions one by one to C codes listed
here. Instead, the instructions are merged into sub-graphs and then
translated to C loops (or CUDA kernels) with much larger bodies, to
make efficient use of computation resource and memory bandwidth
(c.f. \S \ref{sectionTuningTargets}). 

\paragraph{\tt Imm}arity $(0,1)$: load constant value.
Its output can be either {\tt Global} or  {\tt Local} {\tt NValue} node.
For example, for a {\tt Local Temporal} variable {\tt a},
\begin{verbatim}
a <- imm 4.2 
\end{verbatim}
means
\begin{verbatim}
for(int j=0; j<N1; ++j) {
  for(int i=0; i<N0; ++i) {
    a[j][i] = 4.2;
  }
}
\end{verbatim}

\paragraph{\tt Load}arity $(0,1)$: read from static variable to temporal variable.
The realms of the static and temporal variable must match, and
can be either of {\tt Global} or  {\tt Local}.
For example, for a {\tt Local Temporal} variable {\tt a} and {\tt Local Static} variable {\tt density},
\begin{verbatim}
a <- load "density"
\end{verbatim}
means
\begin{verbatim}
for(int j=0; j<N1; ++j) {
  for(int i=0; i<N0; ++i) {
    a[j][i] = density[j][i];
  }
}
\end{verbatim}

\paragraph{\tt Store}arity $(1,0)$: write a temporal variable to a static variable.
The realms of the static and temporal variable must match, and
can be either of {\tt Global} or  {\tt Local}.
\begin{verbatim}
store "density" <- a
\end{verbatim}
means
\begin{verbatim}
for(int j=0; j<N1; ++j) {
  for(int i=0; i<N0; ++i) {
    density[j][i] = a[j][i];
  }
}
\end{verbatim}

\paragraph{\tt Reduce}arity $(1,1)$:  convert a local variable to a global one with a specified reduction operator.
\begin{verbatim}
b <- reduce Min <- a
\end{verbatim}
means
\begin{verbatim}
b = a[0][0];
for(int j=0; j<N1; ++j) {
  for(int i=0; i<N0; ++i) {
    b = min(b,a[j][i]);
  }
}
\end{verbatim}

\paragraph{\tt Broadcast}arity $(1,1)$:  convert a global variable to a local one.
\begin{verbatim}
b <- broadcast <- a
\end{verbatim}
means
\begin{verbatim}
for(int j=0; j<N1; ++j) {
  for(int i=0; i<N0; ++i) {
    b[j][i] = a;
  }
}
\end{verbatim}

\paragraph{\tt Shift}arity $(1,1)$:  takes a constant vector and an input local variable. Move each cell to its neighbor.
\begin{verbatim}
b <- shift (1,5)<- a
\end{verbatim}
means
\begin{verbatim}
for(int j=0; j<N1-5; ++j) {
  for(int i=0; i<N0-1; ++i) {
    b[j+5][i+1] = a[j][i];
  }
}
\end{verbatim}

\paragraph{\tt LoadIndex}arity $(0,1)$:  get coordinate of each cell. 
The output must be a {\tt Local} value.
\begin{verbatim}
b0 <- loadIndex 0
b1 <- loadIndex 1
\end{verbatim}
means
\begin{verbatim}
for(int j=0; j<N1; ++j) {
  for(int i=0; i<N0; ++i) {
    b0[j][i] = i;
  }
}
for(int j=0; j<N1; ++j) {
  for(int i=0; i<N0; ++i) {
    b1[j][i] = j;
  }
}
\end{verbatim}

\paragraph{\tt LoadSize}arity $(0,1)$:  get array size.
The output must be a {\tt Global} value.
\begin{verbatim}
c0 <- loadSize 0
c1 <- loadSize 1
\end{verbatim}
means
\begin{verbatim}
c0 = N0;
c1 = N1;
\end{verbatim}

\paragraph{\tt Arith} perform various arithmetic operations.
The arity of this {\tt NInst} node is inherited from its operator.
The realms of the inputs and outputs must match, and
can be either of {\tt Global} or  {\tt Local}.
If {\tt Local}, array elements at matching index are operated in parallel (i.e. {\tt zipWith}).

For example,
\begin{verbatim}
c <- arith Add a b <- a,b
\end{verbatim}
means
\begin{verbatim}
for(int j=0; j<N1; ++j) {
  for(int i=0; i<N0; ++i) {
    c[j][i] = a[j][i] + b[j][i];
  }
}
\end{verbatim}

\addspan{
\subsection{An Example of Orthotope Machine Data-Flow Graph} 
}

\addspan{
Here we show the use case of the Orthotope Machine operators introduced in \S \ref{sectionOM}
within a simple PDE solver. We solve the following linear wave equation:
\begin{eqnarray}
  \frac{\partial^2 f}{\partial t^2} - c^2  \frac{\partial^2 f}{\partial x^2} = 0, \label{eqLinearWave}
\end{eqnarray}
where $t$ is time, $x$ is one-dimensional space coordinate and $c$ is the signal speed.
}

\addspan{
By introducing the time derivative $g = {\partial f}/{\partial t}$, Eq. (\ref{eqLinearWave}) is rewritten
as following system of first-order PDEs:
\begin{eqnarray}
   \frac{\partial f}{\partial t} = g,  \label{eqLinearWaveF} \\
   \frac{\partial g}{\partial t} = c^2  \frac{\partial^2 f}{\partial x^2}, \label{eqLinearWaveG}
\end{eqnarray}
and satisfies the following conservation law:
\begin{eqnarray}
  \frac{d}{dt} E = 0, \\
  E \equiv \int \left( \frac{c^2}{2}\left(\frac{\partial f}{\partial x}\right)^2 + \frac{g^2}{2} \right) dx.
  \label{eqLinearWaveEnergy}
\end{eqnarray}
We will write a Paraiso code that simulate the discrete system of Eqs. (\ref{eqLinearWaveF},\ref{eqLinearWaveG}) 
and tests the conservation law.
}

\addspan{
Let $f^n[\mathrm i]$ denote the discrete field where $n$ and $\mathrm i$ are the discrete time and space
coordinate, respectively. We choose the following 2nd-order, Lax-Wendorf and Leap Frog scheme to solve 
Eqs. (\ref{eqLinearWaveF},\ref{eqLinearWaveG}) :
\begin{eqnarray}
  f^{n+1}[\mathrm i] = f^{n}[\mathrm i] + \Delta t \, g^{n}[\mathrm i],  \label{eqLWDiscF} \\
  g^{n+1}[\mathrm i] = g^{n}[\mathrm i] + \frac{c^2 \, \Delta t}{\Delta x^2} 
  \left( f^{n+1}[\mathrm i+1] + f^{n+1}[\mathrm i-1] - 2 f^{n+1}[\mathrm i]   \right),  \label{eqLWDiscG}
\end{eqnarray}
and measure the following discrete form of the conserved quantity:
\begin{eqnarray}
  E^{n+1} \equiv \sum_{i=0}^{N-1} \left[ \left(c^2 \frac{f^{n+1}[\mathrm i+1] - f^{n+1}[\mathrm i-1]}{2 \, \Delta x} \right)^2 + 
  \left( \frac{g^{n+1}[\mathrm i] + g^{n}[\mathrm i]}{2} \right)^2 
  \right] \Delta x
  \label{eqLWDE}
\end{eqnarray}
}

\begin{table}
  \begin{tabular}{p{15cm}}
    \rowcolor{grey}
    \small
    \topsep=-1ex\relax
\begin{verbatim}
proceed :: Builder Vec1 Int Annotation ()
proceed = do 
  c  <- bind $ imm 3.43
  n  <- bind $ loadSize TLocal (0::Double) $ Axis 0
  
  f0 <- bind $ load TLocal (0::Double) $ fieldF
  g0 <- bind $ load TLocal (0::Double) $ fieldG
  dx <- bind $ 2 * pi / n
  dt <- bind $ dx / c

  f1 <- bind $ f0 + dt * g0
  fR <- bind $ shift (Vec:~ -1) f1 
  fL <- bind $ shift (Vec:~  1) f1 
  g1 <- bind $ g0 + dt * c^2 / dx^2 * 
        (fL + fR - 2 * f1)
  store fieldF f1
  store fieldG g1
  
  dfdx <- bind $ (fR - fL) / (2*dx)
  store energy $ reduce Reduce.Sum $
    0.5 * (c^2 * dfdx^2 + ((g0+g1)/2)^2) * dx
\end{verbatim}
  \end{tabular}
  \addspan{
    \caption{
      A Paraiso source code for solving discrete Eq. (\ref{eqLWDiscF},\ref{eqLWDiscG}).
    }\label{tableLinearWaveOM}
  } 
\end{table}

\begin{figure}
  \begin{center}
    \includegraphics[height=20cm]{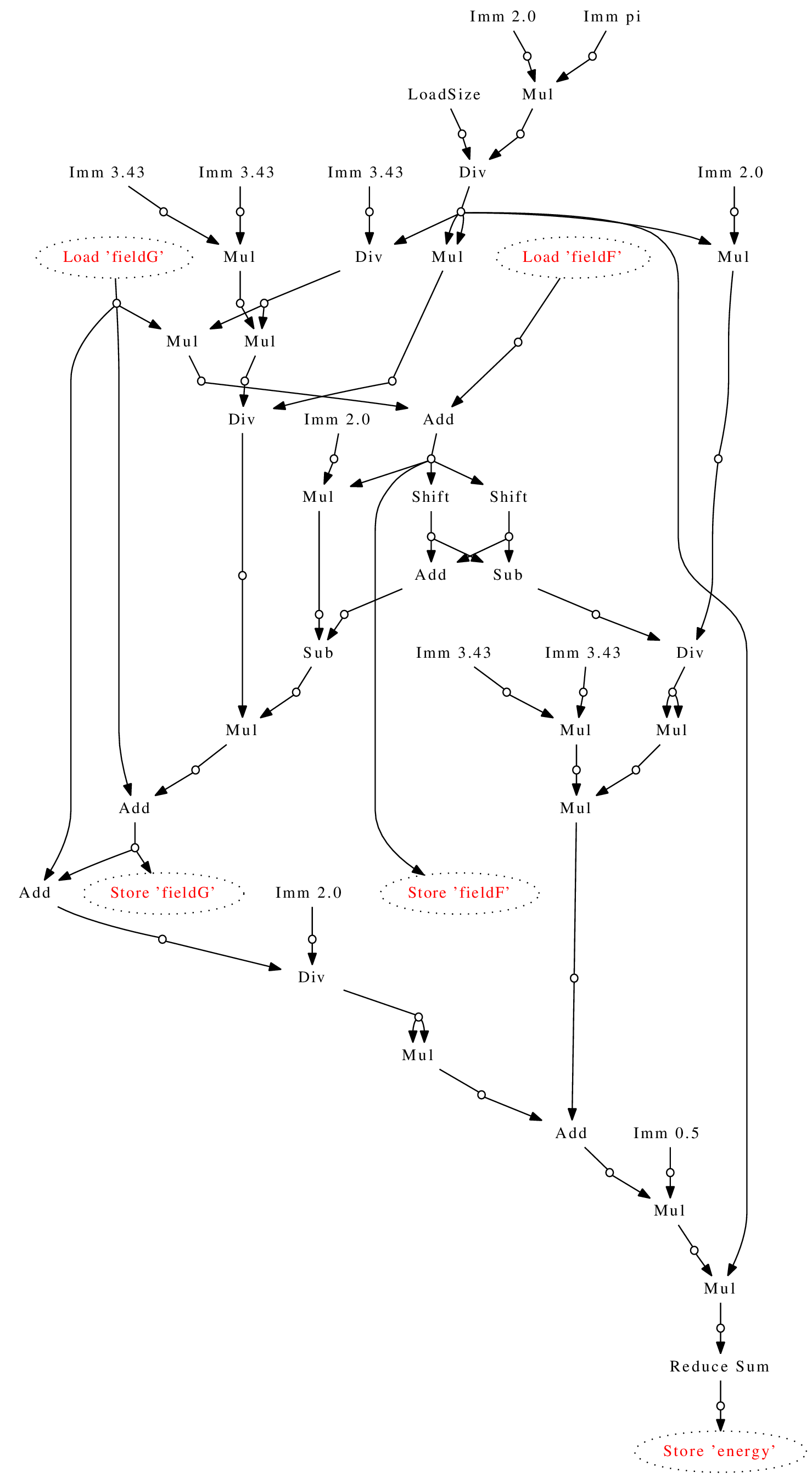}
  \end{center}
  \caption{
    The Orthotope Machine data-flow graph generated from the Paraiso source code listed in 
    Table \ref{tableLinearWaveOM}. For brevity, {\tt NValue} nodes are represented as little circles
    and only {\tt NInst}
    nodes are shown;
    the {\tt Arith} instruction nodes are replaced by arithmetic operators inside them,
    and the instruction arguments are not shown but for {\tt Load},  {\tt Store}
    and {\tt Reduce}.
  }\label{figLinearWaveOM}
\end{figure}

\addspan{
  Table \ref{tableLinearWaveOM} shows the Paraiso implementation for the algorithm
  Eq. (\ref{eqLWDiscF},\ref{eqLWDiscG}). The corresponding OM data-flow graph is visualized in
  Fig. \ref{figLinearWaveOM}. See how the data-flows from {\tt Load} node to {\tt Store} node.
  Here, we have $c=3.43$ and $0 \leq x < 2\pi$. 
  We have performed numerical simulations
  using this program and the following initial condition:
  \begin{eqnarray}
    f(x) = \sin(x),\\
    g(x) = \cos(3x),
  \end{eqnarray}
 with resolution varying from $N=8$ to $N=3072$,
  and have confirmed that the fluctuation of discrete conserved quantity, Eq. (\ref{eqLWDE})
  was smaller than of the order of $10^{-13}$.
}

\subsection{Typelevel Tensor}\label{sectionTLT}

We introduce {\tt typelevel-tensor} library,
to abstract over the dimensions; 
the use of tensor notations allow us to describe the algorithms for different dimensions
in a single source code.
The benefit of having information on tensor dimensions and ranks at the type-level is that 
we can detect erroneous operations like adding two 
tensors of different dimensions at compile time.
Also we have much less need for explicitly mentioning the tensor dimensions
in our programs,
because the dimensions will be type-inferred.

Our approach is similar to that in
\cite{keller_regular_2010}, where two constructors {\tt Z}
and {\tt :.} are used to inductively define multi-dimensional tuples.
Their $n$-dimensional vectors are actually $n$-tuples, so it is possible
that the set of $n$ elements consist of different types. In contrast,
we needed that all the $n$ elements of a vector are of the same type, 
because we wanted to operate on them, for example by applying 
a same function to all of them.

So instead of \cite{keller_regular_2010}'s approach:
\begin{verbatim}
infixl 3 :.
data Z = Z
data tail :. head = tail :. head
\end{verbatim}
We have these:
\begin{verbatim}
infixl 3 :~
data Vec a = Vec
data (n :: * -> *) :~ a = (n a) :~ a
\end{verbatim}

Here, {\tt Vec} is the type constructor for 0-dimensional vector, and
{\tt :\~{}} is a type-level function that 
takes  the type constructor for $n$-dimensional vector as an argument
and returns the type constructor for $n+1$-dimensional vector. We define type synonyms for 
successively higher vector:
\begin{verbatim}
type Vec0 = Vec
type Vec1 = (:~) Vec0 
type Vec2 = (:~) Vec1
type Vec3 = (:~) Vec2
\end{verbatim}
{\tt Vec0}, {\tt Vec1}, {\tt Vec2}, and {\tt Vec3} are all types of kind {\tt * -> *},
meaning that it takes one type (the element type) and returns another (the vector type).
For example, the three-dimensional double precision vector type is {\tt Vec3 Double}.
Higher-rank tensors are defined as nested vectors; for example {\tt Vec3 (Vec3 Int)} is
a $3\times 3$  matrix of integers.

Since we know that all the elements of our tensor are of the same type
{\tt a}, we can make our tensors instances of {\tt Traversable} type
class \cite{mcbride_functional_2008,gibbons_essence_2009}:
\begin{verbatim}
instance Traversable Vec where
  traverse _ Vec = pure Vec 
instance (Traversable n) => Traversable ((:~) n) where
  traverse f (x :~ y) = (:~) <$> traverse f x <*> f y
\end{verbatim}
The benefit of making our tensors instances of {\tt Traversable} is that 
we can {\tt traverse} on them:
\begin{verbatim}
traverse :: Applicative f => (a -> f b) -> t a -> f (t b)
\end{verbatim}
Here, suppose {\tt t} is our tensor type-constructor and {\tt f} is some 
context --- for example, a code generation context. {\tt a} and
{\tt b} are elements of our tensor.
Then the type of the {\tt traverse} function means that
if we have code generators for the computation of one element ({\tt a -> f b}), and we have a  {\tt t a}, a tensor whose elements are of type {\tt a},
then we can deduce the code generator for computation of the entire tensor {\tt f (t b)}.

\subsection{Builder Monad}\label{sectionBuilder}

{\tt Builder} monad is a {\tt State} monad whose state is 
the half-built data-flow graph of the Orthotope Machine.
To represent the data-flow graph, we use
Functional Graph Library (FGL) \cite{erwig_functional_1997, erwig_inductive_2001}.
Authors learned from the {\tt Q} monad \cite{sheard_template_2002}
how to encapsulate the construction process.

The graph carried by the {\tt State} monad has the following type: 

\begin{verbatim}
type Graph (vector :: *->*) (gauge :: *) (anot :: *) 
  = FGL.Gr (Node vector gauge anot) Edge

data Node vector gauge anot =
  = NValue DynValue anot 
  | NInst (Inst vector gauge) anot

data Edge  
  = EUnord 
  | EOrd Int
\end{verbatim}
{\tt Graph} takes three type arguments. 
{\tt vector :: *->*} is a type constructor 
that denotes the dimension of the OM. 
{\tt gauge} is the type for the indices of the arrays, which is usually an {\tt Int}.
{\tt anot} is the type the nodes of the graph are annotated with. Such annotations are
used to analyze and optimize the data-flow graph.

The three types {\tt vector}, {\tt gauge}, and {\tt anot} are passed
to the {\tt Node}s of the graph. {\tt Node}s are either {\tt NValue}
or {\tt NInst}.  Two types {\tt vector} and {\tt gauge} are further
passed to the instruction type {\tt Inst}, because instructions such
as {\tt shift} requires the information on the array dimension and
indices. Every graph nodes are also annotated by type {\tt anot}.

On the other hand, the edges contain none of the three types. They are
just unordered edges {\tt EUnord} or edges ordered by an integer {\tt
  EOrd Int}. For example, we can exchange two edges going into
addition instruction, but cannot exchange those into subtraction.

\begin{figure}
  \begin{center}
    \includegraphics[width=10cm,angle=90]{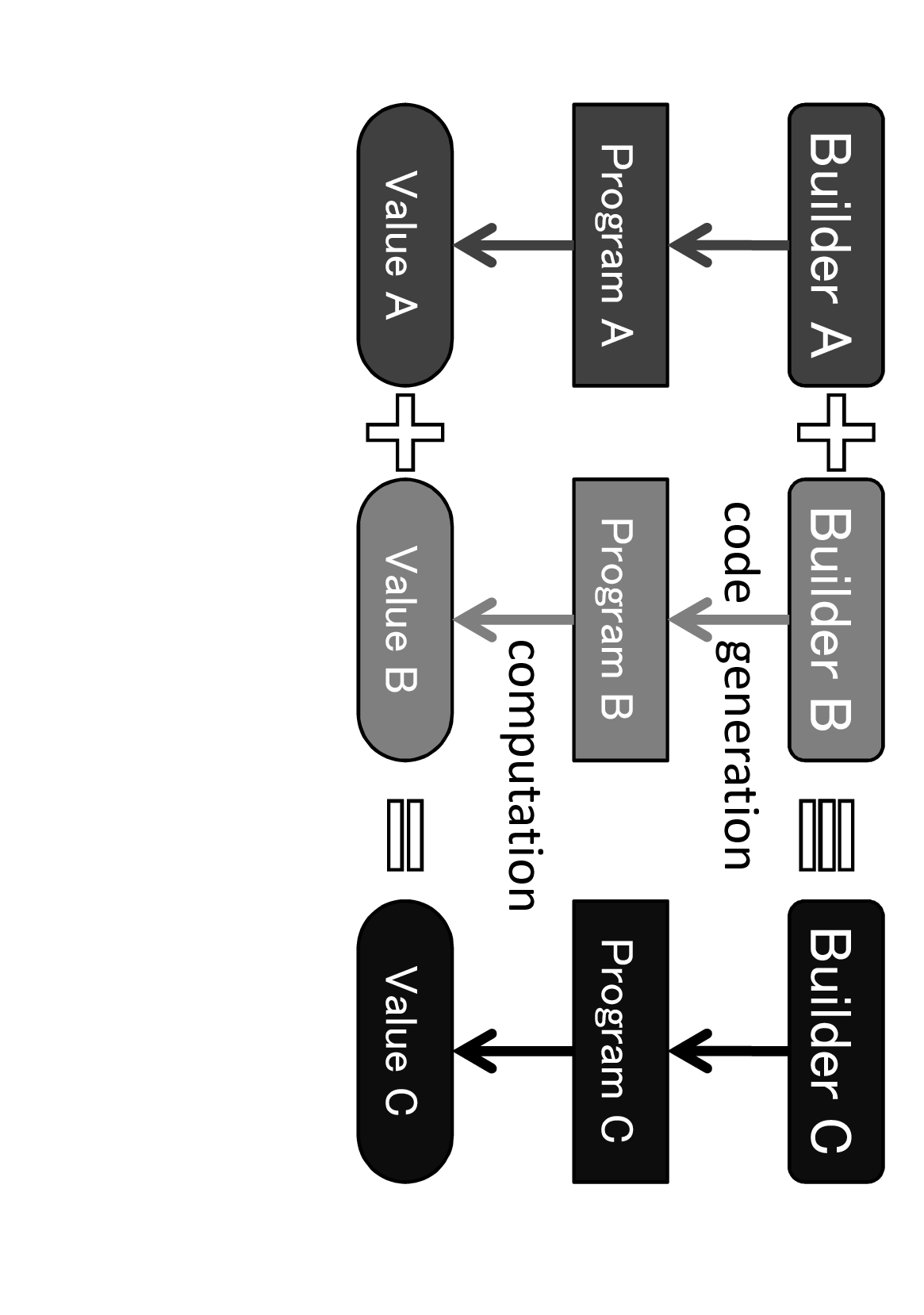}
    \vspace{-3cm}
    \caption {
      A commutative diagram of {\tt Builder Monad} and computation.
    }
    \label{figureCommutative}
  \end{center} 
\end{figure}

We define various mathematical operations between {\tt Builder Monad} in a consistent manner
(c.f. Fig. \ref{figureCommutative}). For any operator $\oplus$,
{\tt Builder A} $\oplus$ {\tt Builder B} $=$ {\tt Builder C} is defined by 
{\tt Value A} $\oplus$ {\tt Value B} $=$ {\tt Value C}, where {\tt Value} $i$
is the value computed by {\tt Program} $i$ which is generated by
{\tt Builder} $i$. For example, a helper function that takes an operator symbol {\tt op}, 
two builders {\tt builder1} and {\tt builder2}, and create a binary operator for builder, is as follows:

\begin{verbatim}
mkOp2 :: (TRealm r, Typeable c) => 
         A.Operator                  -- ^The operator symbol 
      -> (Builder v g a (Value r c)) -- ^Input 1              
      -> (Builder v g a (Value r c)) -- ^Input 2               
      -> (Builder v g a (Value r c)) -- ^Output              
mkOp2 op builder1 builder2 = do
  v1 <- builder1
  v2 <- builder2
  let 
      r1 = Val.realm v1
      c1 = Val.content v1
  n1 <- valueToNode v1
  n2 <- valueToNode v2
  n0 <-  addNodeE [n1, n2] $ NInst (Arith op)
  n01 <- addNodeE [n0] $ NValue (toDyn v1) 
  return $ FromNode r1 c1 n01
\end{verbatim}

We first extract the graph nodes from left hand side and right hand side
builders in {\tt Builder} context, add an {\tt NInst} node that contains {\tt op}
symbol, add an {\tt NValue} node after that, and return the node index.

In Haskell, defining mathematical operators between a data type is
done by declaring the data type as an instance of the {\em type class}
that manages the operator. In this sense type classes in Haskell are
the parallels of algebraic structures such as group, ring and field, that
manages addition, multiplication, and division, respectively.
We use a Haskell package {\tt numeric-prelude} that provides such 
algebraic structures
\cite{thurston_numeric-prelude:_2012}.

For example, an {\tt Additive} instance declaration of {\tt Builder} is as follows:
\begin{verbatim}
instance (TRealm r, Typeable c, Additive.C c) 
  => Additive.C (Builder v g a (Value r c)) where
  zero = return $ FromImm unitTRealm Additive.zero
  (+) = mkOp2 A.Add
  (-) = mkOp2 A.Sub
  negate = mkOp1 A.Neg
\end{verbatim}

These type class instances, together with typelevel-tensor library, allows us 
to write tensor equations used in Paraiso application programs. 
Moreover, such equations are good for arbitrary instances of the type class.
For example, here are the definition of momentum and momentum flux in our 
Euler equations solver: 
\begin{verbatim}
momentum x = compose (\i -> density x * velocity x !i)
momentumFlux x = 
  compose (\i -> compose (\j ->
    momentum x !i * velocity x !j + pressure x * delta i j))
\end{verbatim}
These functions can be used to directly calculate momentum vector and
momentum flux tensor whose components are of type {\tt Double}. The
very same functions are used to generate the solvers for
CPUs and GPUs. At that time, their components are inferred to be {\tt
  Builder} types. In addition to that, these functions can handle
tensors of arbitrary dimensions.

\subsection{Backend}\label{sectionBackend}

\begin{figure}
  \begin{center}
    \includegraphics[width=8cm,angle=0]{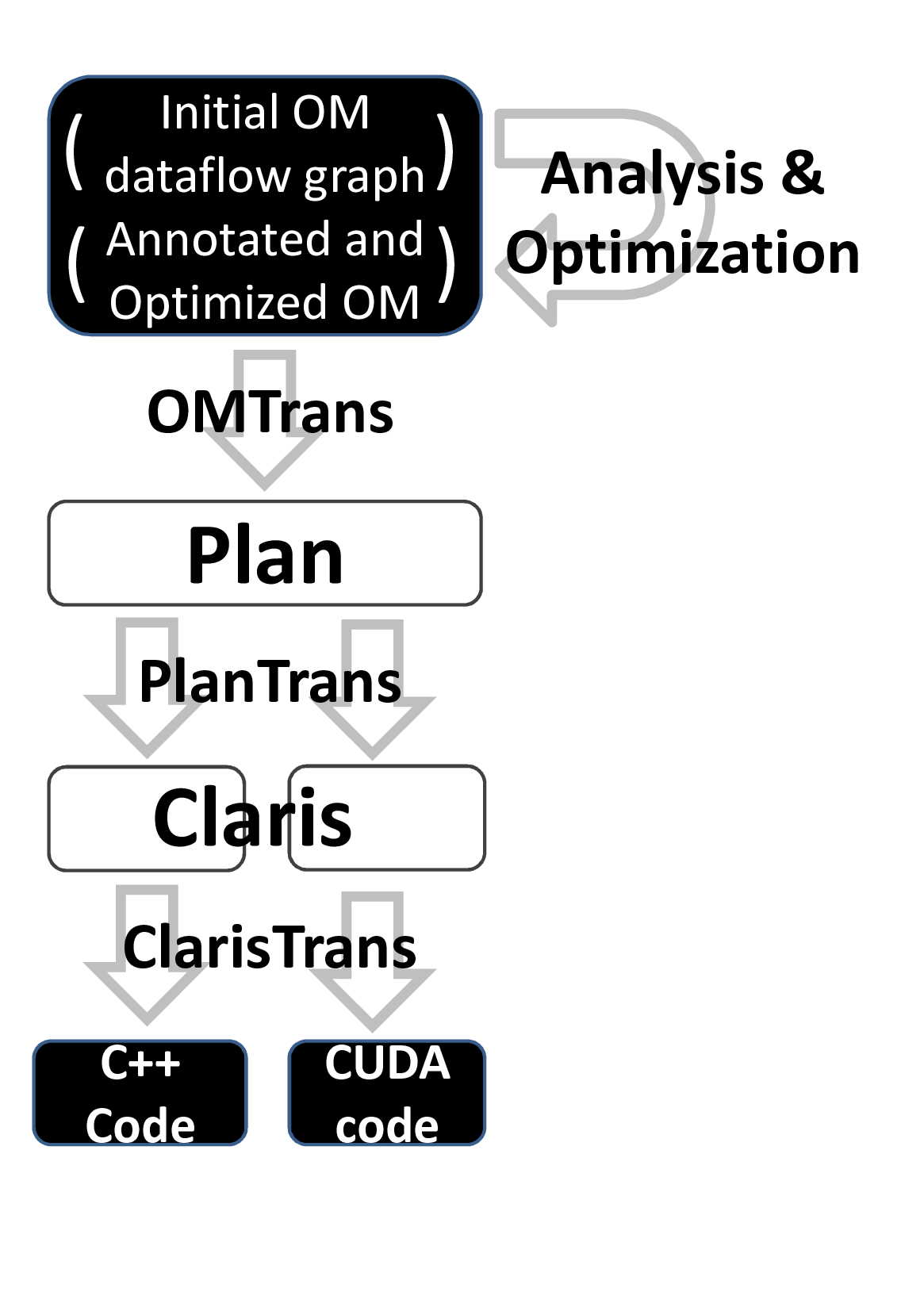}
    \vspace{-1cm}
    \caption {
      Backend.
    }
    \label{figureBackend}
  \end{center} 
\end{figure}

The back end converts the data-flow graph of {\tt Kernel}s native codes,
and create a C++ class corresponding to an OM. The code generation
processes of Paraiso is illustrated in Fig. \ref{figureBackend}.

First, various analysis and optimizations are applied. Analysis and
optimizations are functions that takes an OM and returns an OM, so we
can combine them in arbitrary ways. Though some analysis are
mandatory for code generation.

Paraiso has an omnibus interface for analysis and optimization using
dynamic programming library {\tt Data.Dynamic} in Haskell
\cite{lammel_scrap_2003} :
\begin{verbatim}
type Annotation = [Dynamic]
add :: Typeable a => a -> Annotation -> Annotation
toList :: (Typeable a) => Annotation -> [a]
toMaybe :: (Typeable a) => Annotation -> Maybe a
\end{verbatim}
Here, analyzers as well as human beings can {\tt add} annotations of
arbitrary type {\tt a} to the graph. On the other hand, optimizers can
read out annotations of what type they recognize and perform
transformations on the data-flow graph. The set of annotations can be
serialized to, and deserialized from {\em genomes}, which are binary
strings used in automated tuning phase.

\begin{figure}
  \begin{center}
    \includegraphics[width=10cm,angle=270]{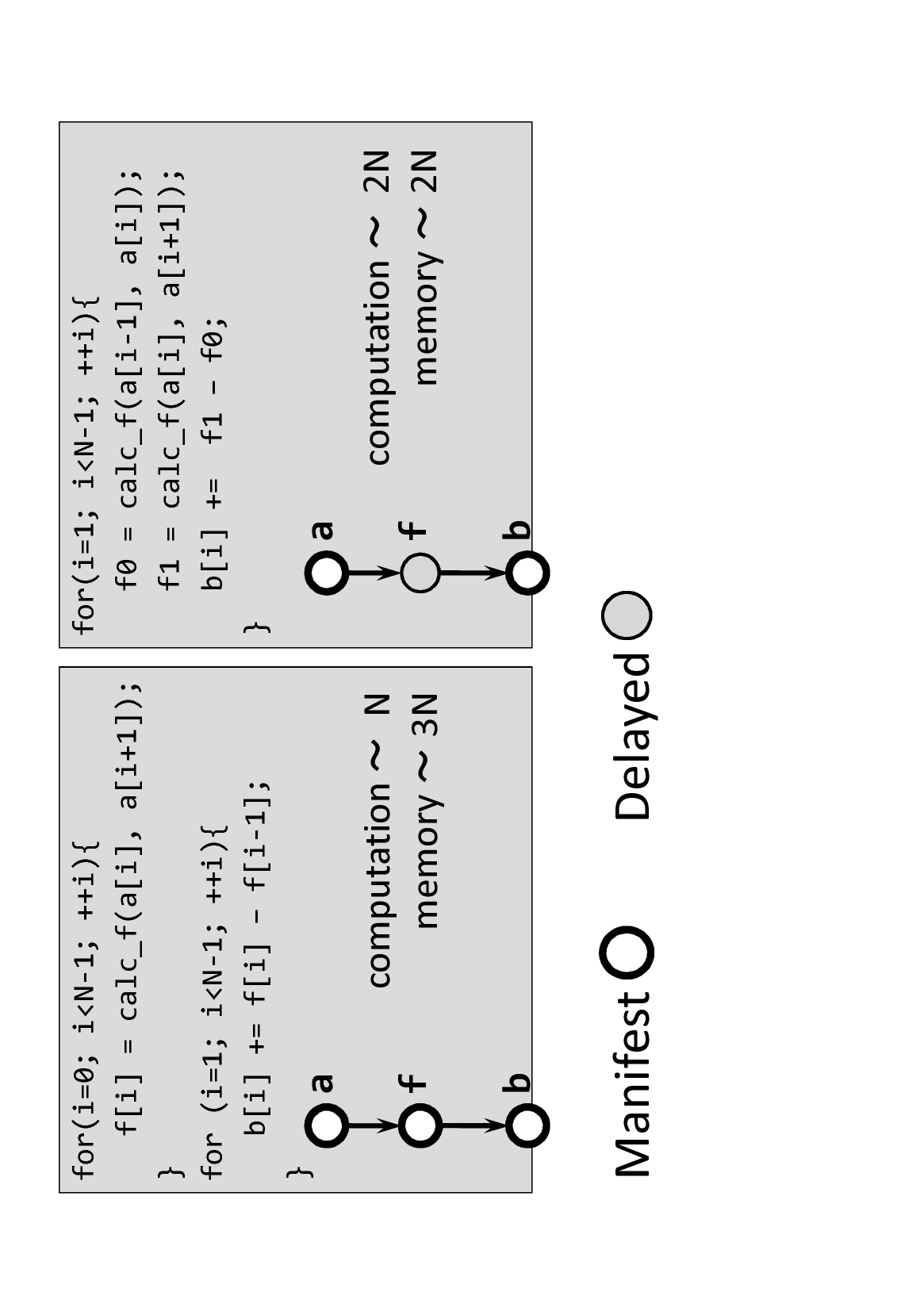}
    \vspace{-2cm}
    \caption {
      A Manifest - Delayed trade off. The left code requires less arithmetic units but
      consumes more memory and its bandwidth; The right code requires less memory and
      bandwidth in cost of more computations.
    }
    \label{figureTradeoff}
  \end{center} 
\end{figure}

Examples of annotations are the choices of whether to store a value on
memory and reuse or not to store and recompute it as is needed
(c.f. Fig. \ref{figureTradeoff}); the boundary analysis result used for
automatically adding ghost cells; dependency analysis; and labels used 
for dead code eliminations.

\begin{figure}
  \begin{center}
    \begin{tabular}{cp{6.0cm}}
      \raisebox{-4.5cm}{\includegraphics[width=5.5cm,angle=90]{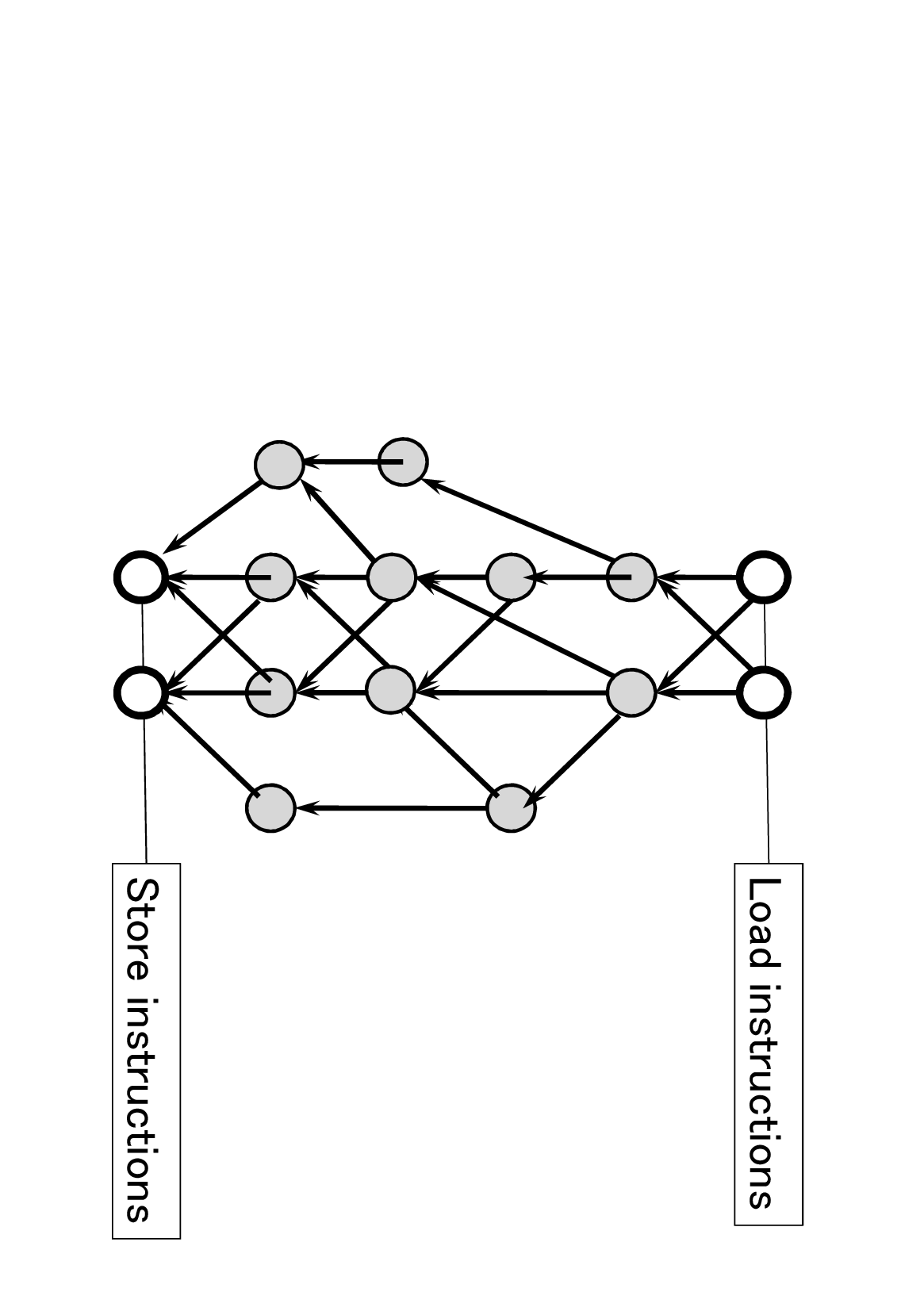}} & 
      (a) Some nodes need to be Manifest.\\
      \raisebox{-4.5cm}{\includegraphics[width=5.5cm,angle=90]{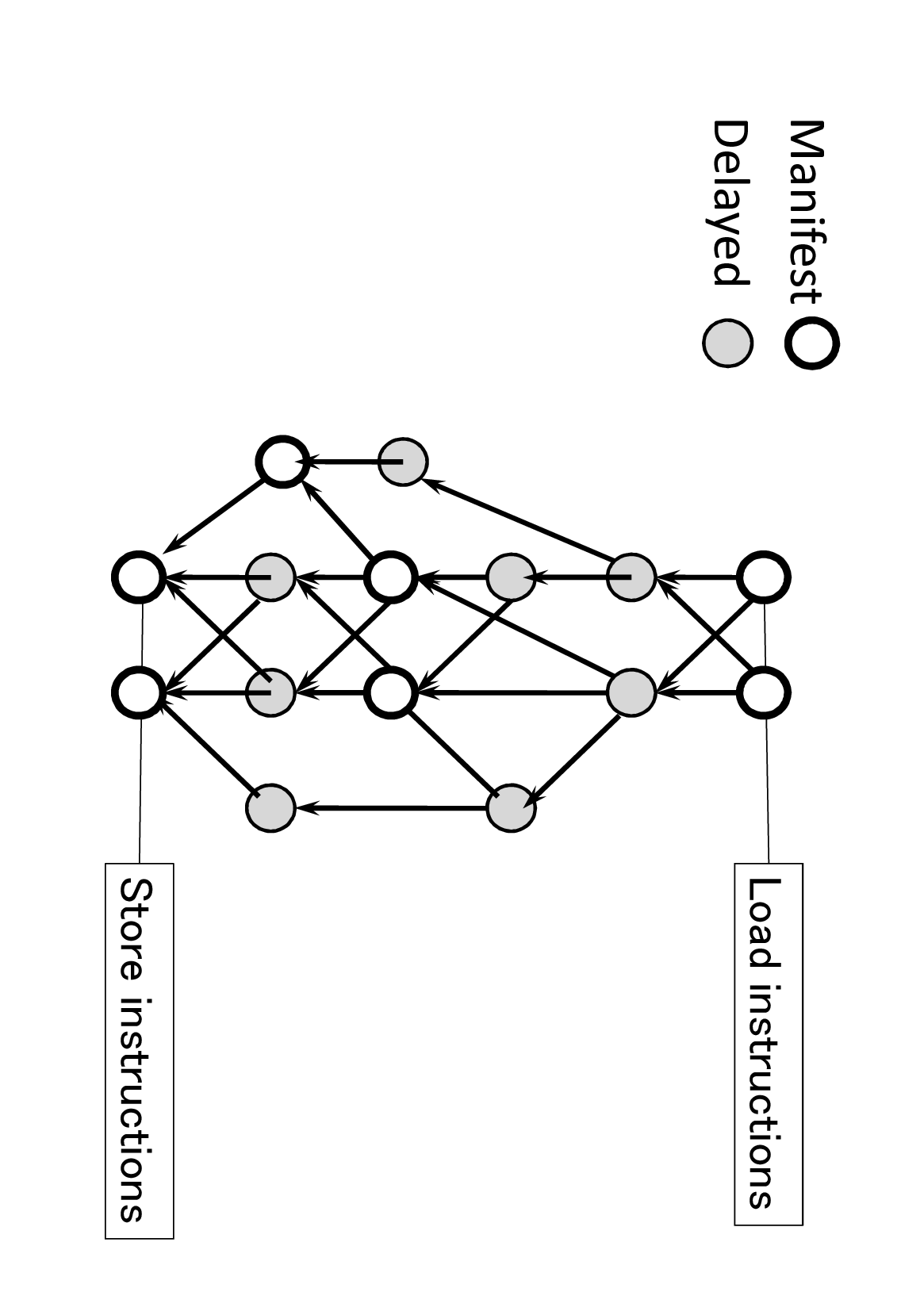}} & 
      (b) For each of other nodes, the genome specifies whether it is Manifest or Delayed.\\
      \raisebox{-4.5cm}{\includegraphics[width=5.5cm,angle=90]{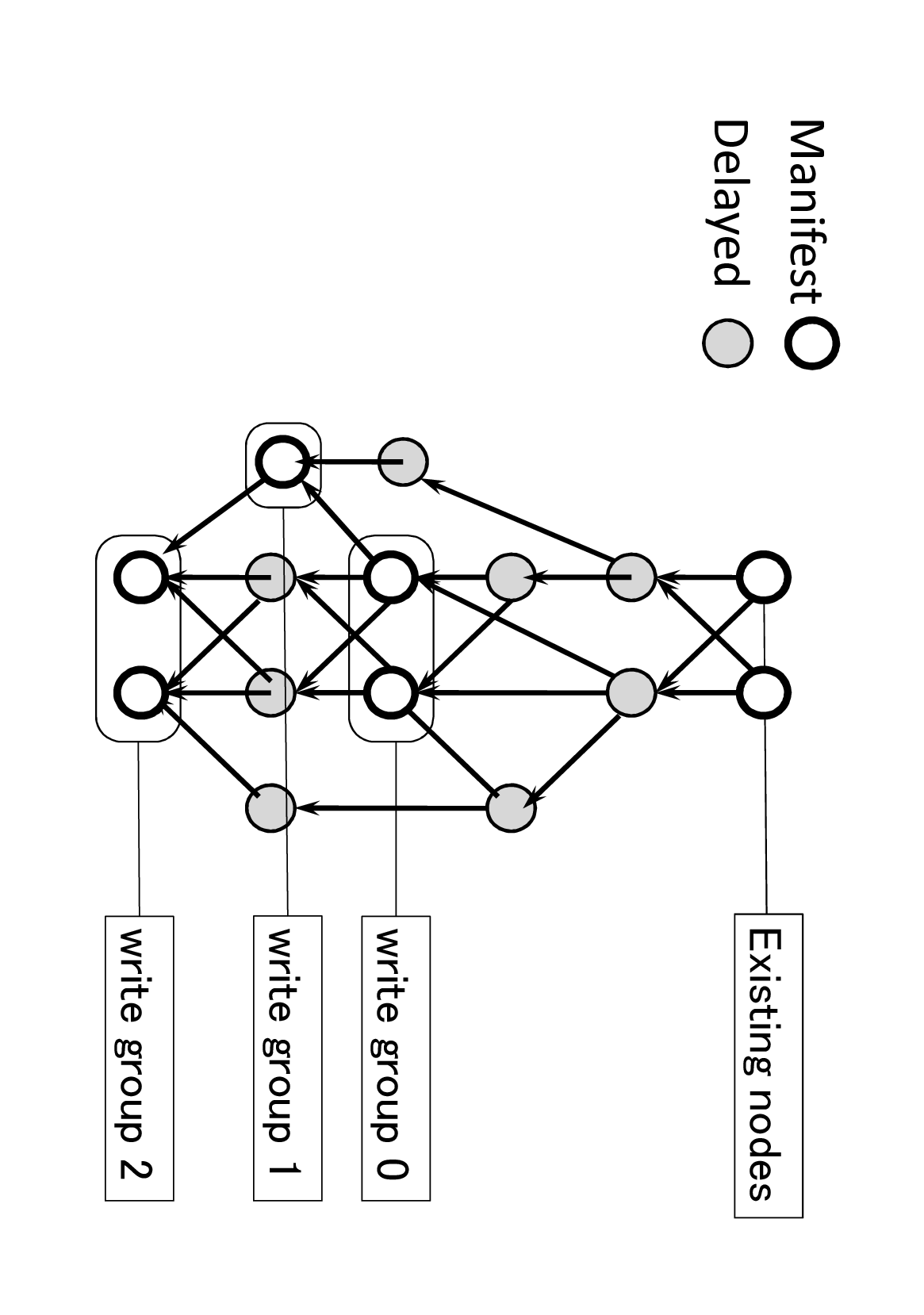}} & 
      (c) The sets of nodes that can be calculated in the same loop are greedily merged into write groups.\\
      \raisebox{-4.5cm}{\includegraphics[width=5.5cm,angle=90]{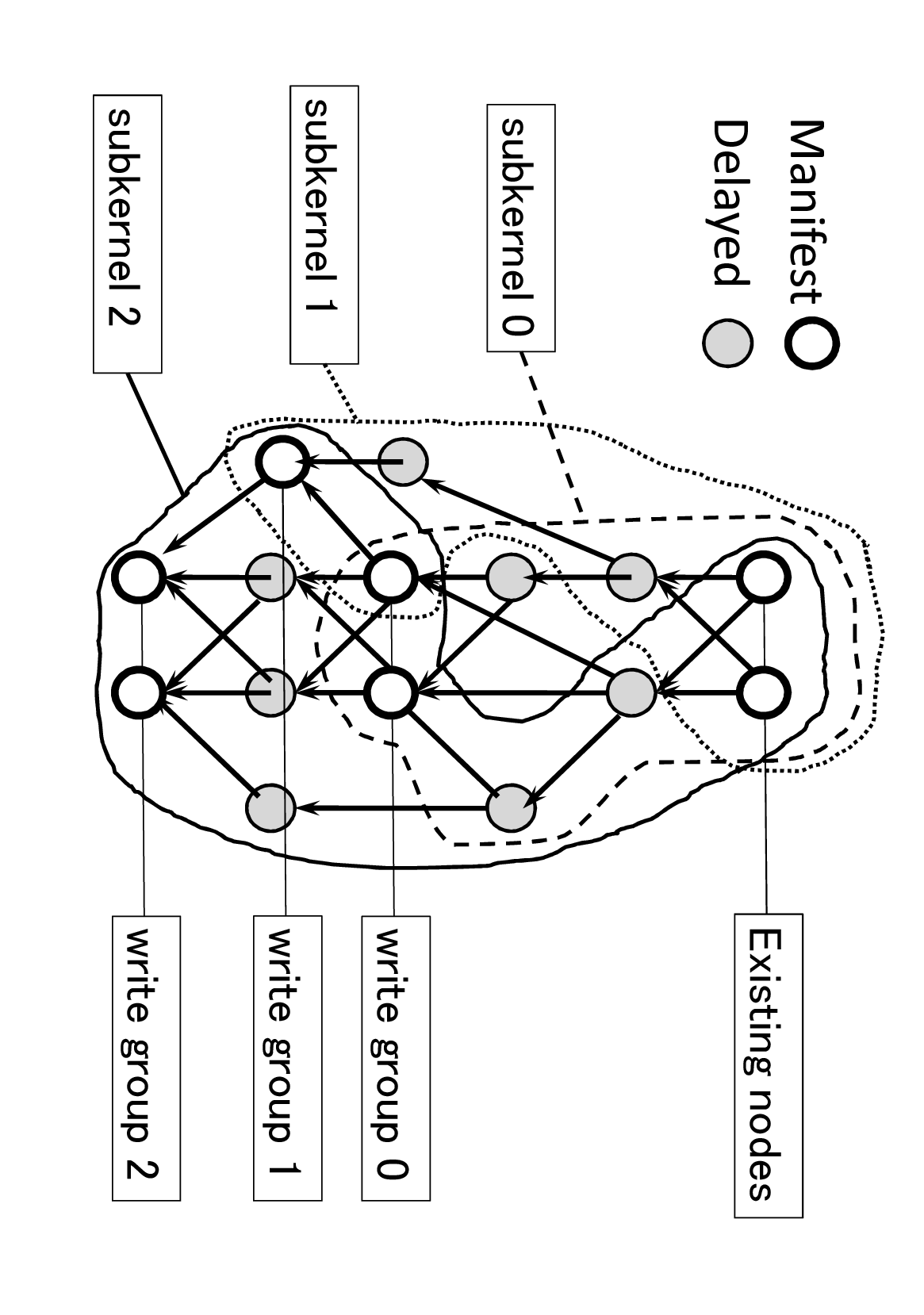}} & 
      (d) The Kernel is divided to multiple subKernels, each of which representing one write group.
        Each subKernel is in turn translated to a function in C or a {\tt \_\_global\_\_} function in CUDA.\\
    \end{tabular}
    \caption {
      \delspan{%
        Code generation mechanism of Paraiso.
      }
      \addspan{
        How Paraiso generates a Plan from the given set of annotations.
      }
    }\label{figureSubKernel}
  \end{center} 
\end{figure}

Once the analysis and optimizations are done, an OM is translated to a
code generation {\tt Plan}. Here, decisions are made for how many
memory are used, what portion of computation goes into a same
subroutine, and so on.  The nodes in data-flow graph are greedily merged as long
as they have no dependence and can be calculated in the same loop 
(Fig. \ref{figureSubKernel}).

The {\tt Plan} is further translated to {\tt CLARIS} (C++-Like
Abstract Representation of Intermediate Syntax).  CLARIS is subset of
C++ and CUDA syntax which is sufficient in generating codes in scope
of Paraiso.

Finally, CLARIS is translated to native C++ or CUDA codes.

\section{Automated Tuning Mechanism} \label{sectionAutoTuning}

\subsection{Tuning Targets} \label{sectionTuningTargets}

The objects we want to optimize are the implementations of a partial
differential equations solving algorithm. We call them {\em
  individuals}, adopting the genetic algorithm terminology. Each
individual has a {\em genome} that encodes how it have chosen to
implement the algorithm. The choices are (1) how much computation
speed and memory bandwidth to use, (2) where to synchronize the
computation, and (3) the CUDA kernel execution configuration.  The
fitness of the genome is the benchmark score of the generated code,
measured in cups (the number of fluid {\bf c}ells {\bf u}pdated {\bf p}er {\bf s}econd)
which we want to maximize.

A simple example program Fig. \ref{figureTradeoff} indicates that we
have implementation choices for intermediate variables: whether to
store its entire contents on the memory ({\tt Manifest}) or not to
store them and recompute them as they're needed ({\tt Delayed}).  The
terms {\tt Manifest} and {\tt Delayed} are inherited from REPA
\cite{keller_regular_2010}.  If we increase the number of {\tt
  Manifest} nodes, we consume less arithmetic units but more memory
and its bandwidth; decreasing the number of {\tt Manifest} nodes have
the opposite effect.  There are two extreme configurations, one is
making as many nodes {\tt Manifest} as possible and the other is
making as few nodes {\tt Manifest} as possible. In most cases both of
them result in poor performance and moderate configurations are
faster.  Paraiso generate codes for many possible combinations of {\tt
  Manifest} / {\tt Delayed} choices (Fig. \ref{figureSubKernel}) and
searches for such configurations by automated tuning.

Another tuning done by Paraiso is the choice of synchronization
points.  In CUDA, inserting {\tt \_\_syncthreads()}, especially before
load/store instructions cause the next instruction to coalesce and
increase the speed of the program. Inserting too much synchronization,
on the other hand, is a waste of time.  Again, Paraiso searches for
better configurations by automated tuning.

The last tuning done by Paraiso is to find the optimal CUDA kernel
execution configuration, i.e. how many CUDA threads and thread blocks
to be launched simultaneously.  These tuning items summed, the genome
size is approximately $6000$ bits for our hydrodynamics solver, which
means that there are $2^{6000}$ possible implementations. Brute-force
searching for {\em the} fastest implementation from this space is
inutile and in the first place impossible.  Instead, the goal of Paraiso is to
stochastically solve such a global optimization problem in this genome
space, where the function to be optimized is the benchmark score of
the PDE solver generated from the genome.

\begin{table}
  \begin{tabular}{lll}
    %\multicolumn{3}{l}{The following pair of Paraiso kernels are the same but for the annotations:}\\
    \begin{tabular}{p{7cm}}
      \rowcolor{grey}
      \footnotesize
      \topsep=-1ex\relax
\begin{verbatim}
proceed = do 
  x <- bind $ loadLD density


  y <- bind $ x * x


  z <- bind $ y + y
  store density z
\end{verbatim}
    \end{tabular} & &
    \begin{tabular}{p{7cm}}
      \rowcolor{grey}
      \footnotesize
      \topsep=-1ex\relax
\begin{verbatim}
proceed = do 
  x <- bind $ 
       Anot.add Sync.Pre <?>
       loadLD density
  y <- bind $ 
       Anot.add Alloc.Manifest <?>
       x * x
  z <- bind $ y + y
  store density z
\end{verbatim}
    \end{tabular} 
    \\ \\
    \multicolumn{3}{l}{Their genome:}\\
{\tt AAAAACAAAAAAACAAAAAA} & &
{\tt AAAAACAAAAAAACAACGAA}
    \\ \\
    \multicolumn{3}{l}{The generated header file, abbreviated:}\\
    \begin{tabular}{p{7cm}}
      \rowcolor{grey}\scriptsize\topsep=-1ex\relax
\begin{verbatim}
class Hello {
public:device_vector<double> static_0_density;
public:device_vector<double> manifest_0_5;


public:void Hello_sub_0 
  (const device_vector <double> &a1,
   device_vector <double> &a5);



public:void proceed ();
};
\end{verbatim}
    \end{tabular} & &
    \begin{tabular}{p{7cm}}
      \rowcolor{grey}\scriptsize\topsep=-1ex\relax
\begin{verbatim}
class Hello {
public:device_vector<double> static_0_density;
public:device_vector<double> manifest_0_3;
public:device_vector<double> manifest_0_5;

public:void Hello_sub_0 
  (const device_vector<double> &a1,
   device_vector < double >&a3);
public:void Hello_sub_1 
  (const device_vector<double> &a3,
   device_vector<double> &a5);
public:void proceed ();
};
\end{verbatim}
    \end{tabular}
    \\ \\
    \multicolumn{3}{l}{The generated {\tt .cu} program file, abbreviated:}\\
    \begin{tabular}{p{7cm}}
      \rowcolor{grey}\scriptsize\topsep=-1ex\relax
\begin{verbatim}
__global__ void Hello_sub_0_inner 
  (const double *a1, double *a5) {
  for (int i = INIT; (i) < (256);
       (i) += STRIDE) {
    int addr_origin = i;
    double a1_0 = (a1)[(addr_origin) + (0)];
    double a3_0 = (a1_0) * (a1_0);
    ((a5)[addr_origin]) = ((a3_0) + (a3_0));
  }
}











void Hello::proceed () {
  Hello_sub_0 (static_0_density, manifest_0_5);

  (static_0_density) = (manifest_0_5);
}
\end{verbatim}
    \end{tabular} & &
    \begin{tabular}{p{7cm}}
      \rowcolor{grey}\scriptsize\topsep=-1ex\relax
\begin{verbatim}
__global__ void Hello_sub_0_inner 
  (const double *a1, double *a3) {
  for (int i = INIT; (i) < (256);
       (i) += STRIDE) {
    int addr_origin = i;
    __syncthreads ();
    double a1_0 = (a1)[(addr_origin) + (0)];
    ((a3)[addr_origin]) = ((a1_0) * (a1_0));
  }
}

__global__ void Hello_sub_1_inner 
  (const double *a3, double *a5) {
  for (int i = INIT; i < (256);
       (i) += STRIDE) {
    int addr_origin = i;
    double a3_0 = (a3)[(addr_origin) + (0)];
    ((a5)[addr_origin]) = ((a3_0) + (a3_0));
  }
}

void Hello::proceed () {
  Hello_sub_0 (static_0_density, manifest_0_3);
  Hello_sub_1 (manifest_0_3, manifest_0_5);
  (static_0_density) = (manifest_0_5);
}
\end{verbatim}

    \end{tabular}
  \end{tabular}
  \caption {An example of changes in the genome and the implemented algorithm caused by annotations.}
  \label{tblAnnotationSample}
\end{table}

\addspan{
Table \ref{tblAnnotationSample} shows a simple Paraiso kernel and 
how its genome and implementation is altered by adding annotations. 
This kernel named {\tt proceed} performs the following calculations:
\begin{eqnarray}
  \begin{array}{l}
    \hspace{-1cm} \mathrm{foreach}\ {\mathbf i}: \\
    x = \mathrm{density}[{\mathbf i}] \\
    y = x \times x \\
    z = y + y \\
    \mathrm{density}[{\mathbf i}] = z
  \end{array}
\end{eqnarray}
It first loads from an array named {\tt density}, performs a multiplication, then an addition,
and then stores the result to {\tt density} again.
}

\addspan{
The header file on the right side of the Table \ref{tblAnnotationSample}, compared to the left one,
allocates an additional {\tt device\_vector} and declares an additional subkernel
as results of a {\tt Manifest} annotation.
The right {\tt .cu} file, compared to the left one, differs in two points: one is that it
performs the multiplication {\tt y = x * x} and the addition {\tt z = y + y} in separate CUDA
kernels and stores the intermediate result to the 
additional {\tt device\_vector}, which is yet another result of the {\tt Manifest} annotation.
Another difference is that it calls {\tt \_\_syncthreads()} just before the load instruction,
which is the result of the synchronization annotation.
}

\subsection{Parallel Asynchronous Genetic Simulated Annealing} \label{sectionPAGSA}

Simulated annealing has been widely used to solve global optimization
problems.  However, a standard simulated annealing method is not
suitable for parallelization, and if the annealing schedule (how we
cool down the temperature as function of time) is too quick, it tends
to fall into local minima.  Replica-exchange Monte Carlo method
\cite{hukushima_exchange_1996} solves these drawbacks by introducing
multiple replicas of heat baths with different temperatures, and by
allowing replica exchange between adjacent heat baths. Thus, replicas
can be computed in parallel, and the annealing schedule is spontaneously
managed by replica migration.

\begin{figure}
  \begin{center}
    \vspace{-2cm}
    \includegraphics[width=12cm,angle=270]{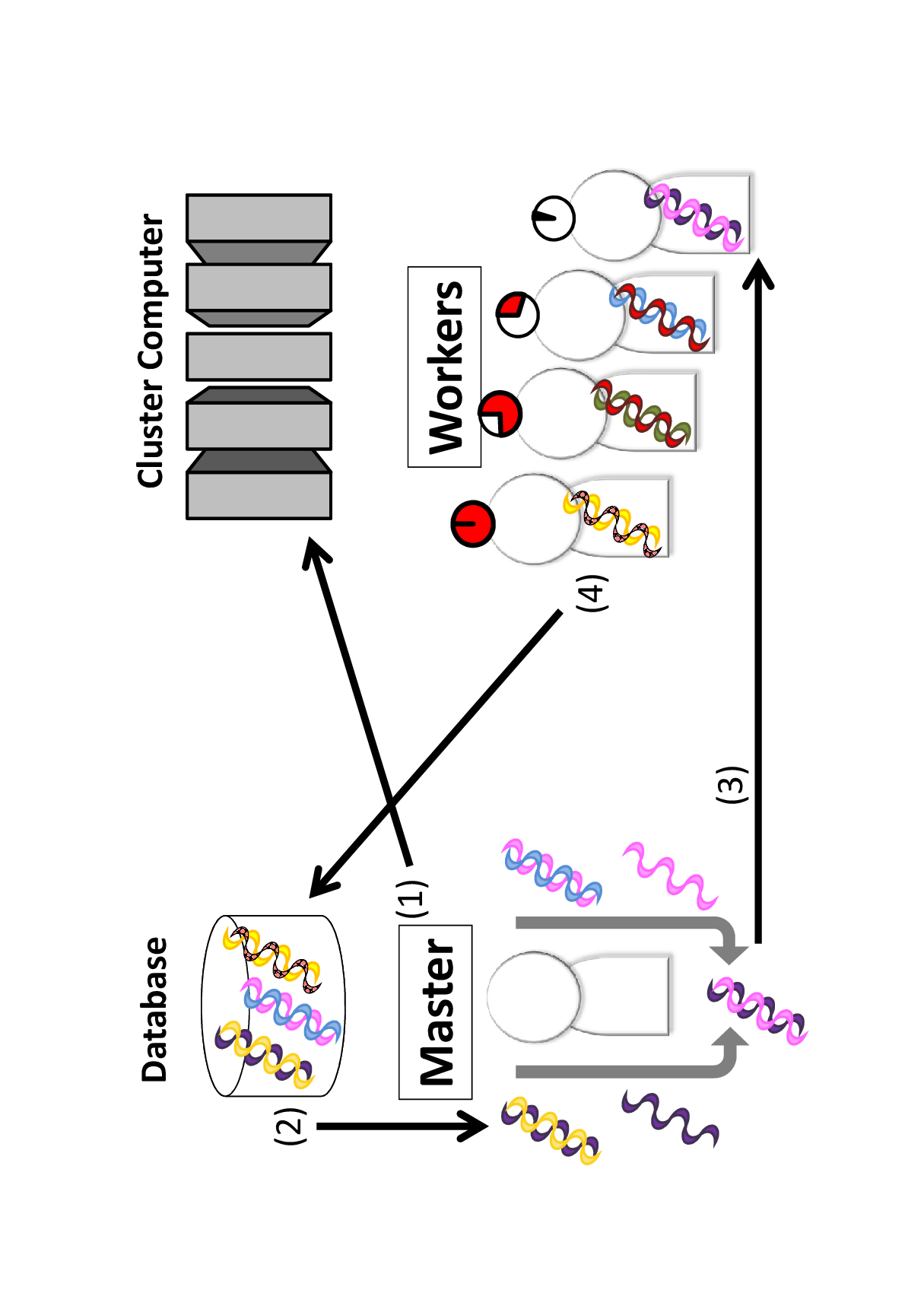}
    \vspace{-2cm}
  \end{center}
  \addspan{
    \caption{
      The master/worker model of parallel asynchronous genetic simulated annealing 
      used for automated tuning in Paraiso. 
      (1) The master periodically monitors for vacancy in the cluster computer system. 
      (2) When a vacancy is found, the master 
      creates a new individual from several genomes
      obtained from the database by performing {\em draws} of temperature $T$.
      (3) The master creates a worker job that generates native program from the genome
      and measures its speed.
      (4) Each worker writes the benchmark result to the database and quit 
      as soon as it finishes its assigned measurements.
    }\label{figureMasterWorker}
  }
\end{figure}

We further extend this method to fit into benchmark based tuning on
shared cluster computer systems. First, in shared systems it is hard
to maintain a fixed number of replicas because the available amount of
nodes changes with time. Second, it is not efficient to perform
replica-exchange synchronously, since the wall clock time required to
calculate the fitness function varies with replicas.  For these two
reasons, we modify replica-exchange Monte Carlo method to a
master/worker model (c.f. Fig. \ref{figureMasterWorker}), 
where a master asynchronously launches varying
number of workers in parallel.

The master holds the genomes and scores of all the past individuals in
a database (DB). The role of the master is to draw individuals from
the DB, to create new individuals using their genomes, and to launch
the workers when the computer resource is available.  The role of the
workers, on the other hand, is to generate codes from the given
genome, to take the benchmark, and to write the results into the DB.

In this way we can launch any number of workers asynchronously as long
as the DB is not a bottleneck. In addition to that, we eliminate a
common weak point of simulated annealing and genetic algorithms ---
that individuals of older generation are overwritten and are inaccessible.

For each individual $I$, the generated code is benchmarked 30 times.
We record the mean $\mu(I)$ and the sample standard deviation $\sigma(I)$
of the score. It is important to record the deviation. When we benchmark
each individual only once, some individuals receive over-evaluated score
by chance, infest the system and tend to stall the evolution.

At the end of each benchmark, the individual is briefly tested if the
state of the simulation has developed substantially from the initial
condition and there is no NaN (not a number). The test is chosen
because not evolving at all and generating NaN are two dominant modes
of failure, and the individual test time needs to be kept smaller than
benchmark itself.  If an individual fails a test, its score is $0$. At
the end of a tuning experiment, the champion individual is extensively
tested if it can reproduce various analytic solutions, which takes
hours.

A {\em draw} is the operation to randomly choose an individual from 
the DB. Each draw has a temperature $T$. In a draw of temperature $T$, the
probability the individual $I$ is chosen is proportional to
\begin{eqnarray}
  \exp\left(\frac{\mu(I_\top) - \mu(I) + \sigma(I_\top) + \sigma(I)}{T + \sigma(I_\top) + \sigma(I)}\right) ,
\end{eqnarray}
where $I_\top$ is the individual with the largest $\mu(I)$.

Low-temperature draws strictly prefer high-score individuals while high-temperature draws does
not care the score too much. And the difference small compared to $\sigma(I_\top) + \sigma(I)$
is gracefully ignored.

Every time master creates a new individual, it chooses the draw
temperature $T$ randomly, so that the probability density of $\log{T}$
is uniformly distributed between $\log(\sigma(I_\top))$ and
$\log(\mu(I_\top))$. 

To summarize, our method is a master/worker variant of
replica-exchange Monte Carlo method \cite{hukushima_exchange_1996},
using genetic algorithms as neighbor generators.
Thus, our method 
\begin{itemize}
\item can utilize parallel and dynamically varying computer resource.
\item can find global maximum without hand-adjusted annealing schedule.
\item can combine independently-found improvements.
\end{itemize}
An extensive review on the applications of the evolutionary computation in
astronomy and astrophysics is found in \cite{gutierrez_evolutionary_2012}.

\subsection {Three Methods of Birth for Generating New Individuals}

\begin{center}
  \begin{figure}
    \includegraphics[width=10cm,angle=90]{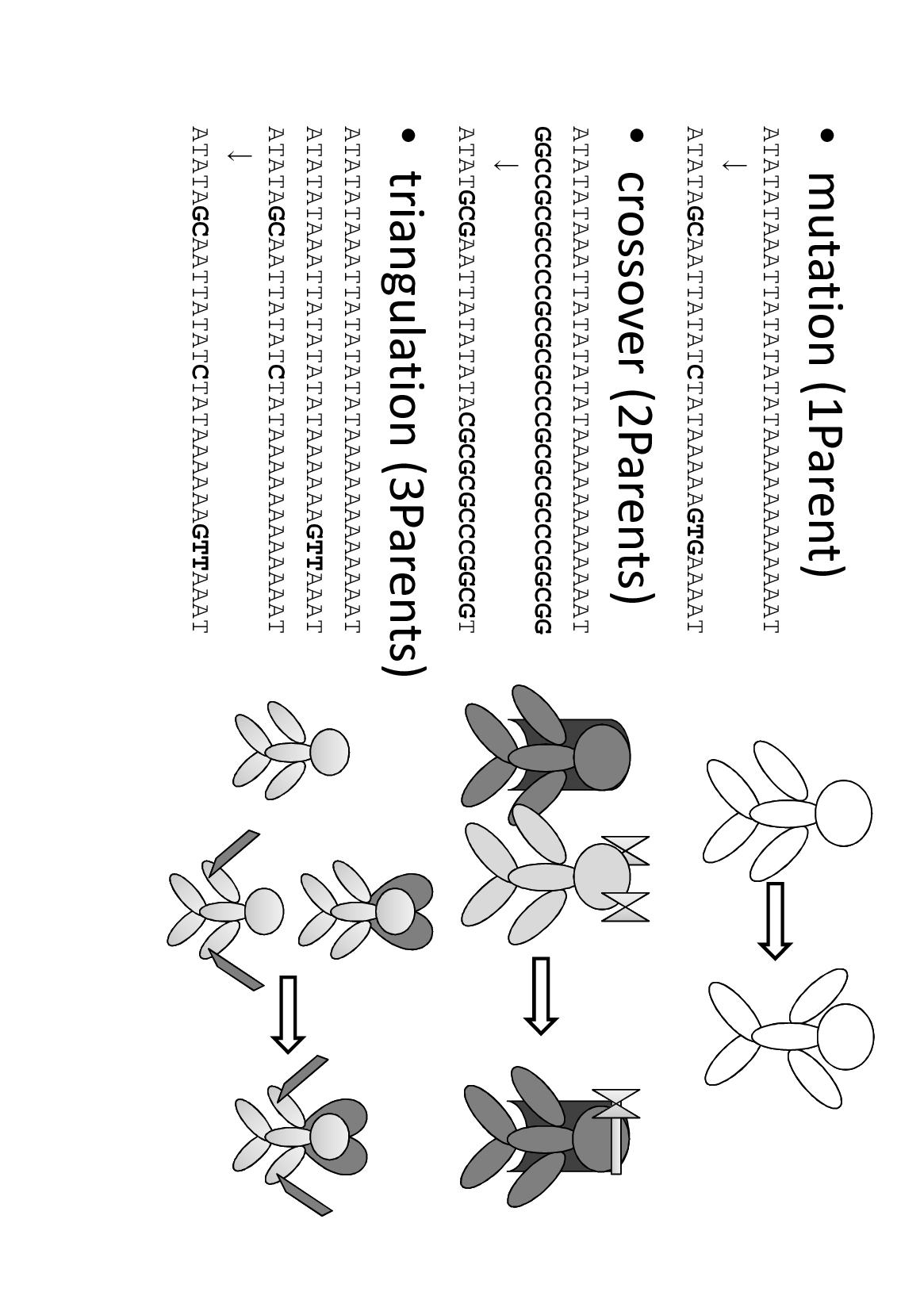}
    \caption{Three ways of generating new individuals.}
    \label{figureMutation}
  \end{figure}
\end{center} 

We use three different methods of birth to generate new individuals (c.f.
Fig. \ref{figureMutation}).
The first method is {\em mutation}. We draw one individual from the
database (DB), take its genome, overwrite it randomly and create a new
genome. 
The second method is {\em crossover}. We draw two individuals from the 
DB, split their genomes at several random points and exchange the segments.
The third method is {\em triangulation}, or three-parent crossover.

Multi-parent crossovers are crossovers that takes more than two
parents and create one or multiple children. The proposed multi-parent
crossover methods include taking bit-wise majority vote
\cite{eiben_genetic_1994, sivanandam_introduction_2007} and exchanging
segments between multiple parents \cite{eiben_orgy_1995}.
Multi-parent crossovers for real-number coded genomes are also studied
\cite{tsutsui_multi-parent_1999}.  The novelty of our three-parent
crossover (triangulation) is to take the scores of the parents into
consideration.  Triangulation is designed to efficiently combine
independent improvement found in sub-problems.

FFTW \cite{frigo_design_2005} uses the divide and conquer approach to
its target problems. It first recursively decomposes large FFT problem
into smaller ones and solves the optimization problem from smaller
part. In contrast, Paraiso deals with monolithic data-flow graphs. It
is not obvious how to decompose them into subgraphs --- decompositions
are actually the target of optimization. Therefore, we optimize
subgraphs in vivo; we optimize subgraphs keeping them embedded into
the entire graph.

\begin{table}
  \begin{center}
    \begin{tabular}{r|cccccccc}
      Base      & 0 & 0 & 0 & 0 & 1 & 1 & 1 & 1 \\
      Secondary & 0 & 0 & 1 & 1 & 0 & 0 & 1 & 1 \\
      Primary   & 0 & 1 & 0 & 1 & 0 & 1 & 0 & 1 \\
      \hline
      Child     & 0 & 1 & 1 & 1 & 0 & 0 & 0 & 1
    \end{tabular}
  \end{center}
  \caption {Binary operation table for triangulation.} \label{tableTriang}
\end{table}

Aiming to merge two distinct optimized subgraphs, we draw three
individuals from the DB. Then we sort them in ascending order of the
score and name them {\tt Base}, {\tt Secondary} and  {\tt Primary}. We
perform bit-wise operation as in Table \ref{tableTriang} to create
their child.  For each bit, if at least one of {\tt Primary} or {\tt
  Secondary} has been changed from {\tt Base}, then we adopt the
change.

When creating a new individual, the master chooses mutation,
crossover, or triangulation at equal probability of $1/3$. Then the
master draws the needed number of parents from the DB. 
If two of the parents are the same the master retries the draws. Otherwise,
the master creates a child with the chosen method of birth. 
Then if the genome of the newly created individual is already
in the DB, the master discards the individual and retries the mutation
until it gets a genome not found in the DB. With each retry, draw
temperature $T$ is multiplied by $1.2$, so that the master eventually
succeed in creating a new genome.

\section{Tuning Experiments} \label{sectionExperiment}

\subsection{The Target Program}

We implemented a 2nd order compressible hydrodynamics solver 
\cite{toro_riemann_2009} in Paraiso, and then optimize it. Here we detail numerical scheme.

Let $A_a, B_{ab}, ...$ denote tensor indices for tensor $A, B, ...$. 
Let $[{\mathbf i}], [{\mathbf j}], ...$ denote array indices which is a $d$-tuple
of integer where $d$ is the dimension. Values integrated over cell volumes are
given integer indices, and values integrated over cell surfaces are given indices
with one half-integer and $d-1$ integers. For example, 
$A_{y}[{\mathbf i}]$ is the volume integral of $y$-component of a vector $A$ over the cell ${\mathbf i}$, 
and 
$B_{xz}[{\mathbf i} + \frac{1}{2}{\mathbf e}_x]$
 is the surface integral of $xz$-component of a rank-2 tensor $B$ over the $+x$ surface of cell ${\mathbf i}$.

The equations of compressible hydrodynamics have $d+2$ degrees of freedom. The primitive variables $\mathbf V$
and the conserved variables $\mathbf U$ are the two representations of the degrees of freedom. The flux variables $\mathbf F$
are calculated from  $\mathbf U$ or  $\mathbf V$, as follows :
\begin{eqnarray}
  {\mathbf V} = \left(
  \begin{array}{c}
    \rho \\ v_a \\ p
  \end{array}
  \right)  , \label{eqPrimitiveVars}
  \\
  {\mathbf U} = \left(
  \begin{array}{c}
    \rho \\ m_a \\ E
  \end{array}
  \right) , \label{eqConservedVars}
  \\
  {\mathbf F}_a = \left(
  \begin{array}{c}
    u_a \\ 
    u_a m_b + p \delta_{ab} \\ 
    u_a E + u_a  p
  \end{array}
  \right) .
\end{eqnarray}
where $\rho$ , $v_a$, $m_a$, $p$ and $E$ are the density, the
velocity, the momentum, the pressure and the total energy,
respectively, The primitive and conserved variables are related by
$m_a = \rho v_a$, $E = E_i(p) + \frac{1}{2}\rho v^2$ where
$E_i$ is the internal energy of the gas. By assuming the adiabatic equation of state
for perfect gas, $E_i(p) = (\gamma -1)^{-1}p $ where $\gamma$ is the
 ratio of specific heats.

We numerically solve the Euler equations $\partial_t {\mathbf U} =
\partial_a {\mathbf F}_a$ , or more specifically:
\begin{eqnarray}
  \partial_t \rho &=& \partial_a (\rho v_a) ,\\
  \partial_t m_b &=& \partial_a (u_a m_b + p \delta_{ab}) ,\\ 
  \partial_t E &=& \partial_a (u_a E + u_a  p) .
\end{eqnarray}

\def\half{{\textstyle \frac{1}{2}}}

The numerical method we used is as follows, based on a function $\mathrm{Flux}_a(\mathbf {U0})$
which calculates the fluxes across the cell surfaces using a Riemann solver.
$\mathrm{Flux}_a(\mathbf {U0})$ is defined as follows. First, $\mathbf {V0}$ are the primitive variables calculated from $\mathbf {U0}$:
\begin{eqnarray}
  \mathbf {V0} = {\mathbf V}(\mathbf {U0}).
\end{eqnarray}
The interpolated primitive variables $\mathbf {VL}$ and $\mathbf {VR}$ are 
\begin{eqnarray}
  (\mathbf {VR}[{\mathbf i} - \half{\mathbf e}_a], \mathbf {VL}[{\mathbf i} + \half{\mathbf e}_a]) = &\nonumber \\
  \hspace{1cm} \mathrm {Interpolate}(\mathbf {V0} [{\mathbf i}-{\mathbf e}_a], \mathbf {V0} [{\mathbf i}], \mathbf {V0} [{\mathbf i}+{\mathbf e}_a]), \label{eqInterpolateV}
\end{eqnarray}
where we use piecewise linear interpolation with {\sc minbee} flux limiter \cite{sweby_high_1984,toro_riemann_2009}:
\begin{eqnarray}
  \begin{array}{lrcl}
    \multicolumn{4}{l}{\mathrm {Interpolate}(\mathit{y0}, \mathit{y1}, \mathit{y2}) = 
      (\mathit{y1} - \half \mathit{dy}, \mathit{y1} + \half \mathit{dy}) \ \ \mathrm{where} }\\
      \ & \mathit{dy01} &=& \mathit{y1} - \mathit{y0} \\
      \ & \mathit{dy12} &=& \mathit{y2} - \mathit{y1} \\
      \ & \mathit{dy} &=& \left\{ \begin{array}{lll} 
        0             &\mathrm{if } & \mathit{dy01}  \cdot \mathit{dy12}  < 0   \\
        \mathit{dy01} &\mathrm{if } & |\mathit{dy01}|  < |\mathit{dy12}|         \\
        \mathit{dy12} & \multicolumn{2}{l}{\mathrm{otherwise}} .
        \end{array} \right.  \label{eqInterpolateSingle} \\
  \end{array}
\end{eqnarray}
 Then, fluxes across the boundaries are 
defined using the HLLC Riemann solver \cite{toro_restoration_1994}.
\begin{eqnarray}
  {\mathbf F}_a[{\mathbf i} + \half{\mathbf e}_a] &=& 
   \mathrm {HLLC}_{a}(\mathbf {VL}[{\mathbf i} + \half{\mathbf e}_a], \mathbf {VR}[{\mathbf i} + \half{\mathbf e}_a]) .
\end{eqnarray}
This is the flux defined by $\mathrm{Flux}_a(\mathbf {U0})$.
Then, linear addition of flux to conserved variable $\mathrm{AddFlux}(\Delta t, \mathbf F_a, \mathbf U)$ is defined by
\begin{eqnarray}
  & & \mathbf {U2} = \mathrm{AddFlux}(\Delta t, \mathbf F_a, \mathbf {U1}) \nonumber \\
  &\Leftrightarrow & \mathbf {U2}[{\mathbf i}] = \mathbf {U1}[{\mathbf i}] 
  + \sum_{a} \frac{\Delta t}{\Delta r_a} \left(\mathbf{F}_a[{\mathbf i} - \half{\mathbf e}_a] - \mathbf{F}_a[{\mathbf i} + \half{\mathbf e}_a]\right), \label{eqAddFlux}
\end{eqnarray}
where $\Delta r_a$ is the mesh-size along the $a$-axis.  Using these
notations, we construct the second-order time marching as follows,
where $\Delta t$ is the time step determined by the CFL-condition:
\begin{eqnarray}
  \mathbf{F0}_a &=& \mathrm{Flux}_a(\mathbf {U0}) \\
  \mathbf{U1} &=& \mathrm{AddFlux}(\half \Delta t, \mathbf {F0}_a, \mathbf {U0}) \\
  \mathbf{F1}_a &=& \mathrm{Flux}_a(\mathbf {U1}) \\
  \mathbf{U2} &=& \mathrm{AddFlux}(\Delta t, \mathbf {F1}_a, \mathbf {U0}) 
\end{eqnarray}
$\mathbf{U2} $ is the set of conserved variables for the next generation.

Notice how we rely on human mind flexibility when we convey the
algorithms in the forms like above.  In languages like Fortran or C,
we are often forced to decompose those expression into elemental
expressions and the source codes tend to become longer. In Haskell, we
can express these ideas in well-defined machine readable forms while
keeping the compactness and flexibility as in above.

We declare both the primitive variables Eq. (\ref{eqPrimitiveVars}) and
conserved variables Eq. (\ref{eqConservedVars}) as instances of the type
class {\tt Hydrable}, the set of variables large enough for
calculating any other hydrodynamic variables.  We used the general
form by default, and {\tt bind} either the primitive or conserved
variables when we need the specific form.  By doing so requisite
minimum number of conversion code between primitive and conserved
variables are generated.

When we defined the {\sc minbee} interpolation for a triplet of real
numbers Eq. (\ref{eqInterpolateSingle}) and then applied it to
primitive variables Eq. (\ref{eqInterpolateV}), we implicitly
extended a function on real numbers to function on set of real
numbers. In Haskell, we can define how any such generalized function
application over the set of hydrodynamic variables should behave, just
by making it an instance of {\tt Applicative} type class.

When we calculate the space-derivatives of the fluxes in
Eq. (\ref{eqAddFlux}), the component of the flux we access (index $a$
in $\mathbf{F}_a$) and the direction in which we differentiate
(indices $a$ in $[{\mathbf i} - \half{\mathbf e}_a]$ and $[{\mathbf i}
  + \half{\mathbf e}_a]$) should match. Although the flux $\mathbf{F}$
and the spatial array indices ${\mathbf i}$ have very different types,
Haskell's type inference guarantees that they are both tensors of the
same dimensions, and we can access both of them by the common tensor
index $a$. And we can sum over $a$, because tensors are {\tt Foldable}
and their components are {\tt Additive}.  Tensors are {\tt
  Traversable} as mentioned before, and any type constructors that are
{\tt Traversable} are also {\tt Foldable}.

\subsection{Annotating By Hand}

\begin{table}
  \begin{center}
    \begin{tabular}
      {l   | c|c|c||r    | r           | r            }
      ID & config & (1)&(2)& lines & subKernel & memory  \\
      \hline
      \Izag  &$32\times32$ &D&D& 13128 &  7 & $52\times N$ \\
      \Izam  &$448\times256$ &D&D& 13128 &  7 & $52\times N$ \\
      \Iwat  &$448\times256$ &M&D& 17494 & 12 & $68\times N$ \\
      \Shin  &$448\times256$ &D&M&  3010 & 11 & $68\times N$ \\
      \Haya  &$448\times256$ &M&M& 3462 & 15 & $84\times N$ \\
    \end{tabular}

    \

    \begin{tabular}
      {l   | r    | r                    }
      ID & score (SP) & score (DP) \\
      \hline
      \Izag & $ 1.551 \pm 0.0005$ & $ 1.138 \pm 0.000$\\ %  \pm 0.00003$\\
      \Izam & $ 5.838 \pm 0.004$  & $ 3.091 \pm 0.002$\\
      \Iwat & $ 5.015 \pm 0.002$  & $ 2.491 \pm 0.001$ \\
      \Shin & $42.682 \pm 0.083$  & $19.831 \pm 0.021$ \\
      \Haya & $34.100 \pm 0.110$  & $15.632 \pm 0.024$ \\
    \end{tabular}
    \caption{ The codes annotated by hand and their
      performances. First four columns are the ID of the individuals,
      the CUDA kernel execution configuration, and the two choices of
      annotation --- whether to {\tt Manifest} or {\tt Delay} (1) the
      result of spatial interpolation (2) the result of the Riemann
      solver.  Next three columns are the properties of the generated
      codes as results of these choices made.  They are namely the
      size of the code (number of lines), the number of subKernels in
      the code, the memory consumption in proportion to fluid cell
      number $N$.  The last two columns are the speed of the generated
      codes for single precision (SP) and double precision (DP). The
      speed of the codes are measured by Mcups ($10^6$ cell update per
      second).  } \label{tableHandGen}
  \end{center}
\end{table}

\begin{table}
  \begin{tabular}{p{15cm}}
    The interpolation code for \Izam:\\
    \rowcolor{grey}
    \small
    \topsep=-1ex\relax
\begin{verbatim}
interpolateSingle order x0 x1 x2 x3 
  | order == 1 = do
    return (x1, x2)
  | order == 2 = do
    d01 <- bind $ x1-x0
    d12 <- bind $ x2-x1
    d23 <- bind $ x3-x2
    let absmaller a b = select ((a*b) `le` 0) 0 $ 
                        select (abs a `lt` abs b) a b
    d1 <- bind $ absmaller d01 d12
    d2 <- bind $ absmaller d12 d23
    l <- bind $ x1 + d1/2
    r <- bind $ x2 - d2/2
    return (l,r)
\end{verbatim}
\\
    For \Iwat, the last line is modified as follows:\\
    \rowcolor{grey}
    \small
    \topsep=-1ex\relax
\begin{verbatim}
    return (Anot.add Alloc.Manifest <?> l,  Anot.add Alloc.Manifest <?> r)
\end{verbatim}
  \end{tabular}
  \addspan{
    \caption{
      An annotation made in Paraiso source code that causes the results of the 
      interpolations to be stored on memory and reused.
    }\label{tableAnotIwat}
  } 
\end{table}

Before we start the automated tuning experiment, we generate several
individuals by hand. The initial individual \Izag\ is the one with the
least number of {\tt Manifest} nodes as possible.  \Izam\ is the same individual
with the CUDA execution configuration suggested by the CUDA occupancy
calculator.  Based on it, we add several {\tt Manifest} annotation and
create new individuals (c.f.  Table \ref{tableHandGen}). Manual
annotations are not blind-search process; each annotation has clear
motivation such as ``let us store the result of the Riemann solvers
because it is computationally heavy'', ``let us store the result of
interpolations because it moves a lot of data,'' etc. 
\addspan{
  For example, the sample code in Table \ref{tableAnotIwat} shows the
  implementation of the piecewise linear interpolation with the {\sc minbee} flux limiter, Eq. (\ref{eqInterpolateSingle})
  in Paraiso, and how to annotate the return values of the limiter as  {\tt Manifest}.
}

We use \Izam\ as
the base line of the benchmark, and use \Izag\ as the initial
individual of some experiment to see if the automated tuning can find
out the optimal CUDA execution configuration by itself.

The codes are benchmarked on TSUBAME 2.0 cluster at
Tokyo Institute of Technology.  Each individual was benchmarked on a
TSUBAME node with two Intel Xeon X5670 CPU(2.93GHz, 6 Cores $\times$
HT = 12 processors) and three M2050 GPU(1.15GHz 14MP x
32Cores/MP=448Cores).

The abstraction power of {\tt Builder Monad} lets us change the code
drastically with little modification.  For example, Paraiso source
code of \Shin\ and \Izam\ differs by just one line of annotation.  This
introduces the 16 bits of difference in their genome, causing \Shin\ 
generate a code that has $1/3$ lines, which contains four more
subroutines, consume $1.31$ times more memory and is $6.42$ times
faster.

\subsection{Automated Tuning}

\begin{table}
  \begin{center}
    \begin{tabular}
      {c    | c    | r          | r        | c                 | r}
      RunID & prec. & initial score & wct & best ID/total & high score \\
      \hline
      GA-1  & DP   &  $ 1.138 \pm 0.000$  & 3870 & % pm 0.0004
      20756 / 20885 & $14.158 \pm 0.002$ \\
      GA-S1 & DP   &  $ 1.138 \pm 0.000$ & 4120 & % pm 0.0004
      33958 / 34328 & $16.247 \pm 0.002$ \\
      GA-DE & DP   &  $19.253 \pm 0.044$ & 7928 &
      41250 / 41386 & $31.015 \pm 0.032$ \\
      GA-D  & DP   &  $19.253 \pm 0.044$ & 8770 &
      59841 / 68138 & $34.968 \pm 0.043$ \\
      GA-4  & DP   &  $19.253 \pm 0.044$  & 5811 &
      39991 / 40262 & $35.303 \pm 0.035$\\
      GA-F  & SP   &  $42.682 \pm 0.083$ & 2740 &
      23019 / 23062 & $53.300 \pm 0.078$ \\
      GA-F2 & SP   &  $42.682 \pm 0.083$  & 4811 &
      22242 / 24887 & $53.656 \pm 0.078$ \\
      \multicolumn{1}{c}{}\\
      GA-3D & SP    & $24.638 \pm 0.001$ & 5702 &
      38146 / 39200 & $45.443 \pm 0.116$
    \end{tabular}
  \end{center}
  \caption{
    The statistics of auto-tuning experiments. The columns are
    RunID, precision, the score of initial individual, the wall-clock time
    for the experiment (in minutes), the ID of the best individual and the number of
    individuals generated, the high score (in Mcups).
    Experiments GA-1 and GA-S1 started with \Izag, others started with \Shin. GA-3D started with 
    \Shin, and solved 3D problems.
  }\label{tableAT}
\end{table}

\begin{table}
  \begin{center}
    \begin{tabular}{l|ccc}
      \      & mutation    & crossover   & triangulation \\
      \hline
      GB-333 & $1/3$ & $1/3$ & $1/3$   \\
      \hline
      GB-370 & $1/3$ & $2/3$ & $  0$   \\
      \hline
      GB-307 & $1/3$ & $  0$ & $2/3$   \\
    \end{tabular}
    \caption{ The probability of the master node attempting each
    method of birth in experiment series GB-*. } \label{tableGBProb}
  \end{center}
\end{table}

\begin{table}
  \begin{center}
    \begin{tabular}
      {c    | c     | r                     | c             | r}
      RunID & prec. & initial score         & best ID/total & high score \\
      \hline
      GB-333-0 & DP   &  $19.253 \pm 0.044$ & 35294 / 40014&  $25.343 \pm 0.028$ \\
      GB-333-1 & DP   &  $19.253 \pm 0.044$ & 40256 / 40031&  $26.402 \pm 0.023$ \\
      GB-333-2 & DP   &  $19.253 \pm 0.044$ & 39206 / 40033&  $32.758 \pm 0.006$ \\
      GB-370-0 & DP   &  $19.253 \pm 0.044$ & 36942 / 39994&  $24.698 \pm 0.045$ \\
      GB-370-1 & DP   &  $19.253 \pm 0.044$ & 37468 / 40045&  $28.339 \pm 0.042$ \\
      GB-370-2 & DP   &  $19.253 \pm 0.044$ & 39719 / 40032&  $31.669 \pm 0.039$ \\
      GB-307-0 & DP   &  $19.253 \pm 0.044$ & 30320 / 40018&  $21.744 \pm 0.035$ \\
      GB-307-1 & DP   &  $19.253 \pm 0.044$ & 39358 / 40002&  $26.349 \pm 0.026$ \\
      GB-307-2 & DP   &  $19.253 \pm 0.044$ & 38107 / 40053&  $27.412 \pm 0.005$ \\
    \end{tabular}
  \end{center}
  \caption{ The statistics of GB-* experiment series. The columns are
    RunID, precision, the score of initial individual, the ID of the
    best individual and the number of individuals generated, the high
    score (in Mcups).  Experiments started with \Shin.  The nine
    experiments were performed in parallel, and took 92252 minutes of
    wall-clock time.  }\label{tableGB}
\end{table}

Next, we performed several automated tuning experiments c.f. Table
\ref{tableAT}.  

To distinguish the contribution of the three methods to generate new
individuals, we also performed automated tuning experiments with
either crossover or triangulation turned off.  Table \ref{tableGBProb}
shows the initial possibility of the master node attempting each
method of birth. Note that the actual frequencies of crossover and
triangulation are smaller than these value because the master may
default to mutation. Table \ref{tableGB} shows the result of
experiments. 
The resolutions were $1024^2$ for GA-1 and GA-S1,
problems, $100^3$ for GA-3D problems and $512^2$ for other GA-*
series.  The resolutions were $512^2$ in GB-* series.

The automated tuning system can generate and benchmark approximately
$10'000$ individuals per day. $20 - 100$ workers were running at the
same time.  It takes a few days to tune up \Izam\ to speed comparable
to \Shin, or speed up \Shin\ by another factor of $2$. The best speed
obtained was $35.3$Mcups for double precision, and $53.7$Mcups for
single precision.  Our automated tuning experiments on 3D solvers mark
$42.4$Mcups SP.  These are competitive performances to hand-tuned codes
for single GPUs; e.g.  Schive et. al. \cite{schive_directionally_2011}
reports $30$Mcups per C2050 card (single precision, note that their
code is 3D. Asunci\'ona et.al. \cite{de_la_asunciona_efficient_????}
reports $6.8$Mcups per GTX580 card (single precision, 2D).

\begin{figure}
  \begin{center}
    \begin{tabular}{ccc}
      (a)&(b)&(c) \\
      \includegraphics[width=5cm]{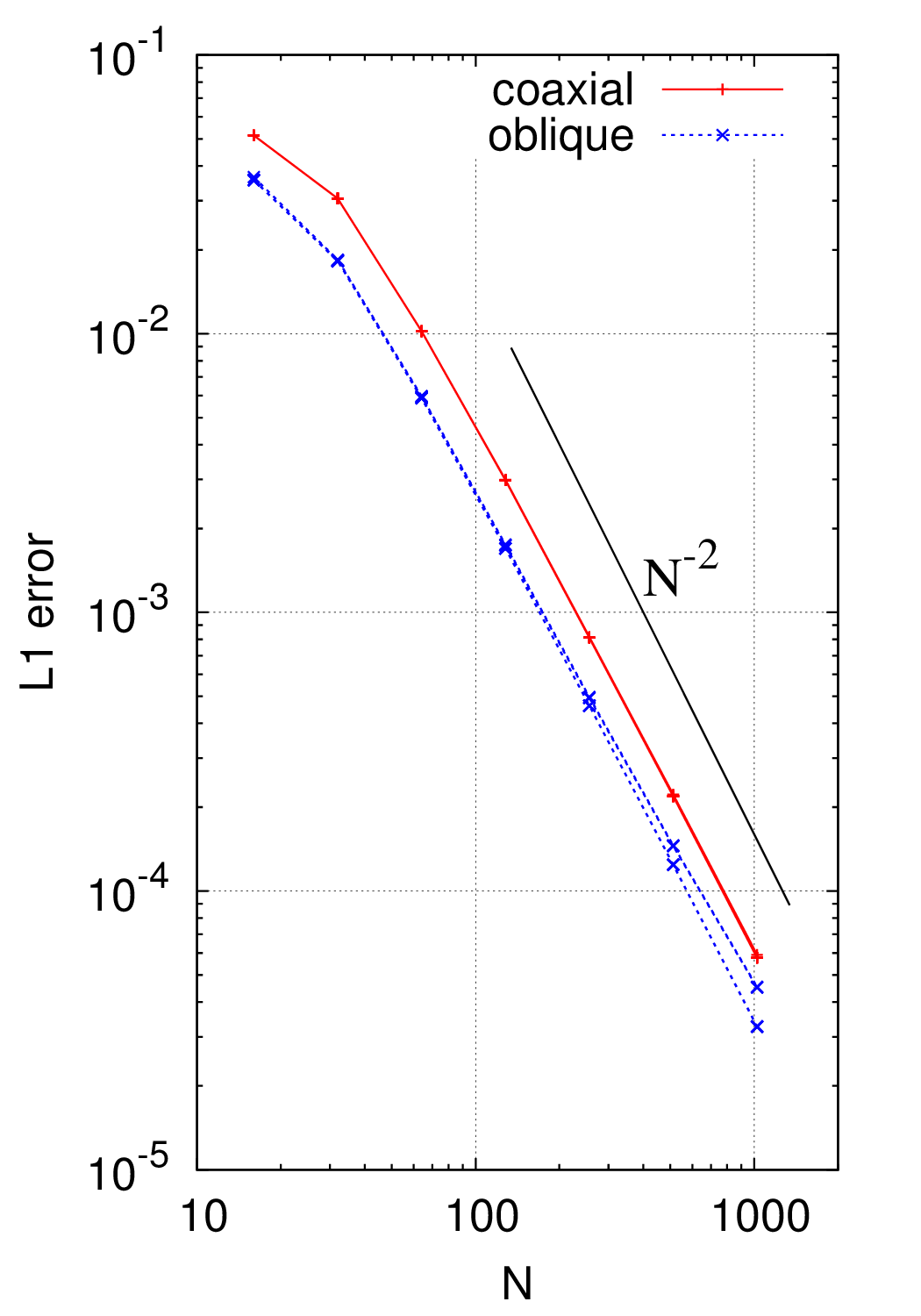}&
      \includegraphics[width=5cm]{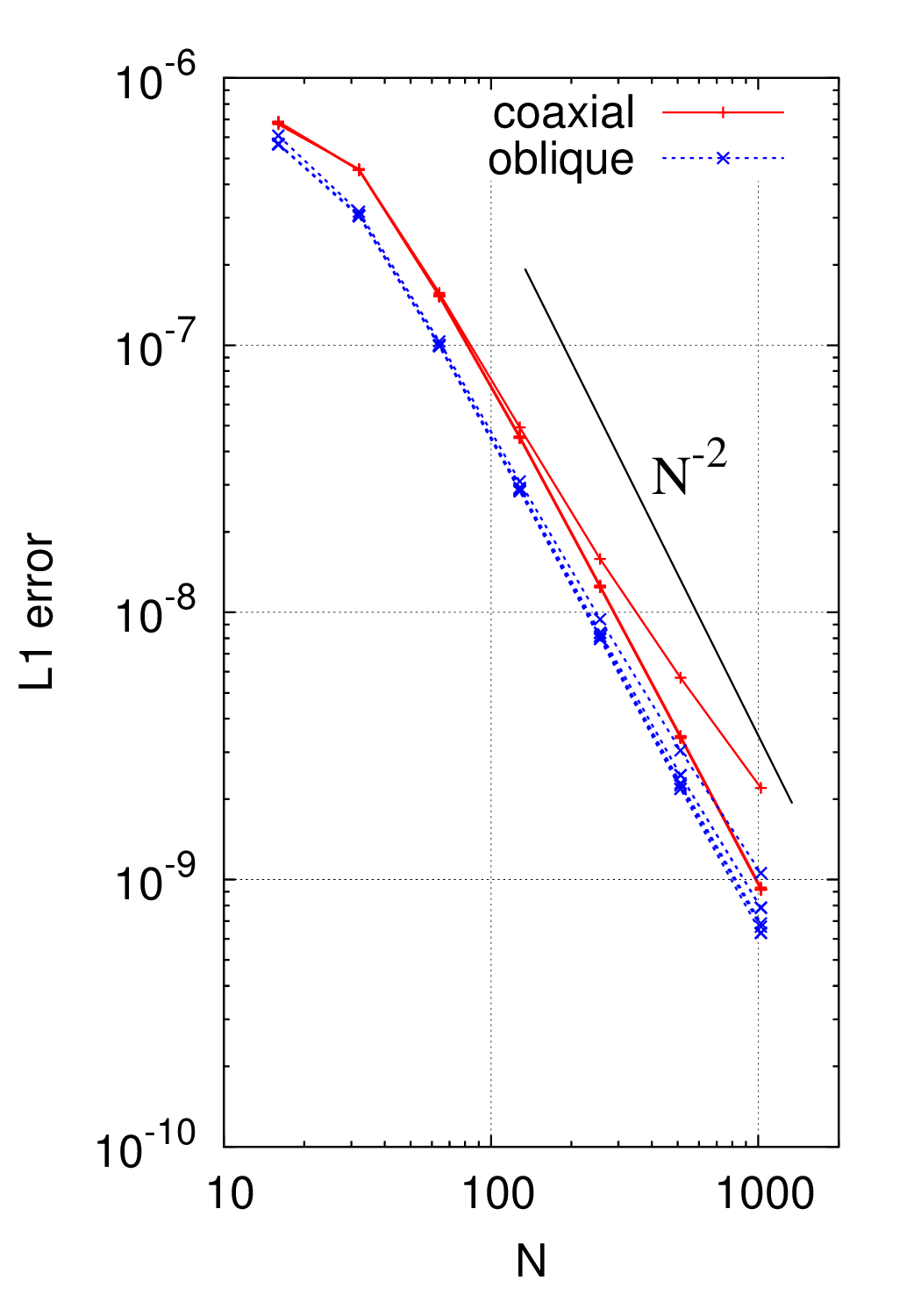}  &  
      \includegraphics[width=5cm]{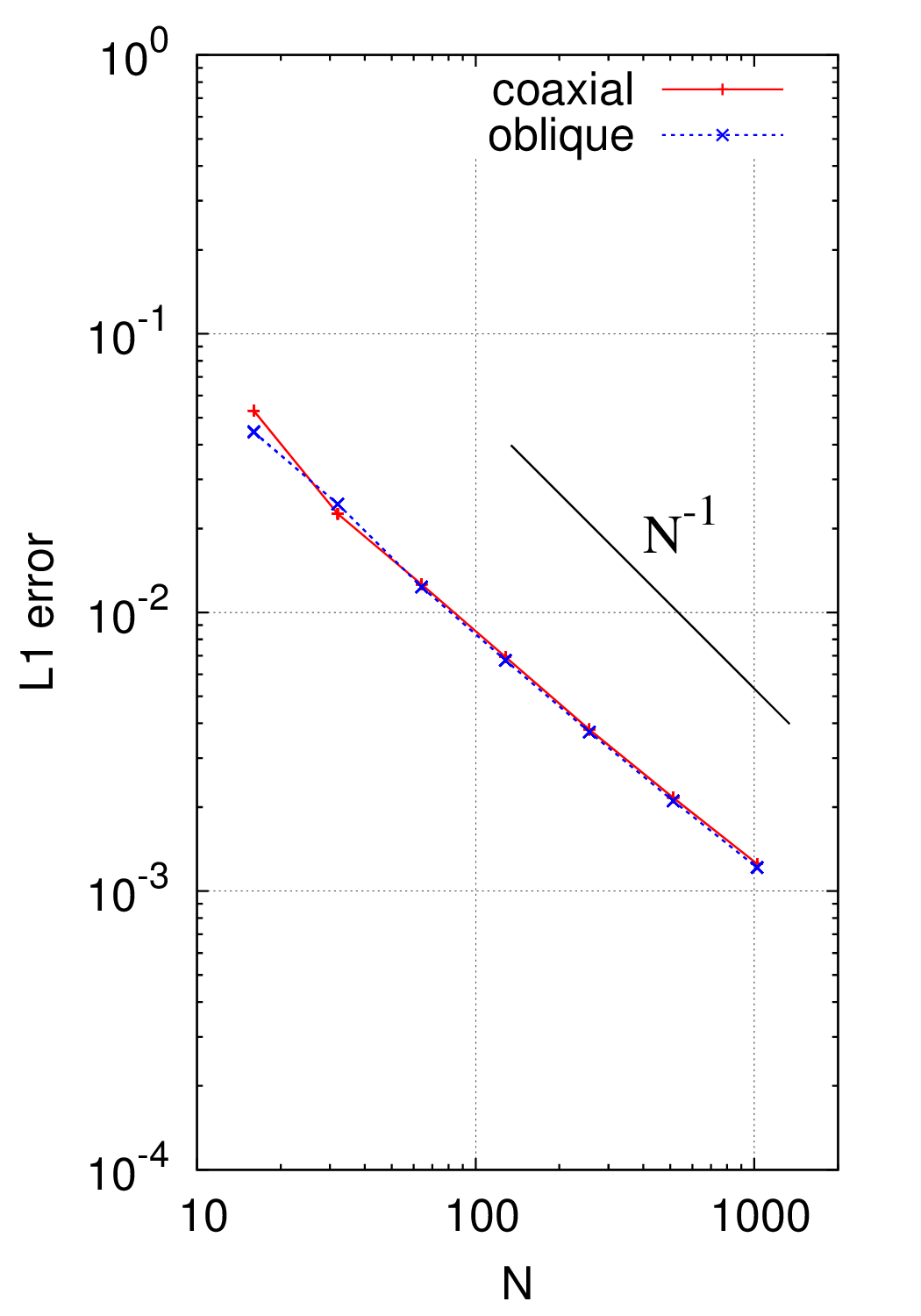}   
    \end{tabular}
  \end{center}
  \caption{Convergence tests of the norm of the $L_1$ error vector of density field
    for (a)entropy waves, (b)sound waves and (c)shock-tube problem.
  }
  \label{figureExam}
\end{figure}

Fig. \ref{figureExam} shows the result of convergence tests for
individuals {\em GA-1.20756}, {\em GA-4.33991}, {\em GA-D.59841}, {\em
  GA-DE.41250}, {\em GA-S1.33958}, \Izam\ and \Shin.  The resolution
$N$ is varied from $16^2$ to $1024^2$. The codes are tested for
entropy wave propagation, sound wave propagation and Sod's shock-tube
problem. The detail of test initial conditions are as follows. For all
tests, numerically solved domains are $(0<x<1, 0<y<1)$, out of which
the analytic solutions are continuously substituted as boundary
conditions. The system was numerically developed until $t=1$ for
entropy wave and sound wave problems, and $t=0.125$ for shock tube
problem, and then the numerical solutions were compared to analytic
solutions using the norm of $L_1$ error vector.

The initial conditions are, for entropy wave problem:
\begin{eqnarray}
  \left\{
  \begin{array}{rcl}
  \rho(x,y) &=& 2 + \sin(2\pi x) ,\\
  v_x(x,y) &=& 1 ,\\
  v_y(x,y) &=& 0 ,\\
  p(x,y) &=& 1,
  \end{array}
  \right.
\end{eqnarray}
for sound wave problem:
\begin{eqnarray}
  \left\{
  \begin{array}{rcl}
  \rho(x,y) &=& \gamma \left( 1+ A \sin(2\pi x)\right) \\
  v_x(x,y) &=&  A \sin(2\pi x) ,\\
  v_y(x,y) &=& 0 ,\\
  p(x,y) &=& 1 + \gamma  A \sin(2\pi x),
  \end{array}
  \right.
\end{eqnarray}
where the amplitude $A = 10^{-5}$, and for Sod's shock-tube problem:
\begin{eqnarray}
  x < 0.5
  \left\{
  \begin{array}{rcl}
  \rho(x,y) &=& 1 ,\\
  v_x(x,y) &=&  0 ,\\
  v_y(x,y) &=& 0 ,\\
  p(x,y) &=& 1 ,\\
  \end{array}
  \right. \\
  x > 0.5
  \left\{
  \begin{array}{rcl}
  \rho(x,y) &=& 0.125 ,\\
  v_x(x,y) &=&  0 ,\\
  v_y(x,y) &=& 0 ,\\
  p(x,y) &=& 0.1 .\\
  \end{array}
  \right.
\end{eqnarray}
While the ``coaxial'' tests used the above initial conditions as they
are, the ``oblique'' tests used the initial conditions rotated about
$(x,y) = (0.5, 0.5)$ for $60^\circ$.

\section{Analysis on the Automated Tuning Experiments} \label{sectionAnalysis}
\subsection{Overview of The Simulated Evolution and Analyses}

\begin{figure}[p]
  \begin{center}
    \includegraphics[width=15cm]{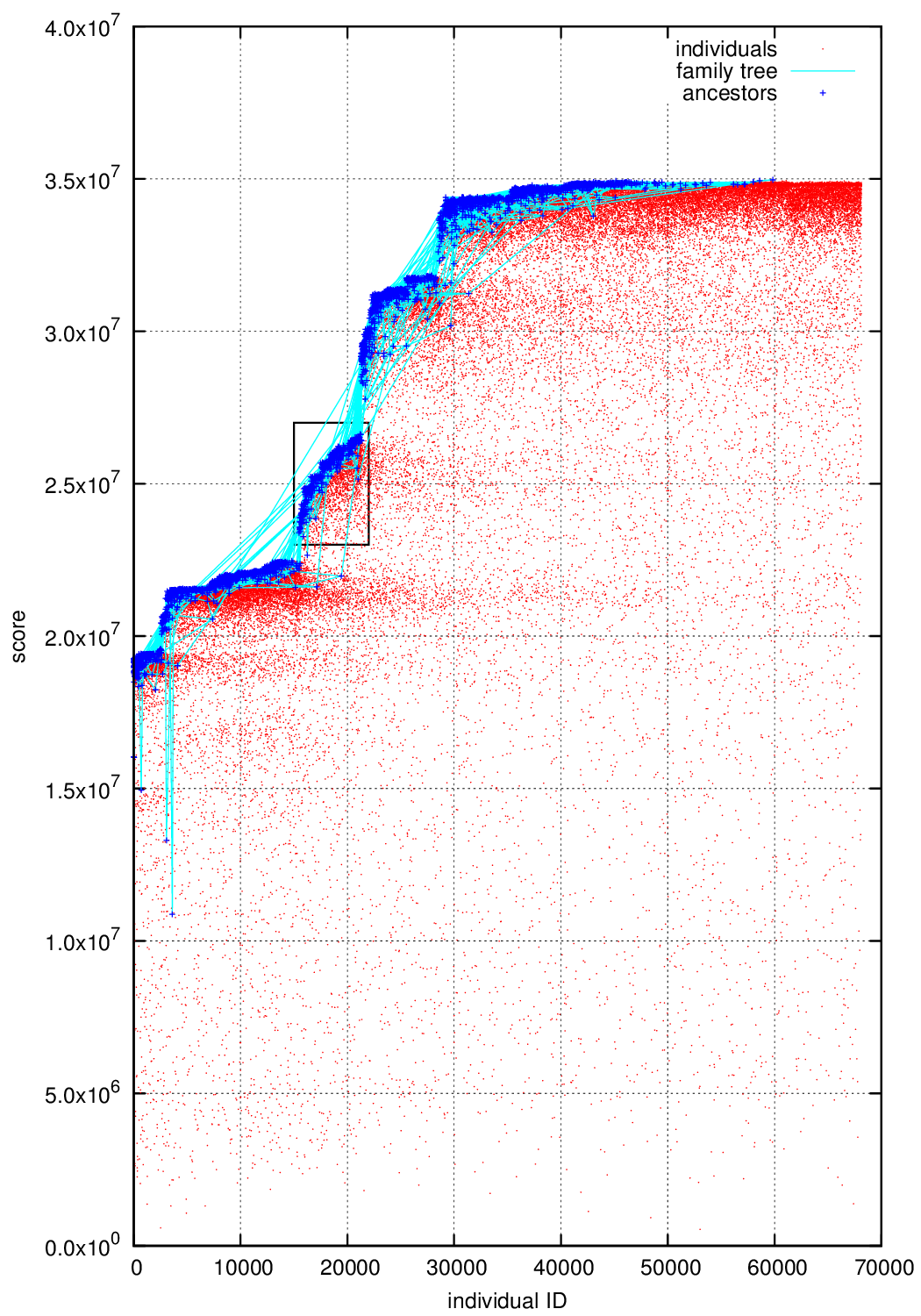}    
  \end{center}
  \caption{ The ID and the scores (in cups) of all the individuals
    generated in automated tuning experiment GA-D. The individual with the
    highest score as well as its ancestors are marked by crosses, and
    each of them is connected to its parents with lines.
  }\label{figureFamilyTreeGA-D}
\end{figure}

\begin{figure}[p]
  \begin{center}
    \includegraphics[height=15cm,angle=270]{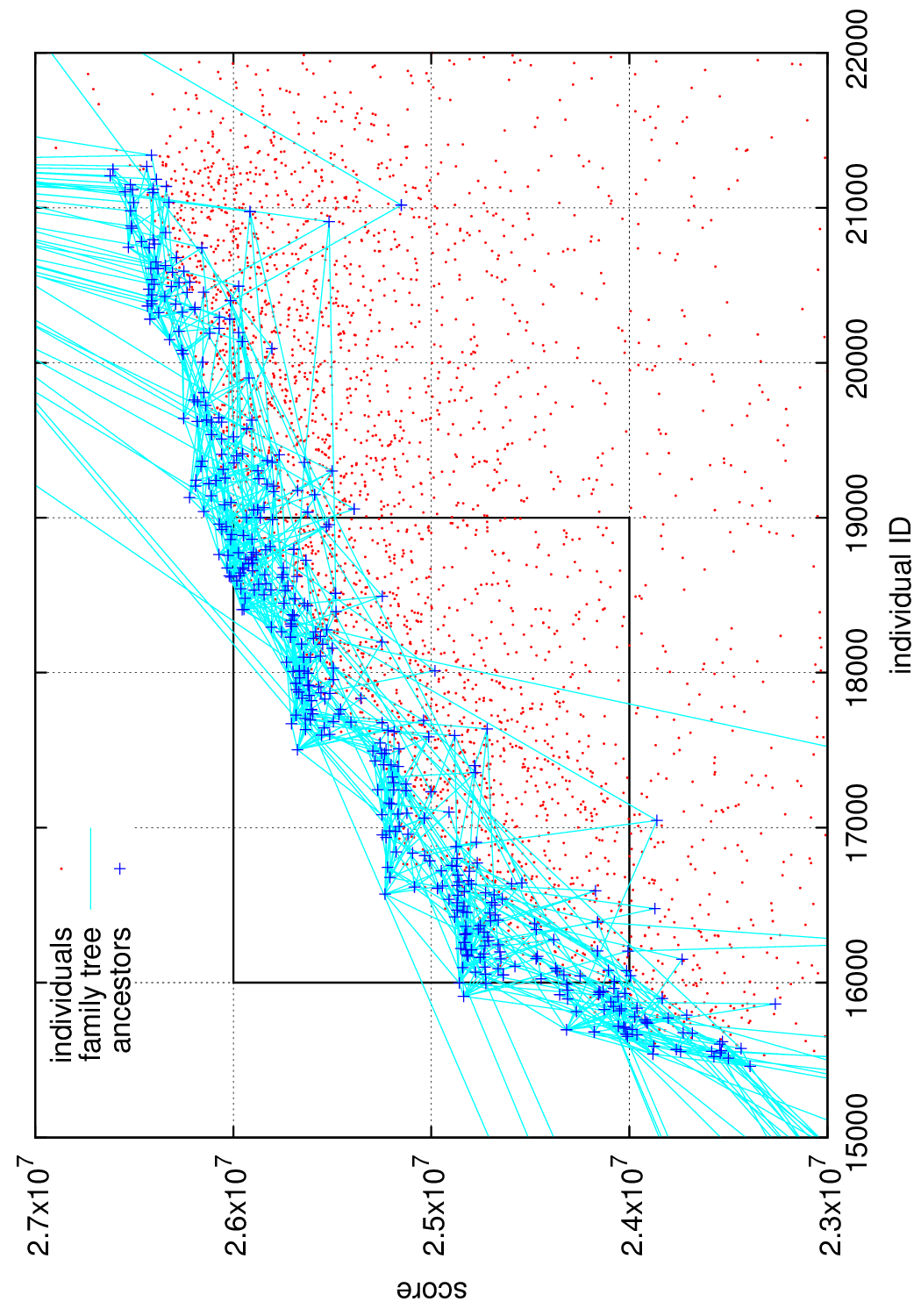}    
    \includegraphics[height=15cm,angle=270]{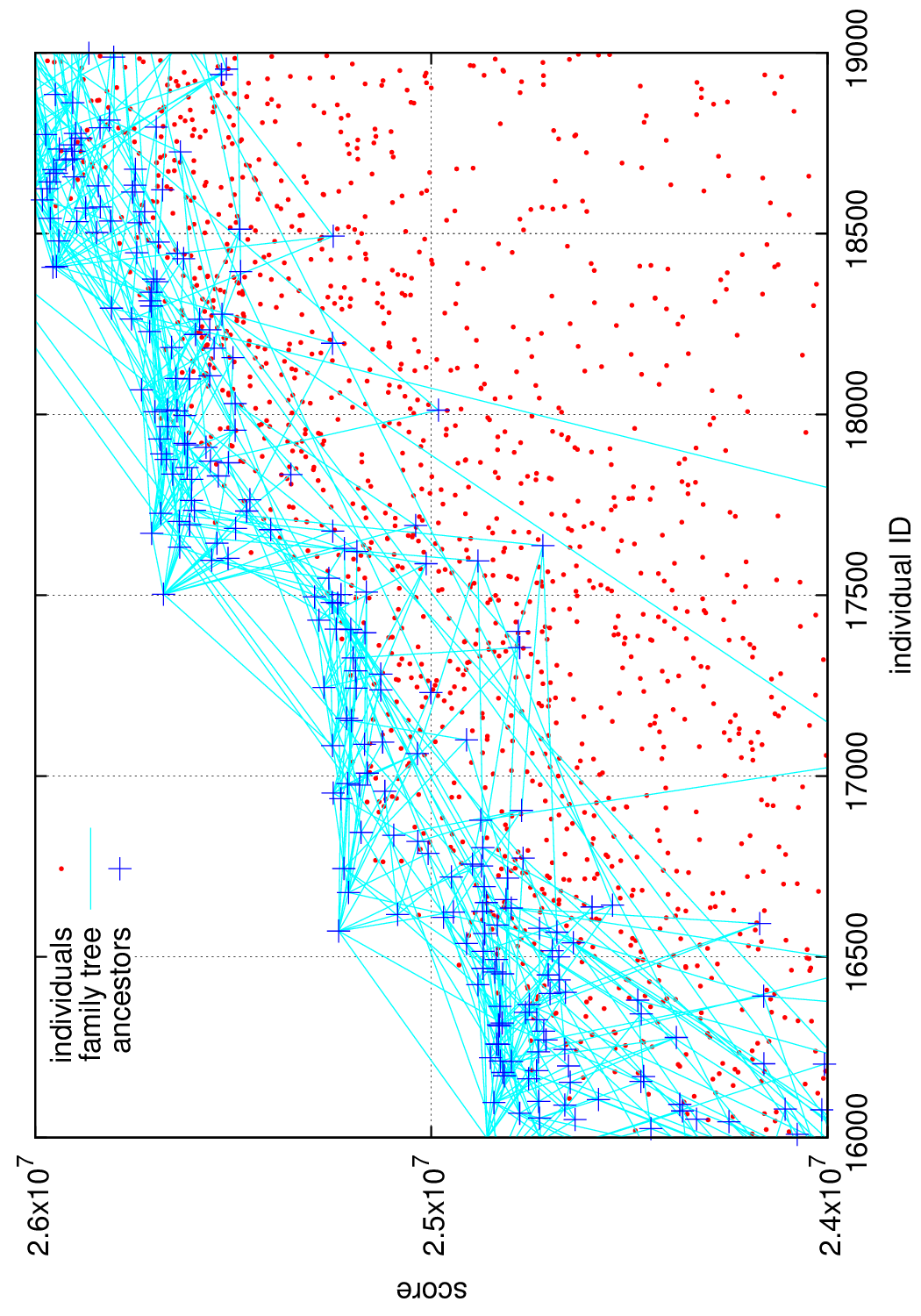}    
  \end{center}
  \caption{
    Two enlarged views of Fig. \ref{figureFamilyTreeGA-D}
  }\label{figureFamilyTreeGA-D-zoom}
\end{figure}

\begin{figure}[p]
  \begin{center}
    \includegraphics[height=15cm,angle=270]{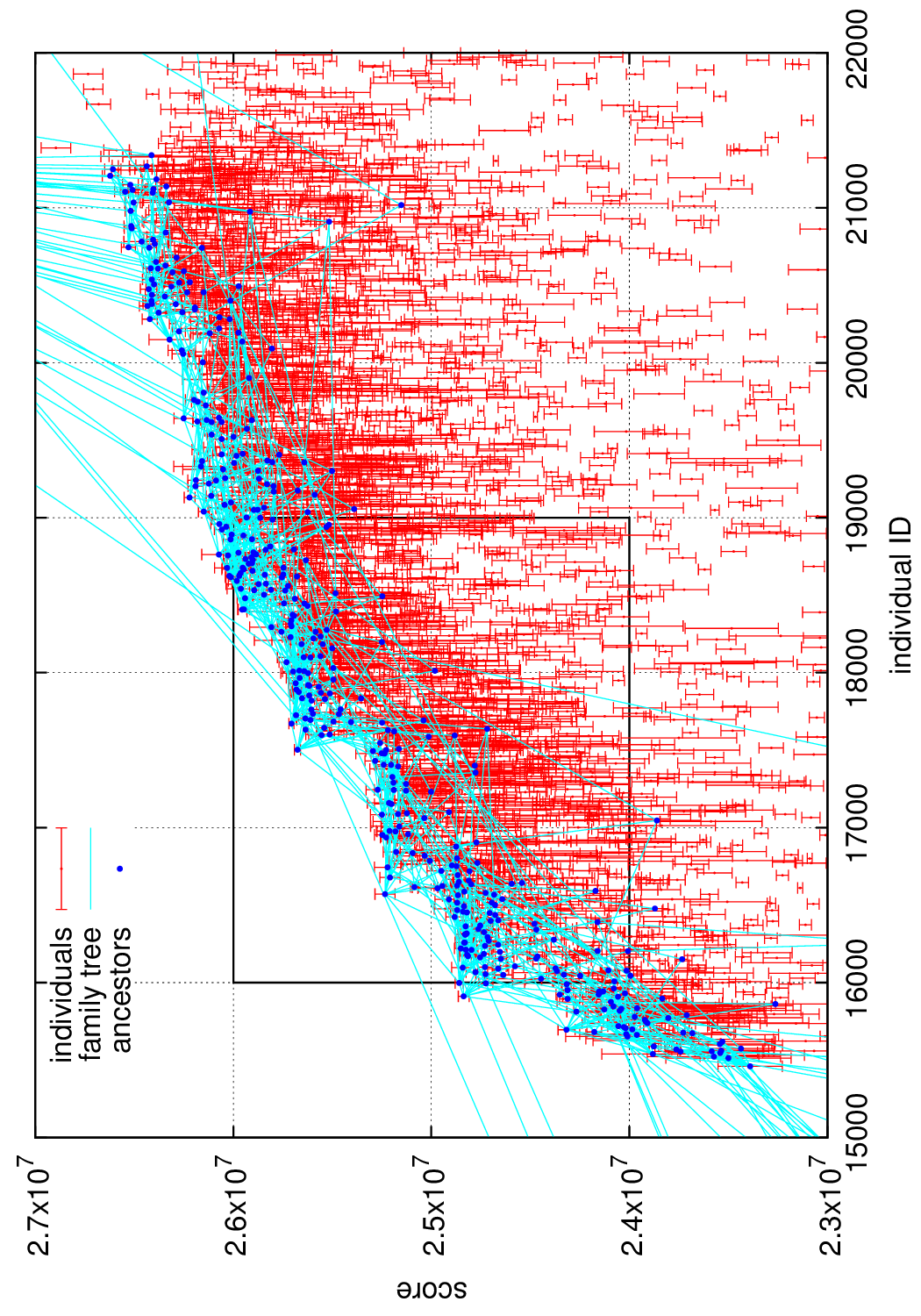}    
    \includegraphics[height=15cm,angle=270]{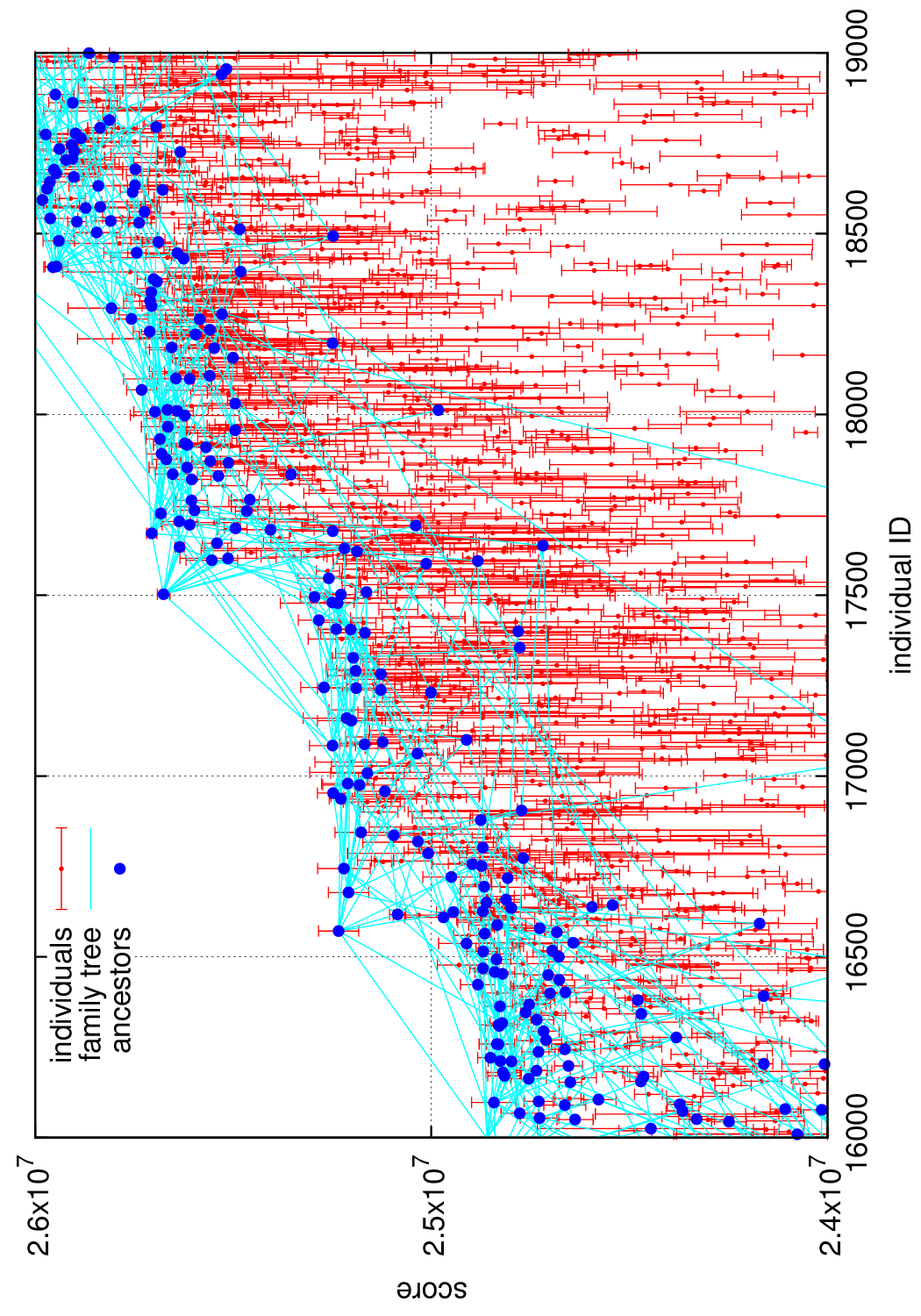}    
  \end{center}
  \caption{
    Two enlarged views of Fig. \ref{figureFamilyTreeGA-D} showing $\mu(I)$ and $\sigma(I)$
    as functions of $I$.
  }\label{figureFamilyTreeGA-D-zoom-2}
\end{figure}

\begin{figure}
  \begin{center}
    \begin{tabular}{cc}
      (a) & (b) \\
      \includegraphics[width=7.5cm]{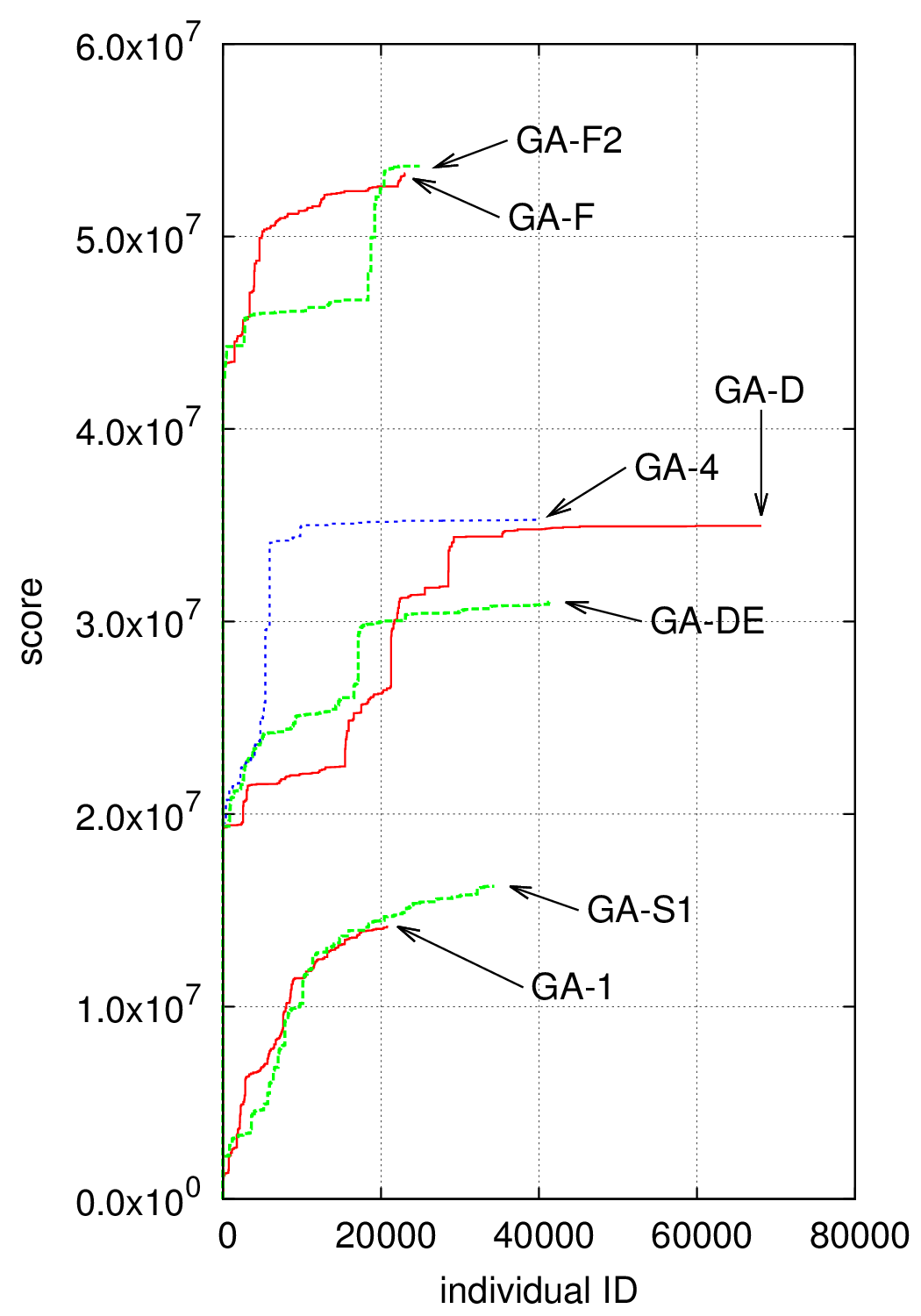} &
      \includegraphics[width=7.5cm]{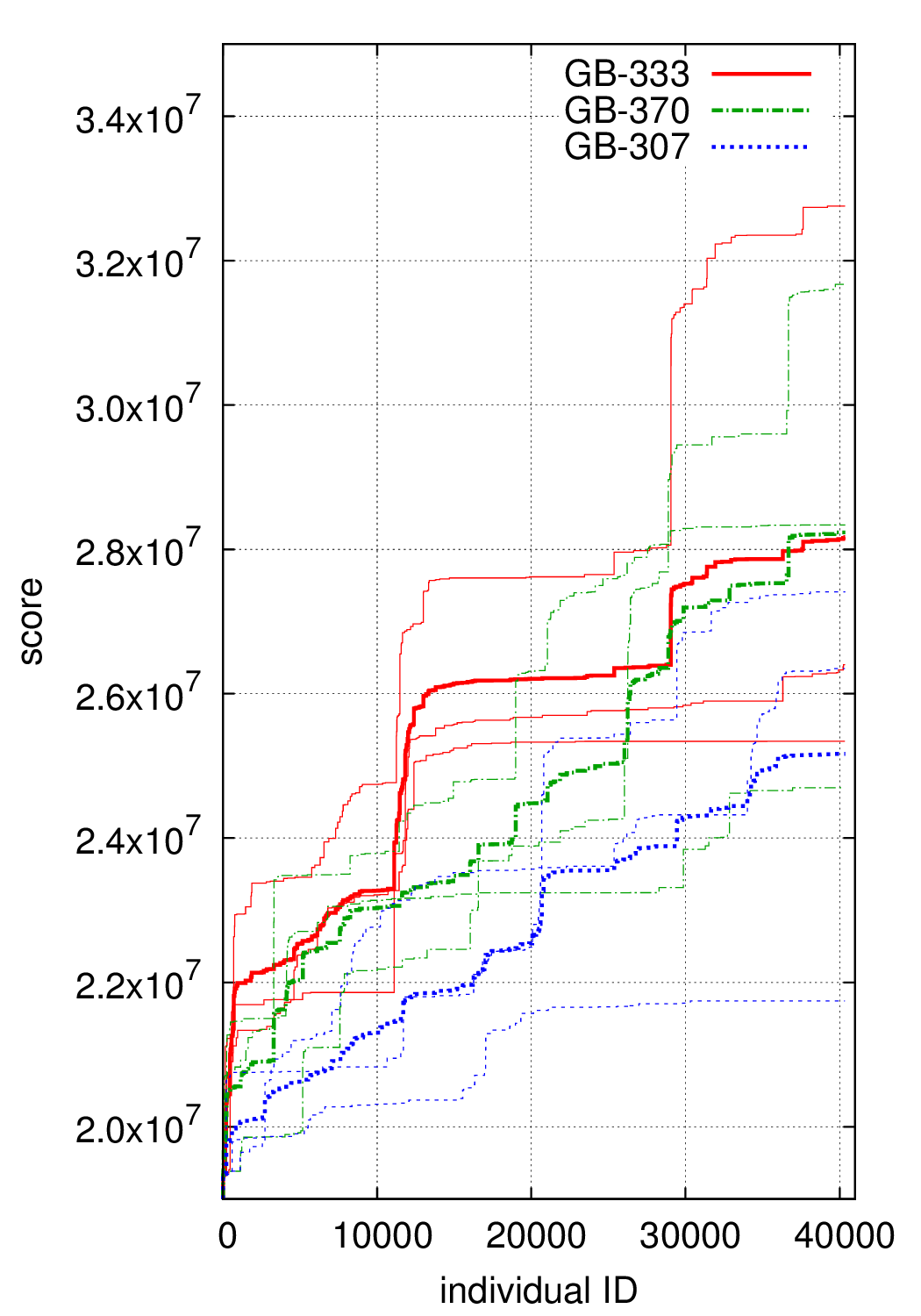}    
    \end{tabular}
  \end{center}
  \caption{ The evolution of the high score (in cups) as functions of ID 
    in (a) experiments GA-* and (b) experiments GB-* . The three thick curves are the 
    average of the GB-333-*, GB-370-* and GB-307-*, respectively. The average of
    GB-370-* was always higher than that of GB-307-*, and the average of 
    GB-333-* was always higher than that of GB-370-* for the 89\% of the time.
  }
  \label{figureGABHiscore}
\end{figure}

The functions of score against evolution progress exhibit self-similar
structure of repeated cliff and plateau. For example, see the
evolution history of experiment GA-D illustrated in
Fig. \ref{figureFamilyTreeGA-D} and its two enlarged views
Fig. \ref{figureFamilyTreeGA-D-zoom} and Fig. \ref{figureFamilyTreeGA-D-zoom-2}, with the family tree of the best
individual superimposed.  The evolution of the high scores in the
experiments are shown in Fig. \ref{figureGABHiscore}.

To design more efficient auto-tuning strategies, we investigate what have
happened and which part of the evolution contributed to create better
individuals in our automated tuning experiments.
In section \ref{sectionParts}, we measure the contributions from three 
different tuning items.
In section \ref{sectionClassIntro}, we classify the individuals by their aspects
such as their method of birth, their distance to the champion in the family tree, and
their fitness relative to their parents.
In section \ref{sectionChiSquare} we study how these classes contribute to the evolution
by performing correlation analyses among these classes.
In section \ref{sectionMarkov}, we study how the method of birth of parents affect their children.
In section \ref{sectionAnalClose} we summarize and conclude the analyses.

\begin{table}
  \begin{center}
    \begin{tabular}{l|ccc|rrr}
      ID          &C&M&S&\multicolumn{1}{c}{score(Mcups)}&\multicolumn{1}{c}{relative score}&
      \multicolumn{1}{c}{logscale}\\
      \hline
      \Izag            &0&0&0& $1.137 \pm 0.003$ & $0.000 \pm 0.000$   & $0.000 \pm 0.001$   \\
                       &0&0&1& $1.122 \pm 0.000$ & $-0.001 \pm 0.000$  & $-0.005\pm 0.000$  \\
                       &0&1&0& $5.400 \pm 0.006$ & $0.300 \pm 0.000$   & $0.599 \pm 0.000$   \\
                       &0&1&1& $5.300 \pm 0.006$ & $0.293 \pm 0.000$   & $0.591 \pm 0.000$   \\
                       &1&0&0& $3.073 \pm 0.002$ & $0.136 \pm 0.000$   & $0.382 \pm 0.000$   \\
                       &1&0&1& $2.946 \pm 0.000$ & $0.127 \pm 0.000$   & $0.366 \pm 0.000$   \\
                       &1&1&0& $15.829 \pm 0.027$ & $1.033 \pm 0.002$  & $1.012 \pm 0.001$   \\
      {\em GA-S1.33958}&1&1&1& $15.354 \pm 0.020$ & $1.000 \pm 0.001$  & $1.000 \pm 0.001$   \\
                                                                                             
      \multicolumn{1}{c}{}\\                                                                 
                                                                                             
      \Shin           &0&0&0&  $19.808 \pm 0.033$ & $0.000 \pm 0.002$  & $0.000 \pm 0.003$   \\
                      &0&0&1&  $19.817 \pm 0.030$ & $0.001 \pm 0.002$  & $0.001 \pm 0.003$   \\
                      &0&1&0&  $32.821 \pm 0.058$ & $0.848 \pm 0.004$  & $0.880 \pm 0.003$   \\
                      &0&1&1&  $32.694 \pm 0.057$ & $0.839 \pm 0.004$  & $0.873 \pm 0.003$   \\
                      &1&0&0&  $19.773 \pm 0.050$ & $-0.002 \pm 0.003$ & $-0.003\pm 0.004$  \\
                      &1&0&1&  $19.859 \pm 0.058$ & $0.003 \pm 0.004$  & $0.005 \pm 0.005$   \\
                      &1&1&0&  $32.994 \pm 0.273$ & $0.859 \pm 0.018$  & $0.889 \pm 0.014$   \\
      {\em GA-4.33991}&1&1&1&  $35.160 \pm 0.082$ & $1.000 \pm 0.005$  & $1.000 \pm 0.004$   \\
    \end{tabular}
  \end{center}
  \caption{ The score of the individuals created by artificial
    crossover between the initial individual $I_0$ and the best
    scoring individual $I_\top$. The second to fourth columns indicate
    which component was taken from which individual. Columns C,M,S
    correspond to CUDA kernel execution configuration, Manifest/Delay
    choice, synchronization timing, respectively.  For each individual
    $I$ the fifth column shows $\mu(I)\pm\sigma(I)$,
    the sixth column shows
    $\displaystyle\frac{\mu(I)}{\mu(I_\top)-\mu(I_0)}\pm\frac{\sigma(I)}{\mu(I_\top)-\mu(I_0)}$
    , and the seventh column shows
    $\displaystyle \frac{ \log\mu(I) -\log \mu(I_0) }{ \log\mu(I_\top)-\log \mu(I_0)}$
    $\displaystyle \pm \frac{\sigma(I) }{ (\log\mu(I_\top)-\log \mu(I_0))\mu(I)}$.
  } \label{tableComponent}
\end{table}

\subsection{Contributions of The Three Genome Parts} \label{sectionParts}

We assign the symbols to genome parts with different functions as
follows: (C): the CUDA kernel execution configuration, (M): which data
to store on the memory (to make them {\tt Manifest}), and (S): when to
synchronize the computation.  We first investigate how these three
components contributed to the score by component-wise artificial
crossover between the initial individual and the best scoring
individual (Table \ref{tableComponent}).  In the case of {\em
  GA-S1.33958}, introducing improvement only in C, M, S part increase
the score by 14\%, 30\%, and 0\%, respectively.  In the case of {\em
  GA-4.33991}, the increase are 0\%, 85\%, 0\%. Both cases exhibit
synergy effect. Introducing several modifications simultaneously have
more effect than the sum of the separate effects, they multiply. So
addition in log-space explains 98\% and 88\% of the progress for {\em
  GA-S1.33958} and {\em GA-4.33991}, respectively, but the score of
the final individuals are still slightly higher than predicted.

Removing only one of C,M,S part from {\em GA-4.33991} decreases the
score by 16\%, 100\%, and 14\%, respectively.  Removing C,M,S part
from {\em GA-S1.33958} decreases the score by 71\%, 87\%, and $-3$\%,
respectively. Again we see the synergy effect, except that 
{\em GA-S1.33958} were faster if S part were removed.

We conclude that the Manifest/Delayed trade-off plays the central part
in improving the score.  Fixing the Manifest/Delayed nodes, determines
the decomposition of the data-flow graph.  Then tuning the synchronization
timing and CUDA kernel execution configuration help further improve
the score.

\subsection{Classifications of Individuals} \label{sectionClassIntro}

To measure how each individual $I$ contributed in generating one of
the best individuals, we define contribution distance $d(I)$, and
classify the individuals to those who contributed to the evolution and
those who didn't.  To begin with, let $\Parent(I)$ be the set of
individual $I$'s parents.  (we use $p$ for probability.)  For
individual $I$ born by mutation, crossover, and triangulation, the
size of the parent set $n(\Parent(I))$ is 1,2, and 3, respectively.

We define $d(I)$ as follows:

\begin{itemize}
\item $d(I_\top) = 0$ where $I_\top$ is the individual whose $\mu(I)$ was the largest in the history.
\item $d(I) = 0$ if $\mu(I) > \mu(I_\top) - \sigma(I_\top)$.
\item $d(I) = 0$ if one of $I$'s children $I_c$ satisfies $d(I_c) = 0$.
\item $d(I) = 1 + \min\{d(I_\Parent) | I_\Parent \in \Parent(I) \}$ otherwise.
\end{itemize}

We say that $I$ is one of the best individuals if $\mu(I) > \mu(I_\top) -
\sigma(I_\top)$.  For them, and their ancestors, $d(I)$ = 0. For other
individuals $d(I)$ is the graph theoretical distance from the family
trees of the best individuals.
Table \ref{tableContributionBegin} - \ref{tableContributionEnd} shows
the distributions of $d(I)$ for individuals born of mutation,
crossover, and triangulation.

Next, we compare the fitness of the children with their parents.  We classify the individuals into four
ranks, namely \RankD, \RankC, \RankB, and \RankA, based upon
comparison of their scores with parents' scores.

\begin{itemize}
\item $I \in$ \RankD \ if \ $\forall I_\parent \in \Parent(I): \mu(I) < \mu(I_\parent) - \sigma(I,I_\parent)$,
\item $I \in$ \RankA \ if \ $\forall I_\parent \in \Parent(I): \mu(I) > \mu(I_\parent) + \sigma(I,I_\parent)$,
\item $I \in$ \RankB \ if \ $ |\mu(I) - \mu(I_{\Parent1})| < \sigma(I,I_{\Parent1})$ where $I_{\Parent1}$ is the member of $\Parent(I)$
with the largest $\mu$, and
\item $I \in$ \RankC otherwise,
\end{itemize}
where $\sigma(I,I_\parent) \equiv (\sigma(I)^2 + \sigma(I_\parent)^2)^{0.5} $.

\RankA are children significantly faster than any of their parents;
\RankB are children whose scores are comparable to their fastest parents; 
\RankD are children significantly slower than any of their parents; and 
\RankC are children significantly slower than their best parent but not significantly slower than their slowest parents.
Note that children of rank \RankC are never born by mutations since there is only one parent.
Table \ref{tableTombiBegin} - \ref{tableTombiEnd} shows the
classifications for experiments.

\subsection{Chi-squared Tests of Correlations between Classes} \label{sectionChiSquare}

To study which set of individuals are contributing to produce best species, we perform
Pearson's chi-squared test on pairs of predicates on individuals;
 see Table \ref{tableIndependenceA} and \ref{tableIndependenceB}.
We use $10^{-3}$ significance, or $X^2 > 10.83$ unless otherwise mentioned.
Our observations are as follows:

\begin{itemize}
\item
  (row 1.) Significant negative correlations are detected between 
  being born of mutation ($n(\Parent(I))=1$) and being $d(I)=0$
  for all experiments.
\item
  (row 2.) Significant positive correlations are detected between 
  being born of crossover ($n(\Parent(I))=2$) and being $d(I)=0$
  for all experiments but GA-S1.
\item
  (row 3.) Significant positive correlations are detected between 
  being born of triangulation ($n(\Parent(I))=3$) and being $d(I)=0$
  for all experiments.
\item
  (row 4, 5.) Significant positive correlations are detected between 
  being ($I\in \mathRankA$) and being $d(I)=0$, and also between
  being ($I\in \mathRankB$) and being $d(I)=0$
  for all experiments.
\item
  (row 6.) Being ($I\in \mathRankC$) and being $d(I)=0$ are negatively
  correlated for all experiments, but out of 10 experiments only 4 are
  $10^{-3}$ significant.
\item
  (row 7.) Significant negative correlations are detected between 
  being ($I\in \mathRankD$) and being $d(I)=0$
  for all experiments.
\item
  (row 8, 9.) Within the population born of either crossover or triangulation,
  being born of triangulation ($n(\Parent(I))=3$) and being $d(I)=0$
  are still positively correlated for all experiments, 
  but significant experiments are 7 out of 10.
\item
  (row 10, 11.) Within the elite population ${\mathbb E} \equiv \{I | n(\Parent(I))\geq2 \}
  \cap (\mathRankA \cup \mathRankB)$, 
  being ($I\in \mathRankA$) and being $n(\Parent(I))=2$
  are significantly and positively correlated.
\item
  (row 12-15.) Within the family tree of the champions $d(I)=0$, a subset of experiments 
  shows that the pairs $(n(\Parent(I))=2, I \in \mathRankA)$ and
  $(n(\Parent(I))=3, I \in \mathRankB)$ are positively correlated, and
  that the pairs $(n(\Parent(I))=2, I \in \mathRankB)$ and
  $(n(\Parent(I))=3, I \in \mathRankA)$ are positively correlated. The number of significant
  examples are 7, 5, 2, and 7 out of 10, respectively, and no significant opposite correlation
  are observed.
\item
  (row 16, 17.) Although 7 is considered to be a lucky number and 13 an unlucky
  number in Western culture, there is no correlation between having 7 at
  the lowest digit and being $d(I)=0$, nor having 13 as the two lowest
  digits and being $d(I)\neq 0$. These are control experiments.
\end{itemize}

\subsection{Correlation along the Family Line} \label{sectionMarkov}

Next, we analyze the correlation along the family line
by interpreting the family tree as  Markov processes (c.f. Table \ref{tableMarkov}). 
For each individual $I$ such that $d(I) = 0$, we trace back its family tree
in all possible ways for $n$ steps and obtain an $n$-letter word. For example,
``312'' means that $I$ is born of crossover, and one of its parent was born of 
mutation from an individual who was born of triangulation. With such bag of words
obtained, we investigate how the last letter of a word is correlated 
with letters in front of them.

\begin{table}
  \begin{center}
    \begin{tabular}{c|rr|rrrr}
      RunID     & 0th order & 1st order & $2\to2$  & $3\to3$  & $22\to2$  & $33\to3$ \\
      \hline
GA-1 &2263.22 &266.28 & $\ominus118.86$ & $\oplus1655.46$ & $\oplus32.54$ & $\oplus71.54$\\
GA-S1 &1387.93 &70.51 & $\ominus23.98$ & $\oplus1075.96$ & $\ominus5.19$ & $\oplus7.84$\\
GA-DE &546.42 &43.31 & $\oplus3.34$ & $\oplus427.88$ & $\ominus9.85$ & $\oplus3.68$\\
GA-D &1038.15 &88.20 & $\ominus42.78$ & $\oplus811.09$ & $\oplus3.90$ & $\oplus1.34$\\
GA-4 &755.63 &39.91 & $\ominus7.98$ & $\oplus580.33$ & $\ominus2.09$ & $\ominus2.60$\\
GA-F &422.08 &22.24 & $\ominus2.07$ & $\oplus333.57$ & $\oplus0.96$ & $\ominus0.25$\\
GA-F2 &490.90 &86.34 & $\ominus23.63$ & $\oplus381.72$ & $\oplus16.29$ & $\oplus6.09$\\
GB-333-0 &666.18 &47.52 & $\ominus12.34$ & $\oplus511.62$ & $\oplus1.36$ & $\ominus2.52$\\
GB-333-1 &930.33 &25.26 & $\ominus48.06$ & $\oplus727.01$ & $\ominus0.86$ & $\ominus0.90$\\
GB-333-2 &1208.20 &68.11 & $\ominus39.34$ & $\oplus937.37$ & $\oplus0.34$ & $\ominus7.59$\\

    \end{tabular}
  \end{center}
  \caption{ Chi-squared test of the family tree being lower-order
    Markov processes.  The each column of the table shows the $X^2$
    statistics of the null hypothesis the family tree being a
    $n$-th order Markov process and having no longer correlation.
  } \label{tableMarkov}
\end{table}

The second columns show $X^2$ for null hypothesis ``The bag of words
is a 0th order Markov chain,'' i.e. in the two-letter words the second
letter is not correlated to the first. Formally written, 
the null hypothesis is:
\begin{eqnarray}
  p(ab) = p(a) p(b),
\end{eqnarray}
where $a$, $b$,
$c$, ...  denote characters and $ab$, $abc$, ... denote words.

The third columns show $X^2$ for null hypothesis ``The bag of words is
a 0th order Markov chain,'' i.e. the probability of the occurrences of
the letters depend only on their immediate predecessors and the
distribution of three-letter words can be determined from the
distribution of two-letter words. Formally written, the null
hypothesis is:
\begin{eqnarray}
  p(abc) = \frac {p(ab) ~ p(bc)}{p(b)}.
\end{eqnarray}

Table \ref{tableMarkov} shows the result of the Markov chain analysis.
Note that, the degree of freedom $k$ for 0th-order and 1st-order
Markov chains are $4$ and $16$, respectively, and the chi-squared
values for $10^{-3}$ significance are $X^2 > 18.47$ and $X^2 > 37.70$,
respectively.  From the table, we can deny the 0th-order Markov chain
model for all of the experiments, and deny the 1st-order Markov chain
model for 8 out of the 10 experiments.  We also studied whether an
occurence of a character is correlated to its prefix word.  ``3'' was
significantly likely to be followed by another ``3'' in all of the
experiments, while ``2'' was significantly unlikely to be followed by
another ``2'' in 6 out of 10 experiments. On the other hand, three
consecutive occurences of the same letter was not significant at least
in 8 out of 10 experiments.

\subsection {Summary of The Analyses} \label{sectionAnalClose}

Mutation is not efficient in directly producing
individuals with good scores. Nevertheless mutations are indispensable part
of the evolution because it is the only way that introduces new
genomes to the genome pools and that have chance of finding new
improvements.

Triangulations and crossover are two different methods to combine
independently-found improvements, and are efficient in directly
producing good individuals.  Triangulations are relatively more
efficient in producing $d(I)=0$ individuals.  However, crossover is
relatively good at making large jumps that belongs to \RankA.
Triangulations are working in another way; the statistics indicate
that one triangulation are likely to be followed by another
triangulation in a family tree, and the sequence of Triangulations
work to accumulate minor improvements that belongs to \RankB to make a
larger one.  Fig. \ref{figureGABHiscore} --- although the number of
samples are too few to make a statistically significant statement ---
suggests that having both mode of combiner is better than having only
one. 

\addspan{
  The main source of performance improvement was
  the change in memory layout and subroutine re-organization 
  caused by making different Manifest/Delayed trade-offs.
  Tuning the synchronization
  timing and CUDA kernel execution configuration provided additional improvements.
}

The case of {\em GA-S1.33958} is an evidence of the current genetic
algorithm not finding the optimal individual. Random mutations tend
to increase the genome entropy even under the selection pressure.
Adding the rule-based mutations such as ``remove all synchronization,''
which are improbable under random
mutations, may contribute to
further optimization.

\section{Conclusion and Discussion} \label{sectionConclusion}

We have designed and implemented Paraiso, a domain specific language
for describing explicit solvers of partial differential equations,
embedded in Haskell.  Using the typelevel-tensor and algebraic type
classes, we can explain our algorithms with simplicity of mathematical
equations.  Then the algorithm is translated to OpenMP code or CUDA
code.  The generated code can be optimized both by applying
annotations by hand, and by automated tuning based on a genetic
algorithm. Although we present just one example here --- a second
order Euler equations solver --- the front end of Paraiso with
typelevel-tensor and {\tt Builder Monad} readily accept various other
algorithms.  The code generation and automated tuning can
revolutionize the way we invent, implement and optimize
various partial differential equation solvers.

The website of Paraiso is {\tt http://www.paraiso-lang.org/wiki/}, 
where you can find codes, documentations, slides and videos.

Making more efficient searches for fast individuals is an important
future extension of Paraiso.  One way is to use machine learning, for
example to make suggestions for genomes to introduce, or to predict
the scores of the individuals before benchmarking them. Another
approach is careful selection of the tuning items, importance of them given either
by hand or by inference from the syntactic structure of Paraiso source
code.  The computing time invested in automated tuning is only
justified if the optimized program is used repeatedly, or if optimized
single-GPU program is used as a ``core'' of a multi-GPU program.

\addspan{
  Making the automated tuning results more flexible and re-usable is another future challenge.
  For example, the size of the array e.g. $(N0, N1)$ in \S \ref{sectionOM} for the two-dimensional OM,
  is fixed before native code generation under the current design of Paraiso.
  To change the size $N0$ or $N1$, the users need to re-run the code generation process. 
  One solution is to improve the representation of the constant values in Paraiso, so that the users can set the values that are constant within a run but may vary across different runs. Possible examples of such constants include
  the OM size and the parameters of the algorithm.
}

\addspan{ 
  On the other hand, the change in the dimension of the OM
  e.g. to three; $(N0, N1, N2)$, or the change in the algorithm described
  by the Builder Monad will change the number of nodes in the OM data
  flow graph and the bits in the genome. Therefore, the automatically
  tuned genomes before the change become useless. The genome must be redesigned 
  to perform more re-usable, general automated tunings.
}

Generating codes for distributed computers / distributed and
accelerated computers is another important future extension of
Paraiso.  To this end, instead of generating MPI codes by ourselves,
we plan to generate parallel languages such as X10
\cite{charles_x10:_2005}, Chapel \cite{chamberlain_parallel_2007},
and XcalableMP \cite{lee_implementation_2010}, or domain specific
languages for stencil computations such as Physis
\cite{maruyama_physis:_2011}.

\addspan{
  Finally, the potential users and developers of Paraiso are
  obstructed by the programming language Haskell, which is a very
  computer-science language and is far from being mainstream in the
  fields of simulation science and high-performance computing.
  This is a disadvantage of Paraiso compared to other embedded DSL approaches
  based on more popular syntax such as Python (e.g. \cite{nilsen_simplifying_2010}.)
  We make two remarks on this.
}

\addspan{ 
  On the one hand, we show in this paper that abstract concepts
  developed by computer scientists and found in Haskell, is
  useful and was almost necessary in implementing a DSL that
  consistently handles the translation from mathematical notations to evolutionary
  computation. We hope that such powerful DSLs appear in many different
  fields, with help of the findings in computer science.
  On the other hand, we want many other researchers to
  become developers and users of Paraiso without requiring much
  knowledge in computer science.
}

\addspan{ 
  For promotion of such development and use of Paraiso, we plan to provide interfaces
  to the code generation and the tuning stages in Paraiso,
  in terms of strings or in lightweight markup languages, along with their specifications.
  Those will allow many other programming languages can access those stages, 
  such as OM data-flow graphs, their annotations, and the genomes.
  Adding a  simulation scientist friendly scripting language layer on top of
  {\tt Builder Monad} is another future work, so that people not familiar with
  Haskell can access Paraiso.
  We also plan to publish a series of tutorials and example programs in the Paraiso website.
  Such collaborations are fundamental to 
  the realization of the future works listed in this section 
  and computer-assisted programming in many fields of simulation science.  
}

\section*{Acknowledgments}
The authors thank the anonymous referees for a number of suggestions 
that improved this paper.
T.M. is supported by The Hakubi Center for Advanced Research.
T.M. is also supported by Grants-in-Aid from the Ministry of Education, Culture, Sports, Science, and Technology (MEXT) of Japan, No. 24103506. 
The use of TSUBAME2.0 Grid Cluster in this project was supported by
JST, CREST through its research program: ``Highly Productive, High
Performance Application Frameworks for Post Petascale Computing.''

\clearpage

\appendix
\section{supplementary data}

\begin{table}[h]
  \begin{center}
    \begin{tabular}{c|rrr|r}
$d(I)$  & \multicolumn{1}{c}{mutation} & \multicolumn{1}{c}{crossover} & \multicolumn{1}{c}{triangulation} 
& \multicolumn{1}{|c}{total} \\
\hline
0       &1627(0.171)    &1875(0.433)    &3011(0.440)    &6513(0.315)\\
1       &4492(0.473)    &1368(0.316)    &2639(0.385)    &8499(0.411)\\
2       &2434(0.256)    &944(0.218)     &1129(0.165)    &4507(0.218)\\
3       &831(0.088)     &141(0.033)     &69(0.010)      &1041(0.050)\\
4       &104(0.011)     &5(0.001)       &2(0.000)       &111(0.005)\\
5       &6(0.001)       &0(0.000)       &0(0.000)       &6(0.000)\\
6       &1(0.000)       &0(0.000)       &0(0.000)       &1(0.000)\\
\hline
sum     &9495(1.000)    &4333(1.000)    &6850(1.000)    &20678(1.000)\\

\end{tabular}

\caption{
  Contribution distance analysis for experiment GA-1 :
  The numbers of individuals with specific $d(I)$ that is
  born of mutation, crossover, and triangulation.
  The last column shows the total numbers of individuals with specific $d(I)$.
  The decimal numbers inside the parentheses are the ratio of the number of
  individuals with specific $d(I)$ of specific birth, divided by 
  total number of individuals of specific birth. Note that
  the grand total is slightly smaller than that in Table \ref{tableAT};
  this is because some worker nodes failed to write into DB.
}\label{tableContributionBegin}
  \end{center}
\end{table}

\begin{table}
  \begin{center}
    \begin{tabular}{c|rrr|r}
$d(I)$  & \multicolumn{1}{c}{mutation} & \multicolumn{1}{c}{crossover} & \multicolumn{1}{c}{triangulation} 
& \multicolumn{1}{|c}{total} \\
\hline

0       &1304(0.082)    &915(0.106)     &1411(0.145)    &3631(0.106)\\
1       &6136(0.386)    &3601(0.416)    &4696(0.481)    &14433(0.421)\\
2       &5461(0.344)    &3211(0.371)    &3161(0.324)    &11833(0.345)\\
3       &2392(0.151)    &837(0.097)     &470(0.048)     &3699(0.108)\\
4       &507(0.032)     &79(0.009)      &25(0.030)      &611(0.018)\\
5       &80(0.005)      &7(0.001)       &0(0.000)       &87(0.003)\\
6       &7(0.000)       &0(0.000)       &0(0.000)       &7(0.000)\\
7       &1(0.000)       &0(0.000)       &0(0.000)       &1(0.000)\\
\hline
sum     &15888(1.000)   &8650(1.000)    &9763(1.000)    &34302(1.000)\\

\end{tabular}
\caption{  Contribution distance analysis for experiment GA-S1. }
  \end{center}
\end{table}

\begin{table}
  \begin{center}
    \begin{tabular}{c|rrr|r}
$d(I)$  & \multicolumn{1}{c}{mutation} & \multicolumn{1}{c}{crossover} & \multicolumn{1}{c}{triangulation} 
& \multicolumn{1}{|c}{total} \\
\hline

0       &480(0.023)     &430(0.047)     &592(0.051)     &1503(0.036)\\
1       &10130(0.495)   &4149(0.458)    &6080(0.519)    &20359(0.494)\\
2       &6952(0.340)    &3830(0.423)    &4640(0.396)    &15422(0.374)\\
3       &2499(0.122)    &622(0.069)     &387(0.033)     &3508(0.085)\\
4       &385(0.019)     &34(0.004)      &12(0.001)      &431(0.010)\\
5       &21(0.001)      &0(0.000)       &0(0.000)       &21(0.001)\\
6       &4(0.000)       &0(0.000)       &0(0.000)       &4(0.000)\\
7       &1(0.000)       &0(0.000)       &0(0.000)       &1(0.000)\\
\hline
sum     &20472(1.000)   &9065(1.000)    &11711(1.000)   &41249(1.000)\\

\end{tabular}
\caption{  Contribution distance analysis for experiment GA-DE. }
  \end{center}
\end{table}

\begin{table}
  \begin{center}
    \begin{tabular}{c|rrr|r}
$d(I)$  & \multicolumn{1}{c}{mutation} & \multicolumn{1}{c}{crossover} & \multicolumn{1}{c}{triangulation} 
& \multicolumn{1}{|c}{total} \\
\hline

0       &785(0.023)     &1099(0.071)    &1680(0.087)    &3565(0.052)\\
1       &16113(0.482)   &6208(0.403)    &9699(0.503)    &32020(0.470)\\
2       &11510(0.344)   &6946(0.451)    &7490(0.389)    &25946(0.381)\\
3       &4509(0.135)    &1139(0.074)    &408(0.021)     &6056(0.089)\\
4       &472(0.014)     &13(0.001)      &1(0.000)       &486(0.007)\\
5       &21(0.001)      &0(0.000)       &0(0.000)       &21(0.000)\\
6       &2(0.000)       &0(0.000)       &0(0.000)       &2(0.000)\\
\hline
sum     &33412(1.000)   &15405(1.000)   &19278(1.000)   &68096(1.000)\\

\end{tabular}
\caption{  Contribution distance analysis for experiment GA-D. }
  \end{center}
\end{table}

\begin{table}
  \begin{center}
    \begin{tabular}{c|rrr|r}
$d(I)$  & \multicolumn{1}{c}{mutation} & \multicolumn{1}{c}{crossover} & \multicolumn{1}{c}{triangulation} 
& \multicolumn{1}{|c}{total} \\
\hline
0       &631(0.037)     &730(0.066)     &940(0.078)     &2302(0.057)\\
1       &7674(0.450)    &5234(0.475)    &6658(0.554)    &19566(0.488)\\
2       &6062(0.356)    &4397(0.399)    &4215(0.350)    &14674(0.366)\\
3       &2385(0.140)    &644(0.058)     &214(0.018)     &3243(0.081)\\
4       &291(0.017)     &8(0.001)       &0(0.000)       &299(0.007)\\
5       &9(0.001)       &0(0.000)       &0(0.000)       &9(0.000)\\
\hline
sum     &17052(1.000)   &11013(1.000)   &12027(1.000)   &40093(1.000)\\

\end{tabular}
\caption{  Contribution distance analysis for experiment GA-4. }
  \end{center}
\end{table}

\begin{table}
  \begin{center}
    \begin{tabular}{c|rrr|r}
$d(I)$  & \multicolumn{1}{c}{mutation} & \multicolumn{1}{c}{crossover} & \multicolumn{1}{c}{triangulation} 
& \multicolumn{1}{|c}{total} \\
\hline

0       &373(0.031)     &329(0.068)     &499(0.082)     &1202(0.052)\\
1       &6440(0.533)    &2137(0.441)    &3302(0.540)    &11879(0.515)\\
2       &3854(0.319)    &2112(0.436)    &2224(0.363)    &8190(0.355)\\
3       &1285(0.106)    &262(0.054)     &95(0.016)      &1642(0.071)\\
4       &137(0.011)     &2(0.000)       &0(0.000)       &139(0.006)\\
5       &4(0.000)       &0(0.000)       &0(0.000)       &4(0.000)\\
\hline
sum     &12093(1.000)   &4842(1.000)    &6120(1.000)    &23056(1.000)\\

\end{tabular}
\caption{  Contribution distance analysis for experiment GA-F. }
  \end{center}
\end{table}

\begin{table}
  \begin{center}
    \begin{tabular}{c|rrr|r}
$d(I)$  & \multicolumn{1}{c}{mutation} & \multicolumn{1}{c}{crossover} & \multicolumn{1}{c}{triangulation} 
& \multicolumn{1}{|c}{total} \\
\hline

0       &332(0.024)     &314(0.076)     &767(0.113)     &1414(0.057)\\
1       &7187(0.516)    &1648(0.397)    &3641(0.538)    &12476(0.502)\\
2       &4641(0.333)    &1884(0.454)    &2264(0.335)    &8789(0.354)\\
3       &1564(0.112)    &295(0.071)     &95(0.014)      &1954(0.079)\\
4       &190(0.014)     &5(0.001)       &0(0.000)       &195(0.008)\\
5       &21(0.002)      &0(0.000)       &0(0.000)       &21(0.001)\\
\hline
sum     &13935(1.000)   &4146(1.000)    &6767(1.000)    &24849(1.000)\\

\end{tabular}
\caption{  Contribution distance analysis for experiment GA-F2. }\label{tableContributionEnd}
  \end{center}
\end{table}

\begin{table}
  \begin{center}
    \begin{tabular}{rrr | rrrr | rrrr}

\multicolumn{3}{c}{mutation} & \multicolumn{4}{|c}{crossover} & \multicolumn{4}{|c}{triangulation} \\
\multicolumn{3}{c}{ 9437(1.000) } & \multicolumn{4}{|c}{ 4335(1.000) } & \multicolumn{4}{|c}{ 6834(1.000) } \\
\RankD  &\RankB &\RankA         &\RankD &\RankC &\RankB &\RankA         &\RankD &\RankC &\RankB &\RankA  \\
\hline

7468    &935    &1022   &934    &1228   &824    &1347   &976    &2161   &2420   &1293\\
(0.787  &0.098  &0.108) &(0.216 &0.283  &0.190  &0.311) &(0.142 &0.315  &0.353  &0.189)\\
202     &516    &856    &77     &392    &517    &889    &20     &573    &1593   &825\\
(0.021  &0.054  &0.090) &(0.018 &0.090  &0.119  &0.205) &(0.003 &0.084  &0.233  &0.120)\\
0.027   &0.552  &0.838  &0.082  &0.319  &0.627  &0.660  &0.020  &0.265  &0.658  &0.638\\

    \end{tabular}
    \caption{ 
      Children relative fitness for Experiment GA-1: 
      The individuals classified based upon how they were born
      and their score with respect to their parents. For each class,
      the first row shows the number of individuals of that class, and
      the third row shows the number of individuals of that class with $d(I)=0$.
      The second and fourth rows are first and third row divided by the total number
      of individuals born in that way.
      The fifth row is the contribution ratio, the third row divided by the first row
      (or fourth divided by second).
       }\label{tableTombiBegin}
  \end{center}
\end{table}

\begin{table}
  \begin{center}
    \begin{tabular}{rrr | rrrr | rrrr}
\multicolumn{3}{c}{mutation} & \multicolumn{4}{|c}{crossover} & \multicolumn{4}{|c}{triangulation} \\
\multicolumn{3}{c}{ 15887(1.000) } & \multicolumn{4}{|c}{ 8655(1.000) } & \multicolumn{4}{|c}{ 9758(1.000) } \\
\RankD  &\RankB &\RankA         &\RankD &\RankC &\RankB &\RankA         &\RankD &\RankC &\RankB &\RankA  \\
\hline

9326    &5818   &743    &1004   &2868   &4246   &532    &579    &2427   &6214   &543\\
(0.587  &0.366  &0.047) &(0.116 &0.332  &0.491  &0.062) &(0.059 &0.249  &0.636  &0.056)\\
109     &990    &205    &19     &97     &739    &60     &5      &111    &1190   &105\\
(0.007  &0.062  &0.013) &(0.002 &0.011  &0.085  &0.007) &(0.001 &0.011  &0.122  &0.011)\\
0.012   &0.170  &0.276  &0.019  &0.034  &0.174  &0.113  &0.009  &0.046  &0.192  &0.193\\

    \end{tabular}
    \caption{ Children relative fitness classification for Experiment GA-S1.}
  \end{center}
\end{table}

\begin{table}
  \begin{center}
    \begin{tabular}{rrr | rrrr | rrrr}

\multicolumn{3}{c}{mutation} & \multicolumn{4}{|c}{crossover} & \multicolumn{4}{|c}{triangulation} \\
\multicolumn{3}{c}{ 20473(1.000) } & \multicolumn{4}{|c}{ 9064(1.000) } & \multicolumn{4}{|c}{ 11711(1.000) } \\
\RankD  &\RankB &\RankA         &\RankD &\RankC &\RankB &\RankA         &\RankD &\RankC &\RankB &\RankA  \\
\hline

16032   &3273   &1167   &3082   &876    &3596   &1511   &3820   &1386   &4777   &1728\\
(0.783  &0.160  &0.057) &(0.340 &0.097  &0.397  &0.167) &(0.326 &0.118  &0.408  &0.148)\\
89      &328    &63     &33     &30     &298    &69     &28     &49     &446    &69\\
(0.004  &0.016  &0.003) &(0.004 &0.003  &0.033  &0.008) &(0.002 &0.004  &0.038  &0.006)\\
0.006   &0.100  &0.054  &0.011  &0.034  &0.083  &0.046  &0.007  &0.035  &0.093  &0.040\\

    \end{tabular}
    \caption{ Children relative fitness classification for Experiment GA-DE.}
  \end{center}
\end{table}

\begin{table}
  \begin{center}
    \begin{tabular}{rrr | rrrr | rrrr}

\multicolumn{3}{c}{mutation} & \multicolumn{4}{|c}{crossover} & \multicolumn{4}{|c}{triangulation} \\
\multicolumn{3}{c}{ 33420(1.000) } & \multicolumn{4}{|c}{ 15412(1.000) } & \multicolumn{4}{|c}{ 19261(1.000) } \\
\RankD  &\RankB &\RankA         &\RankD &\RankC &\RankB &\RankA         &\RankD &\RankC &\RankB &\RankA  \\
\hline

30112   &2510   &788    &4110   &5694   &4657   &944    &3899   &8372   &6382   &625\\
(0.901  &0.075  &0.024) &(0.267 &0.370  &0.302  &0.061) &(0.202 &0.434  &0.331  &0.032)\\
420     &313    &52     &122    &204    &648    &125    &90     &370    &1134   &86\\
(0.013  &0.009  &0.002) &(0.008 &0.013  &0.042  &0.008) &(0.005 &0.019  &0.059  &0.004)\\
0.014   &0.125  &0.066  &0.030  &0.036  &0.139  &0.132  &0.023  &0.044  &0.178  &0.138\\

    \end{tabular}
    \caption{ Children relative fitness classification for Experiment GA-D.}
  \end{center}
\end{table}

\begin{table}
  \begin{center}
    \begin{tabular}{rrr | rrrr | rrrr}

\multicolumn{3}{c}{mutation} & \multicolumn{4}{|c}{crossover} & \multicolumn{4}{|c}{triangulation} \\
\multicolumn{3}{c}{ 17054(1.000) } & \multicolumn{4}{|c}{ 11015(1.000) } & \multicolumn{4}{|c}{ 12017(1.000) } \\
\RankD  &\RankB &\RankA         &\RankD &\RankC &\RankB &\RankA         &\RankD &\RankC &\RankB &\RankA  \\
\hline

13649   &2791   &606    &3992   &2849   &3601   &571    &3523   &3791   &4318   &395\\
(0.800  &0.164  &0.036) &(0.362 &0.259  &0.327  &0.052) &(0.293 &0.315  &0.359  &0.033)\\
217     &363    &51     &96     &97     &459    &78     &69     &161    &654    &56\\
(0.013  &0.021  &0.003) &(0.009 &0.009  &0.042  &0.007) &(0.006 &0.013  &0.054  &0.005)\\
0.016   &0.130  &0.084  &0.024  &0.034  &0.127  &0.137  &0.020  &0.042  &0.151  &0.142\\

    \end{tabular}
    \caption{ Children relative fitness classification for Experiment GA-4.}
  \end{center}
\end{table}

\begin{table}
  \begin{center}
    \begin{tabular}{rrr | rrrr | rrrr}

\multicolumn{3}{c}{mutation} & \multicolumn{4}{|c}{crossover} & \multicolumn{4}{|c}{triangulation} \\
\multicolumn{3}{c}{ 12093(1.000) } & \multicolumn{4}{|c}{ 4842(1.000) } & \multicolumn{4}{|c}{ 6119(1.000) } \\
\RankD  &\RankB &\RankA         &\RankD &\RankC &\RankB &\RankA         &\RankD &\RankC &\RankB &\RankA  \\
\hline

10569   &1200   &323    &1744   &1510   &1300   &288    &1592   &2528   &1742   &258\\
(0.874  &0.099  &0.027) &(0.360 &0.312  &0.268  &0.059) &(0.260 &0.413  &0.285  &0.042)\\
147     &191    &35     &43     &43     &216    &27     &33     &114    &323    &29\\
(0.012  &0.016  &0.003) &(0.009 &0.009  &0.045  &0.006) &(0.005 &0.019  &0.053  &0.005)\\
0.014   &0.159  &0.108  &0.025  &0.028  &0.166  &0.094  &0.021  &0.045  &0.185  &0.112\\

    \end{tabular}
    \caption{ Children relative fitness classification for Experiment GA-F.}
  \end{center}
\end{table}

\begin{table}
  \begin{center}
    \begin{tabular}{rrr | rrrr | rrrr}

\multicolumn{3}{c}{mutation} & \multicolumn{4}{|c}{crossover} & \multicolumn{4}{|c}{triangulation} \\
\multicolumn{3}{c}{ 13933(1.000) } & \multicolumn{4}{|c}{ 4151(1.000) } & \multicolumn{4}{|c}{ 6759(1.000) } \\
\RankD  &\RankB &\RankA         &\RankD &\RankC &\RankB &\RankA         &\RankD &\RankC &\RankB &\RankA  \\
\hline

12753   &875    &302    &1286   &1533   &980    &347    &1705   &3096   &1765   &201\\
(0.915  &0.063  &0.022) &(0.310 &0.370  &0.236  &0.084) &(0.252 &0.458  &0.261  &0.030)\\
203     &103    &26     &30     &68     &173    &43     &63     &198    &461    &45\\
(0.015  &0.007  &0.002) &(0.007 &0.016  &0.042  &0.010) &(0.009 &0.029  &0.068  &0.007)\\
0.016   &0.118  &0.086  &0.023  &0.044  &0.177  &0.124  &0.037  &0.064  &0.261  &0.224\\

    \end{tabular}
    \caption{ Children relative fitness classification for Experiment GA-F2.} \label{tableTombiEnd}
  \end{center}
\end{table}

\clearpage

\begin{landscape}
\begin{table}

\begin{tabular} {rl|l|l||r|r|r|r|r|r|r }
& $f_1(I)$&$f_2(I)$&$f_B(I)$&\multicolumn{1}{|c}{GA-1}&\multicolumn{1}{|c}{GA-S1}&\multicolumn{1}{|c}{GA-DE}&\multicolumn{1}{|c}{GA-D}&\multicolumn{1}{|c}{GA-4}&\multicolumn{1}{|c}{GA-F}&\multicolumn{1}{|c}{GA-F2}\\
\hline
 1.&$n(\Parent(I))=1$                       &$d(I)=0$            &True                & $1678.37\ominus$& $176.82\ominus$& $195.35\ominus$& $1101.14\ominus$& $228.43\ominus$& $233.27\ominus$& $646.90\ominus$\\
 2.&$n(\Parent(I))=2$                       &$d(I)=0$            &True                & $352.27\oplus$& $0.00\ominus$& $40.03\oplus$& $144.68\oplus$& $22.07\oplus$& $31.02\oplus$& $32.89\oplus$\\
 3.&$n(\Parent(I))=3$                       &$d(I)=0$            &True                & $736.92\oplus$& $215.63\oplus$& $92.78\oplus$& $656.16\oplus$& $136.57\oplus$& $145.75\oplus$& $552.01\oplus$\\
 4.&$I\in\mathRankA$                        &$d(I)=0$            &True                & $3086.23\oplus$& $193.47\oplus$& $11.85\oplus$& $172.65\oplus$& $109.81\oplus$& $50.53\oplus$& $97.78\oplus$\\ 
 5.&$I\in\mathRankB$                        &$d(I)=0$            &True                & $2384.30\oplus$& $1766.64\oplus$& $1429.42\oplus$& $3566.00\oplus$& $1745.43\oplus$& $1513.70\oplus$& $1698.94\oplus$\\
 6.&$I\in\mathRankC$                        &$d(I)=0$            &True                & $8.22\ominus$& $291.79\ominus$& $0.08\ominus$& $47.13\ominus$& $50.13\ominus$& $16.82\ominus$& $0.05\oplus$\\
 7.&$I\in\mathRankD$                        &$d(I)=0$            &True                & $6373.92\ominus$& $1482.62\ominus$& $1314.95\ominus$& $2233.94\ominus$& $1283.65\ominus$& $923.74\ominus$& $1162.42\ominus$\\
 8.&$n(\Parent(I))=2$                       &$d(I)=0$            &$n(\Parent(I))\geq2$& $0.50\ominus$& $62.38\ominus$& $1.06\ominus$& $29.02\ominus$& $12.05\ominus$& $7.15\ominus$& $40.75\ominus$\\
 9.&$n(\Parent(I))=3$                       &$d(I)=0$            &$n(\Parent(I))\geq2$& $0.50\oplus$& $62.38\oplus$& $1.06\oplus$& $29.02\oplus$& $12.05\oplus$& $7.15\oplus$& $40.75\oplus$\\
10.&$n(\Parent(I))=2$                       &$I\in\mathRankA$    &$I \in {\mathbb E}$ & $410.39\oplus$& $31.79\oplus$& $13.00\oplus$& $179.86\oplus$& $64.28\oplus$& $18.81\oplus$& $144.85\oplus$\\
11.&$n(\Parent(I))=3$                       &$I\in\mathRankA$    &$I \in {\mathbb E}$ & $410.39\ominus$& $31.79\ominus$& $13.00\ominus$& $179.86\ominus$& $64.28\ominus$& $18.81\ominus$& $144.85\ominus$\\
12.&$n(\Parent(I))=2$                       &$I\in\mathRankA$    &$d(I)=0$            & $69.73\oplus$& $17.64\ominus$& $3.72\oplus$& $37.14\oplus$& $10.15\oplus$& $0.26\oplus$& $17.27\oplus$\\
13.&$n(\Parent(I))=2$                       &$I\in\mathRankB$    &$d(I)=0$            & $177.77\ominus$& $0.11\oplus$& $1.20\ominus$& $0.03\oplus$& $0.72\ominus$& $4.60\oplus$& $1.43\oplus$\\
14.&$n(\Parent(I))=3$                       &$I\in\mathRankA$    &$d(I)=0$            & $340.94\ominus$& $19.05\ominus$& $2.49\ominus$& $23.71\ominus$& $9.29\ominus$& $3.77\ominus$& $10.90\ominus$\\
15.&$n(\Parent(I))=3$                       &$I\in\mathRankB$    &$d(I)=0$            & $368.68\oplus$& $22.80\oplus$& $7.69\oplus$& $100.03\oplus$& $20.56\oplus$& $5.72\oplus$& $42.80\oplus$\\
16.&$I\equiv7\ \mathrm{mod}\ 10$            &$d(I)=0$            &True                & $1.73\ominus$& $1.38\ominus$& $0.41\ominus$& $1.00\ominus$& $0.98\oplus$& $0.00\ominus$& $0.02\ominus$\\
17.&$I\equiv13\ \mathrm{mod}\ 100$          &$d(I)=0$            &True                & $1.05\oplus$& $0.69\oplus$& $0.06\oplus$& $0.08\ominus$& $0.43\ominus$& $0.08\oplus$& $1.11\oplus$\\
\end{tabular}

\caption{Chi-squared test of statistical independence of predicates.
  For each pair of experiment (columns) and three predicates $f_1(I),
  f_2(I), f_B(I)$, the table shows the $X^2$ statistics of the null
  hypothesis ``predicates $f_1(I)$ and $f_2(I)$ are statistically
  independent for the population of individuals that satisfy predicate
  $f_B(I)$.  '' Here, ${\mathbb E} \equiv \{I | n(\Parent(I))\geq2 \}
  \cap (\mathRankA \cup \mathRankB)$.  $\oplus$ denotes the positive correlation and $\ominus$
  denotes the negative correlation.}\label{tableIndependenceA}
\end{table}

\begin{table}

\begin{tabular} {rl|l|l||r|r|r }
&$f_1(I)$&$f_2(I)$&$f_B(I)$&\multicolumn{1}{|c}{GB-333-0}&\multicolumn{1}{|c}{GB-333-1}&\multicolumn{1}{|c}{GB-333-2}\\
\hline
 1.&$n(\Parent(I))=1$                       &$d(I)=0$            &True                & $189.24\ominus$& $309.78\ominus$& $374.03\ominus$\\
 2.&$n(\Parent(I))=2$                       &$d(I)=0$            &True                & $17.50\oplus$& $1.38\oplus$& $15.90\oplus$\\
 3.&$n(\Parent(I))=3$                       &$d(I)=0$            &True                & $107.73\oplus$& $284.28\oplus$& $255.24\oplus$\\
 4.&$I\in\mathRankA$                        &$d(I)=0$            &True                & $19.56\oplus$& $28.84\oplus$& $41.97\oplus$\\
 5.&$I\in\mathRankB$                        &$d(I)=0$            &True                & $455.57\oplus$& $517.15\oplus$& $1024.86\oplus$\\
 6.&$I\in\mathRankC$                        &$d(I)=0$            &True                & $0.05\ominus$& $1.67\ominus$& $8.49\ominus$\\
 7.&$I\in\mathRankD$                        &$d(I)=0$            &True                & $333.89\ominus$& $514.23\ominus$& $794.08\ominus$\\
 8.&$n(\Parent(I))=2$                       &$d(I)=0$            &$n(\Parent(I))\geq2$& $9.69\ominus$& $67.58\ominus$& $38.17\ominus$\\
 9.&$n(\Parent(I))=3$                       &$d(I)=0$            &$n(\Parent(I))\geq2$& $9.69\oplus$& $67.58\oplus$& $38.17\oplus$\\
10.&$n(\Parent(I))=2$                       &$I\in\mathRankA$    &$I \in {\mathbb E}$ & $207.54\oplus$& $70.91\oplus$& $138.99\oplus$\\
11.&$n(\Parent(I))=3$                       &$I\in\mathRankA$    &$I \in {\mathbb E}$ & $207.54\ominus$& $70.91\ominus$& $138.99\ominus$\\
12.&$n(\Parent(I))=2$                       &$I\in\mathRankA$    &$d(I)=0$            & $43.23\oplus$& $12.05\oplus$& $70.57\oplus$\\
13.&$n(\Parent(I))=2$                       &$I\in\mathRankB$    &$d(I)=0$            & $1.47\oplus$& $17.39\ominus$& $0.22\ominus$\\
14.&$n(\Parent(I))=3$                       &$I\in\mathRankA$    &$d(I)=0$            & $44.86\ominus$& $18.46\ominus$& $60.99\ominus$\\
15.&$n(\Parent(I))=3$                       &$I\in\mathRankB$    &$d(I)=0$            & $3.15\oplus$& $7.75\oplus$& $0.12\ominus$\\
16.&$I\equiv7\ \mathrm{mod}\ 10$            &$d(I)=0$            &True                & $3.77\ominus$& $0.47\oplus$& $2.74\oplus$\\
17.&$I\equiv13\ \mathrm{mod}\ 100$          &$d(I)=0$            &True                & $0.63\ominus$& $4.27\ominus$& $0.23\ominus$\\
\end{tabular}

  \caption{The same table for GB-333-*.}
\label{tableIndependenceB}
\end{table}
\end{landscape}

\section*{References}
\bibliography{mainbib}

\begin{thebibliography}{10}

\bibitem{frigo_design_2005}
M.~Frigo and {S.G.} Johnson.
\newblock The design and implementation of {FFTW3}.
\newblock {\em Proceedings of the {IEEE}}, 93(2):216–231, 2005.

\bibitem{clint_whaley_automated_2001}
R.~Clint~Whaley, A.~Petitet, and {J.J.} Dongarra.
\newblock Automated empirical optimizations of software and the {ATLAS}
  project.
\newblock {\em Parallel Computing}, 27(1-2):3–35, 2001.

\bibitem{puschel_spiral:_2005}
M.~Puschel, {J.M.F.} Moura, {J.R.} Johnson, D.~Padua, {M.M.} Veloso, {B.W.}
  Singer, J.~Xiong, F.~Franchetti, A.~Gacic, Y.~Voronenko, et~al.
\newblock {SPIRAL:} code generation for {DSP} transforms.
\newblock {\em Proceedings of the {IEEE}}, 93(2):232–275, 2005.

\bibitem{datta_stencil_2008}
Kaushik Datta, Mark Murphy, Vasily Volkov, Samuel Williams, Jonathan Carter,
  Leonid Oliker, David Patterson, John Shalf, and Katherine Yelick.
\newblock Stencil computation optimization and auto-tuning on state-of-the-art
  multicore architectures.
\newblock In {\em Proceedings of the 2008 {ACM/IEEE} conference on
  Supercomputing}, {SC} '08, page 4:1–4:12, Piscataway, {NJ}, {USA}, 2008.
  {IEEE} Press.

\bibitem{maruyama_physis:_2011}
Naoya Maruyama, Tatsuo Nomura, Kento Sato, and Satoshi Matsuoka.
\newblock Physis: an implicitly parallel programming model for stencil
  computations on large-scale {GPU-accelerated} supercomputers.
\newblock In {\em Proceedings of 2011 International Conference for High
  Performance Computing, Networking, Storage and Analysis}, {SC} '11, page
  11:1–11:12, New York, {NY}, {USA}, 2011. {ACM}.

\bibitem{trinder_parallel_2002}
{P.W.} Trinder, {H.W.} Loidl, and {R.F.} Pointon.
\newblock Parallel and distributed haskells.
\newblock {\em Journal of Functional Programming}, 12(5):469–510, 2002.

\bibitem{chakravarty_nepalnested_2001}
M.~Chakravarty, G.~Keller, R.~Lechtchinsky, and W.~Pfannenstiel.
\newblock Nepal—nested data parallelism in haskell.
\newblock {\em {Euro-Par} 2001 Parallel Processing}, page 524–534, 2001.

\bibitem{jones_harnessing_2008}
{S.P.} Jones, R.~Leshchinskiy, G.~Keller, and {M.M.T.} Chakravarty.
\newblock Harnessing the multicores: Nested data parallelism in haskell.
\newblock In {\em {FSTTCS}}, volume~2, page 383–414, 2008.

\bibitem{chakravarty_accelerating_2011}
{M.M.T.} Chakravarty, G.~Keller, S.~Lee, {T.L.} {McDonell}, and V.~Grover.
\newblock Accelerating haskell array codes with multicore {GPUs}.
\newblock In {\em Proceedings of the sixth workshop on Declarative aspects of
  multicore programming}, page 3–14, 2011.

\bibitem{mainland_nikola:_2010}
G.~Mainland and G.~Morrisett.
\newblock Nikola: embedding compiled {GPU} functions in haskell.
\newblock In {\em Proceedings of the third {ACM} Haskell symposium on Haskell},
  page 67–78, 2010.

\bibitem{devito_liszt:_2011}
Zachary {DeVito}, Niels Joubert, Francisco Palacios, Stephen Oakley, Montserrat
  Medina, Mike Barrientos, Erich Elsen, Frank Ham, Alex Aiken, Karthik
  Duraisamy, Eric Darve, Juan Alonso, and Pat Hanrahan.
\newblock Liszt: a domain specific language for building portable mesh-based
  {PDE} solvers.
\newblock In {\em Proceedings of 2011 International Conference for High
  Performance Computing, Networking, Storage and Analysis}, {SC} '11, page
  9:1–9:12, New York, {NY}, {USA}, 2011. {ACM}.

\bibitem{shimokawabe_80-fold_2010}
Takashi Shimokawabe, Takayuki Aoki, Chiashi Muroi, Junichi Ishida, Kohei
  Kawano, Toshio Endo, Akira Nukada, Naoya Maruyama, and Satoshi Matsuoka.
\newblock An {80-Fold} speedup, 15.0 {TFlops} full {GPU} acceleration of
  {Non-Hydrostatic} weather model {ASUCA} production code.
\newblock In {\em Proceedings of the 2010 {ACM/IEEE} International Conference
  for High Performance Computing, Networking, Storage and Analysis}, {SC} '10,
  page 1–11, Washington, {DC}, {USA}, 2010. {IEEE} Computer Society.

\bibitem{shimokawabe_peta-scale_2011}
Takashi Shimokawabe, Takayuki Aoki, Tomohiro Takaki, Toshio Endo, Akinori
  Yamanaka, Naoya Maruyama, Akira Nukada, and Satoshi Matsuoka.
\newblock Peta-scale phase-field simulation for dendritic solidification on the
  {TSUBAME} 2.0 supercomputer.
\newblock In {\em Proceedings of 2011 International Conference for High
  Performance Computing, Networking, Storage and Analysis}, {SC} '11, page
  3:1–3:11, New York, {NY}, {USA}, 2011. {ACM}.

\bibitem{hoshino_pax_1989}
T.~Hoshino, {H.S.} Stone, and S.~Goldman.
\newblock {\em {PAX} Computer; {High-Speed} Parallel Processing and Scientific
  Computing}.
\newblock {Addison-Wesley} Longman Publishing Co., Inc., 1989.

\bibitem{lee_implementation_2010}
J.~Lee and M.~Sato.
\newblock Implementation and performance evaluation of xcalablemp: A parallel
  programming language for distributed memory systems.
\newblock In {\em Parallel Processing Workshops {(ICPPW)}, 2010 39th
  International Conference on}, page 413–420, 2010.

\bibitem{mcbride_functional_2008}
C.~{McBride} and R.~Paterson.
\newblock Functional pearl: Applicative programming with effects.
\newblock {\em Journal of functional programming}, 18(1):1–13, 2008.

\bibitem{keller_regular_2010}
G.~Keller, {M.M.T.} Chakravarty, R.~Leshchinskiy, S.~Peyton~Jones, and
  B.~Lippmeier.
\newblock Regular, shape-polymorphic, parallel arrays in haskell.
\newblock {\em {ACM} {SIGPLAN} Notices}, 45(9):261–272, 2010.

\bibitem{gibbons_essence_2009}
J.~Gibbons and {B.C.S.} Oliveira.
\newblock The essence of the iterator pattern.
\newblock {\em Journal of Functional Programming}, 19(3-4):377–402, 2009.

\bibitem{erwig_functional_1997}
M.~Erwig.
\newblock Functional programming with graphs.
\newblock In {\em {ACM} {SIGPLAN} Notices}, volume~32, page 52–65, 1997.

\bibitem{erwig_inductive_2001}
M.~Erwig.
\newblock Inductive graphs and functional graph algorithms.
\newblock {\em Journal of Functional Programming}, 11(05):467–492, 2001.

\bibitem{sheard_template_2002}
T.~Sheard and {S.P.} Jones.
\newblock Template meta-programming for haskell.
\newblock In {\em Proceedings of the 2002 {ACM} {SIGPLAN} workshop on Haskell},
  page 1–16, 2002.

\bibitem{thurston_numeric-prelude:_2012}
Dylan Thurston, Henning Thielemann, and Mikael Johansson.
\newblock numeric-prelude: An experimental alternative hierarchy of numeric
  type classes.
\newblock {\em http://hackage.haskell.org/package/numeric-prelude}, 2012.

\bibitem{lammel_scrap_2003}
Ralf Lämmel and Simon~Peyton Jones.
\newblock Scrap your boilerplate: a practical design pattern for generic
  programming.
\newblock {\em {SIGPLAN} Not.}, 38(3):26–37, January 2003.

\bibitem{hukushima_exchange_1996}
Koji Hukushima and Koji Nemoto.
\newblock Exchange monte carlo method and application to spin glass
  simulations.
\newblock {\em Journal of the Physical Society of Japan}, 65:1604--1608, 1996.

\bibitem{gutierrez_evolutionary_2012}
José A.~García Gutiérrez, Carlos Cotta, and Antonio~J {Fernández-Leiva}.
\newblock Evolutionary computation in astronomy and astrophysics: A review.
\newblock {\em {arXiv:1202.2523}}, February 2012.

\bibitem{eiben_genetic_1994}
A.~Eiben, P.~Raué, and Z.~Ruttkay.
\newblock Genetic algorithms with multi-parent recombination.
\newblock {\em Parallel Problem Solving from {Nature—PPSN} {III}}, page
  78–87, 1994.

\bibitem{sivanandam_introduction_2007}
{SN} Sivanandam and {SN} Deepa.
\newblock {\em Introduction to genetic algorithms}.
\newblock Springer Verlag, 2007.

\bibitem{eiben_orgy_1995}
A.~Eiben, C.~Van~Kemenade, and J.~Kok.
\newblock Orgy in the computer: Multi-parent reproduction in genetic
  algorithms.
\newblock {\em Advances in Artificial Life}, page 934–945, 1995.

\bibitem{tsutsui_multi-parent_1999}
S.~Tsutsui, M.~Yamamura, and T.~Higuchi.
\newblock Multi-parent recombination with simplex crossover in real coded
  genetic algorithms.
\newblock In {\em Proceedings of the Genetic and Evolutionary Computation
  Conference}, volume~1, page 657–664, 1999.

\bibitem{toro_riemann_2009}
Eleuterio~F. Toro.
\newblock {\em Riemann Solvers And Numerical Methods for Fluid Dynamics: A
  Practical Introduction}.
\newblock {Springer-Verlag}, 3 edition, June 2009.

\bibitem{sweby_high_1984}
{P.K.} Sweby.
\newblock High resolution schemes using flux limiters for hyperbolic
  conservation laws.
\newblock {\em {SIAM} journal on numerical analysis}, page 995–1011, 1984.

\bibitem{toro_restoration_1994}
E.~F. Toro, M.~Spruce, and W.~Speares.
\newblock Restoration of the contact surface in the {HLL-Riemann} solver.
\newblock {\em Shock Waves}, 4(1):25--34, July 1994.

\bibitem{schive_directionally_2011}
{H.Y.} Schive, {U.H.} Zhang, and T.~Chiueh.
\newblock Directionally unsplit hydrodynamic schemes with hybrid
  {MPI/OpenMP/GPU} parallelization in {AMR}.
\newblock {\em Arxiv preprint {arXiv:1103.3373}}, 2011.

\bibitem{de_la_asunciona_efficient_????}
M.~de~la Asuncióna, {M.J.} Castroc, {ED} {Fernández-Nietob}, {J.M.} Mantasa,
  {S.O.} Acostac, and {J.M.} {González-Vidad}.
\newblock Efficient {GPU} implementation of a two waves {WAF} method for the
  two-dimensional one layer shallow water system on structured meshes.

\bibitem{charles_x10:_2005}
Philippe Charles, Christian Grothoff, Vijay Saraswat, Christopher Donawa, Allan
  Kielstra, Kemal Ebcioglu, Christoph von Praun, and Vivek Sarkar.
\newblock X10: an object-oriented approach to non-uniform cluster computing.
\newblock {\em {SIGPLAN} Not.}, 40(10):519–538, October 2005.

\bibitem{chamberlain_parallel_2007}
B.l. Chamberlain, D.~Callahan, and H.p. Zima.
\newblock Parallel programmability and the chapel language.
\newblock {\em International Journal of High Performance Computing
  Applications}, 21(3):291 --312, 2007.

\bibitem{nilsen_simplifying_2010}
Jon~K Nilsen, Xing Cai, Bjørn Høyland, and Hans~Petter Langtangen.
\newblock Simplifying the parallelization of scientific codes by a
  function-centric approach in python.
\newblock {\em Computational Science \& Discovery}, 3(1):015003, September
  2010.

\end{thebibliography}

\end{document}